\documentclass[%
 reprint,
superscriptaddress,
nofootinbib,
 aps,
]{revtex4-2}
\usepackage[centertags]{amsmath}
\usepackage{amssymb}
\usepackage{latexsym}
\usepackage{enumerate}
\usepackage{graphicx}
\usepackage{mathrsfs}
\usepackage{ragged2e}
\usepackage{stmaryrd}
\usepackage{import}
\usepackage{tensor}
\usepackage{xcolor}
\usepackage{hyperref}
\usepackage{bm}
\usepackage{multirow}
\usepackage{breakurl}
\usepackage{subcaption}
\usepackage{float}
\usepackage{lipsum}
\usepackage{siunitx}
\usepackage{comment}
\usepackage{mathtools}
\usepackage{empheq}

\usepackage{autobreak}
\usepackage{orcidlink}
\mathtoolsset{centercolon}
\usepackage{comment}
\usepackage[normalem]{ulem}

\newcommand{\xmin}{x_\text{min}}
\newcommand{\xp}{x_+}
\newcommand{\xm}{x_-}

\newcommand{\dd}{\mathrm{d}}
\newcommand{\de}{\mathrm{e}}
\newcommand{\di}{\mathrm{i}}
\newcommand{\ii}{\mathrm{i}}

\allowdisplaybreaks
\begin{document}

\preprint{APS/123-QED}

\title{Towards ultrarelativistic scattering near the light ring: \\asymptotic divergence of gravitational-wave fluxes}

\author{David Trestini\,\orcidlink{0000-0002-4140-0591}}
\email{trestini@iap.fr}
\affiliation{GReCO, Institut d’Astrophysique de Paris, UMR 7095, CNRS \& Sorbonne Université, 98bis boulevard Arago, 75014 Paris, France}
\affiliation{Centre national d’études spatiales (CNES), 
2, Place Maurice Quentin, 75039, Paris, France}
\affiliation{School of Mathematical Sciences and STAG Research Centre, University of Southampton, Southampton SO17 1BJ, United Kingdom}

\author{Riccardo Gonzo\,\orcidlink{0000-0001-7285-6295}}
\email{r.gonzo@qmul.ac.uk}
\affiliation{School of Mathematical Sciences and STAG Research Centre, University of Southampton, Southampton SO17 1BJ, United Kingdom}
\affiliation{Queen Mary University of London, Mile End Road, London E1 4NS,
United Kingdom}

\author{Benjamin Leather\,\orcidlink{0000-0001-6186-7271}}
\email{bjl1u25@soton.ac.uk}
\affiliation{School of Mathematical Sciences and STAG Research Centre, University of Southampton, Southampton SO17 1BJ, United Kingdom}

\date{\today} 

\begin{abstract}
Near the threshold between scattering and capture by a Schwarzschild black hole, a particle spends a long time close to an unstable circular orbit, which in the ultrarelativistic limit approaches the light ring. Here we study the behavior of the gravitational wave fluxes emitted by such an ultrarelativistic particle at first order in the mass ratio.
The fluxes radiated to infinity and absorbed by the horizon diverge as the inverse distance from the light ring multiplied by its logarithm. 
Using the Wentzel, Kramers, and Brillouin~(WKB) approximation to solve the Regge-Wheeler equation 
and treating the infinite harmonic sum analytically, we derive an exact expression for the coefficient of this logarithmic enhancement and show that it is identical for both fluxes. We then estimate the coefficients of the subleading pole contributions by combining numerical calculations at low multipoles with the exact asymptotics at large multipoles. 
We further conjecture an extension to equatorial circular orbits in Kerr.  
Interestingly, our results also inform the analytical structure of the energy and angular momentum losses of ultrarelativistic scattering orbits near the light ring. 
More broadly, our work illustrates how WKB methods can provide the asymptotic behavior of self-force observables in regimes where numerical mode sums become increasingly challenging.
\end{abstract}
\maketitle

\section{Introduction}
\label{sec:intro}

The radiation emitted by a particle orbiting a black hole near the light ring attracted a burst of interest in the early 1970s, when it was known as `gravitational synchrotron radiation'~\cite{Misner:1972jf,Davis:1972dm,Bardeen:1972fi,Chitre:1972fv,Breuer:1972xs,Breuer:1973kt,Breuer:1973kp,Breuer:1973ktj,Chrzanowski:1974nr,Breuer:1975}.  It was named this way because the angular distribution of the gravitational waves emitted by such a system was sharply beamed around the equatorial plane, similarly to the case of electromagnetic synchrotron radiation~\cite{Schwinger:1949,Jackson:1999}. The motivation for studying this type of source was Joseph Weber's claim that he had detected gravitational waves using resonant bars~\cite{Weber:1966zz,Weber:1969bz} with strains of the order of $\sim 10^{-16}$ (to be compared with the maximum strains of $\sim 10^{-21}$ detected by LIGO~\cite{LIGOScientific:2016aoc}). To explain such a large signal, theorists scrambled for a credible astrophysical source. Misner \textit{et al.}~\cite{Misner:1972jf} were the first in 1972 to propose that a highly relativistic particle orbiting a black hole near the light ring might provide such a source, noticing that the flux was in fact a formally \textit{divergent} quantity in the test particle limit. Of course it was known that this divergence was an artifact of the small mass ratio limit, but this raised the possibility of a large emitted flux at finite mass ratio. That same year, several groups rushed to assess whether this was a credible possibility. On~March~10,~1972, Davis, Ruffini, Tiomno and Zerilli submitted a letter titled `Can Synchrotron Gravitational Radiation Exist?', emphasizing that such an effect was restricted to unstable circular orbits~\cite{Davis:1972dm}. On~April~3, Chitre~and~Price submitted a letter titled  `Nature of Gravitational Synchrotron Radiation'~\cite{Chitre:1972fv}. On June~5, Bardeen, Press and Teukolsky submitted a paper concluding that they were `pessimistic about the applicability of Misner's interesting synchrotron concept to any realistic astrophysical model'~\cite{Bardeen:1972fi}. Campbell and Matzner also investigated the problem~\cite{Campbell:1973}, while Khriplovich and Shuryak considered ultrarelativistic radiation from both circular and unbound scattering trajectories~\cite{Khriplovich:1974}.

Initially, these works were rather qualitative~\cite{Misner:1972jf} and focused on scalar radiation, which offered a simpler setting in which to study the beaming and high frequency spectrum~\cite{Breuer:1972xs,Breuer:1973ktj,Bardeen:1972fi}. Later, the gravitational case was explored more in detail, first in a Schwarzschild~\cite{Davis:1972dm, Breuer:1973kt} and then in a Kerr background~\cite{Bardeen:1972fi,Chrzanowski:1974nr,Breuer:1975}. However, as independent resonant bar experiments failed to reproduce Weber's reported events~\cite{Tyson:1973ra,Drever:1973,Douglass:1975,Braginsky:1972,Billing:1975,Bonazzola:1974}, the problem of gravitational synchrotron radiation largely fell into oblivion.

About a decade later, WKB methods reappeared in black hole perturbation theory, although in a different application: the calculation of quasinormal modes. In~1984 Ferrari and Mashhoon related quasinormal modes to bound states of the inverted black hole potential~\cite{Ferrari:1984zz} and, the following year, Schutz and Will applied the WKB approximation directly to the Regge-Wheeler equation~\cite{Schutz:1985km}. Interestingly, however, the authors were seemingly unaware of the connection with the earlier literature on synchrotron radiation. A major achievement was the development of a systematic higher-order WKB expansion by Iyer and Will~\cite{Iyer:1986np}, with applications to Schwarzschild~\cite{Iyer:1986nq,Guinn:1989bn}, Reissner-Nordström~\cite{Kokkotas:1988fm} and Kerr black holes~\cite{Seidel:1989bp}. This extended the analysis beyond the leading `geometric optics' treatment of the earlier synchrotron calculations,  which required a higher-order expansion around the peak of the potential barrier~\cite{Will:1988zz,Ford:1959}. Later large-multipole expansions made the relation between quasinormal modes and unstable null geodesics explicit~\cite{Cardoso:2008bp,Dolan:2009nk}. The usefulness of WKB methods was noticed again by Bini and Damour~\cite{Bini:2015mza}, who applied related techniques to derive self-force regularization parameters from the radial and angular perturbation equations. More recently, these methods have also been extended to the calculation of quasinormal excitation factors~\cite{Chen:2024hum}.

The logarithmic enhancement of the radiative flux near the light ring has also received renewed attention~\cite{Taracchini:2013wfa,Barausse:2021xes}. Introducing the black hole mass $M$ and the Schwarzschild radius $r_0$ of the circular orbit (with $c=G=1$), modern waveform resummations~\cite{Damour:2007yf,Damour:2007xr,Damour:2008gu,Cipriani:2026xmx} factor out from the waveform amplitudes the particle's energy or angular momentum, exploiting their $(r_0/M-3)^{-1/2}$ divergence as the orbit approaches the light ring for $r_0 \rightarrow 3 M$. Since the flux is quadratic in the amplitude, this factorization captures the $(r_0/M-3)^{-1}$ divergence of each mode of the flux. Summing over modes introduces a further logarithmic enhancement $\sim -\ln(r_0/M-3)/(r_0/M-3)$, however, as increasingly large multipoles contribute when $r_0\to3 M$. More precise numerical evidence for this behavior was obtained in Ref.~\cite{Barack:2026izc}, where calculations of the modes up to $\ell=500$ allowed the fluxes to be evaluated for orbits as close to the light ring as $r_0=3.00576 M$.  

The scattering problem gives a further reason to understand this divergence. Scattering amplitudes and worldline methods have led to analytical calculations of the radiated energy and angular momentum for binary systems in the post-Minkowskian expansion both at infinity~\cite{Kosower:2018adc,Damour:2020tta,Herrmann:2021tct,Mougiakakos:2021ckm,Riva:2021vnj,Jakobsen:2021smu,Manohar:2022dea,DiVecchia:2022piu,Dlapa:2022lmu,Jakobsen:2023hig,Damgaard:2023ttc,Driesse:2024feo,Heissenberg:2024umh,Heissenberg:2025ocy,Jakobsen:2026ayw} and at the horizon~\cite{Goldberger:2020wbx,Jones:2023ugm,Bautista:2024emt,Cipriani:2026myb}. Complementary results for these observables in combined post-Newtonian and post-Minkowskian expansions have also been obtained using traditional multipolar methods~\cite{Cho:2021onr,Bini:2021gat,Bini:2021qvf,Cho:2022pqy,Bini:2022enm} and the Mano-Suzuki-Takasugi approach~\cite{Mano:1996vt,Mano:1996gn} to the Teukolsky equation~\cite{Geralico:2025rof,Geralico:2026kbm,Geralico:2026efi}. As more effort is ongoing to build effective field theories for the self-force expansion~\cite{Adamo:2023cfp,Kosmopoulos:2023bwc,Cheung:2023lnj,Barack:2023oqp,Cheung:2024byb,Akpinar:2025huz}, novel particle physics inspired approaches have been developed to revisit black hole perturbation theory~\cite{Bautista:2021wfy,Bautista:2022wjf,Ivanov:2024sds,Correia:2024jgr,Caron-Huot:2025tlq,Bjerrum-Bohr:2025bqg,Ivanov:2025ozg,Bautista:2026qse,Ivanov:2026icp,Chang:2026eti,Correia:2026utp,Bjerrum-Bohr:2026fhs,Bautista:2026fcp,Brunello:2026rdk,Brunello:2026lzf}.

\begin{figure}
    \centering
    \includegraphics[width=0.9\linewidth]{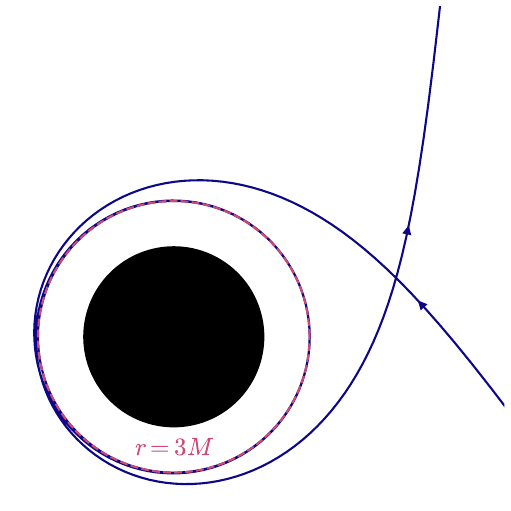}
    \caption{\justifying The black line illustrates an ultrarelativistic scattering orbit with $E_{\textrm{p}}/\mu\simeq57.74$ and closest approach $r_{\min}=3.00011M$. The dashed red circle denotes the light ring at $r_0=3M$ while the black disk marks the region bounded by the event horizon at $r=2M$.}
    \label{fig:scattering_orbit}
\end{figure}

Our interest here is in ultrarelativistic scattering in the small mass ratio regime\footnote{This differs from the ultrarelativistic Regge (or forward scattering) limit, where the center-of-mass energy is large compared with both masses and the scattering angle is small. For related developments in Regge and shockwave approaches to high-energy gravitational scattering and radiation, see Refs.~\cite{Raj:2023irr,Rothstein:2024nlq,Alessio:2026bdi,Bohnenblust:2026ujk,Saavedra:2026ttk,Blackstad:2026eoa,Rosi:2026zjp,Damgaard:2026zqj,Pietrini:2026mcg}.}. Denoting the particle's rest mass by $\mu$ and its incoming energy in the black hole rest frame by $E_{\textrm{p}}$, the ultrarelativistic scattering regime is characterized by the following hierarchy:
\begin{equation}
\mu\ll E_{\textrm{p}}\ll M.
\label{eq:power_counting}
\end{equation}
As the particle approaches the separatrix between scattering and plunging orbits, it exhibits a `zoom-whirl' behavior~\cite{Glampedakis:2002ya,Healy:2009zm}. This is characterized by three phases: (i) the approach from spatial infinity towards the vicinity of the black hole; (ii) a certain number of near-circular orbits around the black hole; and (iii) escape from the black hole back to spatial infinity---see Fig.~\ref{fig:scattering_orbit}.
Near the capture threshold, the particle spends an increasingly long time close to the critical unstable circular orbit of radius $r_0$, with $3M<r_0<4M$. The duration of the whirl grows logarithmically as the closest approach tends to this radius, $r_{\min}\to r_0^+$, or equivalently as the incoming angular momentum approaches its capture threshold value, $L_{\rm p}\to L_{\rm crit}^+(E_{\textrm{p}})$~\cite{Barack:2026izc}---see Eq.~\eqref{eq:Lcrit_of_E} for the explicit expression. 
At leading order in such an expansion, the divergent parts of the energy and angular momentum losses are therefore determined by the corresponding fluxes from the circular orbit. Moreover, in the ultrarelativistic limit of Eq.~\eqref{eq:power_counting}, the critical orbit approaches the light ring at $r_0=3M$, where these fluxes themselves diverge.

Our work thus aims at pushing the `gravitational synchrotron radiation' approach a step further: for the first time, we provide precise statements about the asymptotics of the fluxes near the light ring. Introducing the mass ratio $\nu\equiv\mu/M$, it is natural to normalize the physical energy fluxes by $\mathcal F^X\equiv\mathcal F_{\mathrm{phys}}^X/\nu^2$.
Then, defining $\delta=r_0/M-3$, we find that the energy fluxes emitted at infinity~($\infty$) and down the horizon~($\mathcal{H}$) behave like
\begin{align}\label{eq:F_infty_H_total_alpha_beta}
   \hspace{-5pt}\mathcal{F}^{\infty/\mathcal{H}} = \frac{\alpha\ln(1/\delta)+\beta^{\infty/\mathcal H}}{\delta}
    +o\left(1/\delta\right)
\end{align}
as $\delta\to0^+$. We find that the logarithmic divergence is entirely controlled by the WKB calculation, can be written in an analytical closed form---see Eq.~\eqref{eq:alpha_k_formula}---and reproduces the numerical estimate of $\alpha$ found in Ref.~\cite{Barausse:2021xes}. We also confirm that this term is universal: it is the same coefficient for the flux to infinity and down the horizon, which takes the value $\alpha \approx 0.021217676364$. Combining numerical results for the low-$\ell$ modes exactly on the light ring (provided in the ancillary file \texttt{data\_light\_ring.csv}) and analytical WKB results for the high-$\ell$ modes, we also compute numerical values for the subleading pole contribution, which differ between the infinity and horizon fluxes and are \mbox{$\beta^\infty \approx  0.0009117(6)$} and \mbox{$\beta^\mathcal{H} \approx -0.0352830(5)$}. This prediction agrees very well with our numerical data for unstable circular orbits ($3<r_0\le6$) which we provide in the ancillary file \texttt{flux\_UCO.csv}---see Fig.~\ref{fig:comparison-error-total-flux}.

Additionally, these fluxes determine the divergent parts of the scattering losses, both at infinity and at the horizon, near the separatrix between scattering and capture. At leading order in the mass ratio, we obtain for the total energy and angular momentum losses~\cite{Barack:2026izc} 
\begin{subequations}\begin{align}
    \Delta E^{\infty/\mathcal{H}}
    &=
    \frac{\nu^2 r_0^2 \mathcal F^{\infty/\mathcal{H}}(r_0)}{\sqrt{M(6M-r_0)}}
    \ln\left(\frac{L_{\rm crit}(r_0)}{L_{\rm p}-L_{\rm crit}(r_0)}\right)+\mathcal{O}(1), 
    \\
    \Delta L^{\infty/\mathcal{H}}
    &=
    \sqrt{\frac{r_0^3}{M}}\,\Delta E^{\infty/\mathcal{H}}+\mathcal{O}(1),
\end{align}\end{subequations}
where the $\mathcal{O}(1)$ is to be understood as a bounded remainder 
as $L_{\rm p}\to L_{\rm crit}^+$ at fixed $E_{\textrm{p}}$. Our main result for $\mathcal F^{\infty/\mathcal{H}}(r_0)$ in Eq.~\eqref{eq:F_infty_H_total_alpha_beta} provides the leading terms in the ultrarelativistic limit~\eqref{eq:power_counting}: a common $\delta^{-1}\ln(1/\delta)$ term determined analytically by $\alpha$, and distinct subleading $\delta^{-1}$ terms set by $\beta^\infty$ and $\beta^{\mathcal H}$. Note, however, that the $\mathcal{O}(1)$ remainder is in general not bounded uniformly in $\delta$ and could also diverge in the $\delta \rightarrow 0$ limit. We leave a full analysis of the ultrarelativistic limit of scattering losses outside the separatrix limit for future work.

\section{Energy fluxes}
\label{sec:flux_formalism}

We consider a point particle of rest mass $\mu$ orbiting a Schwarzschild black hole of mass $M$ and work at leading order in the mass ratio $\nu \equiv \mu / M$. For convenience, we henceforth work in units where $M=1$. 
At this order the particle follows a geodesic of the background spacetime and sources a linear metric perturbation, which carries energy and angular momentum to future null infinity and through the future event horizon.
We take the orbit to be circular, lying in the equatorial plane at radius $r_{0} = 3 + \delta$, with angular frequency in Schwarzschild time $t$ given by
\begin{equation}\label{eq:omega0_def}
    \omega_0=r_0^{-3/2}=(3+\delta)^{-3/2}.
\end{equation}
Its energy and angular momentum per unit rest mass are
\begin{subequations}
    \begin{align}
    {\cal E} &= \frac{1 - 2/r_{0}}{\sqrt{1 - 3/r_{0}}} = f_{0}\sqrt{\frac{3 + \delta}{\delta}},\\
    {\cal L} &= \frac{\sqrt{r_{0}}}{\sqrt{1 - 3/r_{0}}} = \frac{3 + \delta}{\sqrt{\delta}},
\end{align}
\end{subequations}
where $f_{0} = 1 - 2/r_{0}$.
Circular geodesics exist for all $r_{0} > 3$; they are unstable for $r_{0} < 6$ and are unbound (${\cal E} > 1$) for $r_{0} < 4$. As shown in the Introduction, the unbound unstable circular orbits are approached by scattering orbits near the separatrix.

Indeed, a scattering orbit conserves its incoming energy, $E_{\textrm{p}}$, and angular momentum, $L_{\rm p}$, along the geodesic. On the separatrix, $L_{\rm p} = L_{\rm crit}(E_{\textrm{p}})$, the orbit asymptotes to the unstable circular orbit that carries the same values. 
We thus identify
\begin{equation}\label{eq:Ep_Lcrit}
    E_{\textrm{p}} = \mu\,{\cal E}(r_{0}),
    \qquad
    L_{\rm crit}(E_{\textrm{p}}) = \mu\,{\cal L}(r_{0}).
\end{equation}
The first relation fixes the radius of the critical orbit in terms of incoming energy.
Inverting Eq.~\eqref{eq:Ep_Lcrit}, we find
\iffalse
\begin{equation}\label{eq:r0_of_E}
    r_{0} = \frac{3{\cal E}^{2} - 4 + {\cal E}\sqrt{9{\cal E}^{2} - 8}}{2\left({\cal E}^{2} - 1\right)},
\end{equation}
\fi
\begin{equation}\label{eq:r0_of_E}
    r_0=
    \frac{2\left(3\mathcal E+\sqrt{9\mathcal E^2-8}\right)}
         {\mathcal E+\sqrt{9\mathcal E^2-8}},
\end{equation}
which decreases monotonically from $r_{0} = 4$ for ${\cal E} = 1$ to $r_{0} = 3$ as ${\cal E} \rightarrow \infty$. Eq.~\eqref{eq:Ep_Lcrit} then gives $L_{\rm crit}$ as a function of $E_{\textrm{p}}$: substituting Eq.~\eqref{eq:r0_of_E} into \mbox{${\cal L}^{2} = r_{0}^{2}/(r_{0} - 3)$} yields  
\begin{equation}\label{eq:Lcrit_of_E}
    \frac{L_{\rm crit}^2}{\mu^2}
    =
    \frac{\left(3\mathcal E+\sqrt{9\mathcal E^2-8}\right)^3}
         {2\left(\mathcal E+\sqrt{9\mathcal E^2-8}\right)},
\end{equation}
\iffalse
\begin{equation}\label{eq:Lcrit_of_E}
    \frac{L_{\rm crit}^{2}}{\mu^{2}} = \frac{27{\cal E}^{4} - 36{\cal E}^{2} + 8 + {\cal E}\left(9{\cal E}^{2} - 8\right)^{3/2}}{2\left({\cal E}^{2} - 1\right)},
\end{equation}
\fi
which increases monotonically from $L_{\rm crit} = 4\mu$ for a particle released from rest at infinity, ${\cal E} = 1$.
As $\delta \rightarrow 0^{+}$, the orbit tends to the null circular geodesic at the light ring and both ${\cal E}$ and ${\cal L}$ diverge like $\delta^{-1/2}$, with
\begin{equation}\label{eq:delta_of_Ep}
    \delta \simeq \frac{1}{3{\cal E}^{2}} = \frac{\mu^{2}}{3E_{\textrm{p}}^{2}},
    \qquad
    L_{\rm crit} \simeq 3\sqrt{3}\,E_{\textrm{p}}.
\end{equation}
The ultrarelativistic condition of Eq.~\eqref{eq:power_counting} is therefore equivalent to $\delta \ll 1$, whilst the ratio ${\cal L}/{\cal E} = L_{\rm crit}/E_{\textrm{p}}$ tends to the critical impact parameter of a photon, $3\sqrt{3}$.

The divergence in Eq.~\eqref{eq:delta_of_Ep} already anticipates the behavior of the fluxes.
The stress energy tensor of the particle is proportional to $\mu u^{t}$, where the time component of the four velocity, $u^{t} = \sqrt{(3 + \delta)/\delta}$, such that the perturbation scales as $\sim \mu \delta^{-1/2}$ and each multipolar contribution scales as $\sim \mu^{2} / \delta$.
Equivalently, since $\delta = f^{2}_{0} (3 + \delta)/{\cal E}^{2} \simeq (3{\cal E}^{2})^{-1}$ near the light ring, the fluxes are finite when measured in units of the squared energy of the particle, $E_{\textrm{p}}^{2} = \mu^{2}{\cal E}^{2}$, rather than of its squared rest mass.
This is the natural normalization in the ultrarelativistic regime \eqref{eq:power_counting}, in which the rest mass plays no role.
The $1/\delta$ pole in Eq.~\eqref{eq:F_infty_H_total_alpha_beta} is therefore kinematical and is present mode by mode, whereas the additional logarithm can only arise from the sum over modes.

\subsection{Fluxes in the Teukolsky formalism}

For a circular orbit, the perturbation is invariant along the helical Killing vector $k^{\alpha}\partial_{\alpha} = \partial_{t} + \omega_{0}\partial_{\varphi}$: it depends on $t$ and $\varphi$ only through the combination $\varphi - \omega_{0}t$.
After decomposing it into spherical harmonics $(\ell, m)$, each mode thus oscillates at the frequency $\omega = m \omega_{0}$, and the problem reduces to a set of ordinary differential equations in the radial coordinate, which we treat in the frequency domain.
We describe the radiation with the Weyl scalar, $\psi_{4}$, which satisfies the Teukolsky equation with spin weight $s = -2$ \cite{Bardeen:1972fi, Teukolsky:1973ha}.
Its asymptotic behavior at infinity directly gives the outgoing flux of energy, while its behavior at the horizon gives the flux absorbed by the black hole through the Teukolsky-Starobinsky identities \cite{Teukolsky:1974yv}; see Refs.~\cite{Poisson:1993vp, Poisson:1994yf, Sasaki:2003xr} for details.
Both fluxes are entirely determined by two complex amplitudes per mode, $Z^{\rm in}_{\ell m}$ and $Z^{\rm up}_{\ell m}$, which are obtained by integrating the source against the homogeneous solutions of the radial equation.
Since the source is supported at $r = r_{0}$ only, this integral is trivial and the amplitudes reduce to the homogeneous solutions and their derivatives evaluated at the location of the particle.
This allows us to obtain closed-form expressions for the fluxes once the homogeneous solutions are known analytically.

A few simplifications are specific to circular orbits.
First, the modes with $m = 0$ are static and do not radiate, and the modes with $m < 0$ contribute the same as those with $m > 0$, so that we only need to sum over positive values of $m$. Second, the fluxes are time independent and no orbit average is required.
Finally, the helical symmetry implies that the fluxes of angular momentum are related mode by mode to the fluxes of energy by a factor $m /\omega = 1/\omega_{0}$, so the total fluxes of angular momentum are simply ${\cal F}^{\infty/{\cal H}}/\omega_{0}$, with $\omega_{0} \rightarrow 1/(3\sqrt{3})$ at the light ring.
We therefore restrict our attention to the fluxes of energy, defined with respect to coordinate time, $t$.

With the rescaling by $\nu^{2}$ introduced in Sec.~\ref{sec:intro}, the total fluxes of energy through infinity and the horizon are 
\begin{align}
    \label{eq:F_of_F_lm}
    \mathcal{F}^\infty  = \sum_{\ell=2}^\infty\sum_{m=1}^\ell\mathcal{F}_{\ell m}^\infty&&\text{and}&&\mathcal{F}^{\mathcal{H}} = \sum_{\ell=2}^\infty\sum_{m=1}^\ell\mathcal{F}_{\ell m}^{\mathcal{H}},
\end{align}
where
\begin{align}
    \label{eq:Flm_of_Zlm}
   \hspace{-5pt}\mathcal{F}^{\infty}_{\ell m} =    \frac{|Z_{\ell m}^{\text{in}}|^2}{2\pi (m\omega_0)^2}&&\text{and}&&\mathcal{F}^\mathcal{H}_{\ell m} = \frac{  |\chi_{\ell m}|^2 |Z_{\ell m}^{\text{up}}|^2}{2\pi (m\omega_0)^2}.
\end{align}
The sums over $m$ are restricted to positive values, the contribution of $m < 0$ being already included in Eq.~\eqref{eq:Flm_of_Zlm}.
The factor $\chi_{\ell m}$ in the horizon flux arises from the Teukolsky-Starobinsky identities and reads
\begin{align}\label{eq:chi_lm}
    \chi_{\ell m} = - \frac{2^{6} (m \omega_0)^2 (1-2\di m \omega_0)(1- 4\di m \omega_0)}{c_0^{\ell m}} \,,
\end{align}
where $c_0^{\ell m}=(\ell-1)\ell(\ell+1)(\ell+2)-12 \di m \omega_0$ is the Teukolsky-Starobinsky constant.

The amplitudes $Z^{\rm in/up}_{\ell m}$ are constructed from the homogeneous solutions of the radial Teukolsky equation, which in the Schwarzschild case reduces to
\begin{multline}
\label{eq:Teukolsky_equations}
   \Big[r^2f(r) \frac{\dd^2}{\dd r^2} -2(r-1)\frac{\dd}{\dd r}
    +\frac{(\omega r)^2-4\di \omega(r-3)}{f(r)} 
     \\*
    - (\ell-1)(\ell+2)\Big]R_{\ell m}(r;\omega) =0 \,.
\end{multline}
Here $f(r)=1-2/r$, and $r_*=r+2\ln(r/2-1)$ denotes the tortoise coordinate, such that $\dd r/\dd r_*=f(r)$ and $f'(r)=2/r^2$.
A convenient basis of solutions is formed by the \emph{in} solution, which is purely ingoing at the horizon, and the \emph{up} solution, which is purely outgoing at infinity.
Their normalization is fixed by setting the amplitude of the transmitted wave to unity:
\begin{subequations}
\label{eq:teukolsky_bcs}
\begin{align}
    &\hspace{-3pt}R^{\rm in}_{\ell m}(r;\omega)\sim
    \begin{cases}
        r^4f^2(r)\,\de^{-\ii\omega r_*} & \text{as } r_*\to-\infty,\\[0.5ex]
        \begin{aligned}[t]&Q^{\rm in}_{\ell m}\,r^{-1}\de^{-\ii\omega r_*}\\&\quad+Q^{\rm out}_{\ell m}\,r^{3}\de^{+\ii\omega r_*}\end{aligned} & \text{as } r_*\to+\infty,
    \end{cases}
    \\
    &\hspace{-3pt}R^{\rm up}_{\ell m}(r;\omega)\sim
    \begin{cases}
        \begin{aligned}[t]&P^{\rm in}_{\ell m}\,r^4f^2(r)\,\de^{-\ii\omega r_*}\\&\quad+P^{\rm out}_{\ell m}\,\de^{+\ii\omega r_*}\end{aligned} & \text{as } r_*\to-\infty,\\[0.5ex]
        r^{3}\de^{+\ii\omega r_*} & \text{as } r_*\to+\infty.
    \end{cases}
\end{align}
\end{subequations}
The powers of $r$ and $f$ multiplying each exponential are those imposed by the Teukolsky equation for $s=-2$, and these relations define the constants $Q^{\rm in/out}_{\ell m}$ and $P^{\rm in/out}_{\ell m}$.
Of these, only $Q^{\rm in}_{\ell m}$, the amplitude of the incident wave of the ``in" solution, will be needed in what follows.
For a circular orbit, the amplitudes are then given explicitly in terms of the homogeneous solutions and their first two derivatives at $r=r_0$~\cite{Poisson:1993vp, Poisson:1996ya}: 
\begin{widetext}
    \begin{align}   \label{eq:Zlm_of_R}
    Z_{\ell m}^{\text{in}/\text{up}} = \frac{\pi}{\di \omega r_0^2 {Q}_{\ell m}^\text{in}}\Bigg\{&\Bigg[{\,}_0 b_{\ell m}  +  2 \di \Bigg(1+ \frac{\di \omega r_0}{2 f_0}\Bigg){\,}_{-1}b_{\ell m} +  \frac{\di\omega}{f_0^2}  \Bigg(1-r_0  - \frac{\di \omega r_0^2}{2}\Bigg) {\,}_{-2}b_{\ell m} \Bigg]R_{\ell m}^{\text{in}/\text{up}}(r_0;\omega)\nonumber\\
    & + r_0 \Bigg[\Bigg(1+ \frac{\di \omega r_0}{f_0}\Bigg){\,}_{-2}b_{\ell m} - \di ({\,}_{-1}b_{\ell m})\Bigg]\frac{\dd R_{\ell m}^{\text{in}/\text{up}}}{\dd r}(r_0;\omega)  - \frac{r_0^2}{2}  ({\,}_{-2}b_{\ell m}) \frac{\dd^2 R_{\ell m}^{\text{in}/\text{up}}}{\dd r^2}(r_0;\omega) \Bigg\},
\end{align}
\end{widetext}
where the coefficients ${}_{s}b_{\ell m}$ depend on $r_0$ and are given explicitly in App.~\ref{app:coeffs_flux}.
They are proportional to the spin-weighted spherical harmonics ${}_{s}Y_{\ell m}(\pi/2,0)$ and to the energy or angular momentum of the particle, and therefore carry the factor $\delta^{-1/2}$ discussed above.

\subsection{Reduction to the Regge-Wheeler equation}

Rather than working with the Teukolsky equation directly, it is advantageous to construct its homogeneous solutions from those of the Regge-Wheeler equation by means of the \emph{Chandrasekhar transformation} \cite{Chandrasekhar:1975nkd,Poisson:1996ya}.
The Regge-Wheeler equation has a real, short-ranged potential and takes the form of a one-dimensional Schr\"{o}dinger equation in the tortoise coordinate, which makes it ideal for the WKB analysis of Sec.~\ref{sec:WKB}. 
It reads
\begin{align}
    &\Big[f^2(r)\frac{\dd^2}{\dd r^2}+  f(r)f'(r)  \frac{\dd}{\dd r} \nonumber \\
    &\qquad\qquad\qquad+\Big(\omega^2-V(r)\Big)\Big]X_{\ell m}(r;\omega)=0,
    \label{eq:RW_equation}
\end{align}
i.e., $\dd X_{\ell m}/\dd r_{*}^{2} + [\omega^{2} - V(r)]X_{\ell m} = 0$, with the potential
\begin{align}\label{eq:V_def}
    V(r) = f(r)\Bigg(\frac{\ell(\ell+1)}{r^2} - \frac{6}{r^3}\Bigg) \,.
\end{align}
We denote the asymptotic behavior of a generic homogeneous solution to the radial Regge-Wheeler equation as
\begin{subequations}\label{eq:X_asymptotic}\begin{align}
     X_{\ell m}(r;\omega) &\underset{r_*\rightarrow +\infty}{\sim} \mathcal{A}_{\ell m}^\text{in} \de^{-\di \omega r_*}+\mathcal{A}_{\ell m}^\text{out}\de^{\di  \omega r_*} \,, \\
     X_{\ell m}(r;\omega) &\underset{r_*\rightarrow -\infty}{\sim}\mathcal{B}_{\ell m}^\text{in} \de^{\di  \omega r_*}+\mathcal{B}_{\ell m}^\text{out} \de^{-\di  \omega  r_*} \,. 
\end{align}\end{subequations}
Since the equation is of second order, only two of these four coefficients are independent. Because the potential is real, the Wronskian of a solution with its complex conjugate is conserved, and the relation between the coefficients can be parametrized by a unitary S-matrix:\footnote{Note that we denote complex conjugation with a superscripted asterisk ($z \rightarrow z^*$): this should not be confused with the subscripted asterisk appearing in the tortoise coordinate $r_*$.}
\begin{align}\label{eq:S_matrix}
    \begin{pmatrix} \mathcal{B}^\text{out}_{\ell m}\\[4pt] \mathcal{B}^\text{in}_{\ell m} \end{pmatrix} = \begin{pmatrix} \frac{\mathcal{R}_{\ell m}}{\mathcal{T}_{\ell m}}& \frac{1}{\mathcal{T}_{\ell m}^*} \\[7pt]
    \frac{1}{\mathcal{T}_{\ell m}} & \frac{\mathcal{R}_{\ell m}^*}{\mathcal{T}_{\ell m}^*}\end{pmatrix}  \begin{pmatrix} \mathcal{A}^\text{out}_{\ell m} \\[4pt] \mathcal{A}^\text{in}_{\ell m} \end{pmatrix},
\end{align}
where $\mathcal{T}_{\ell m}(r_0)$ and $\mathcal{R}_{\ell m}(r_0)$ are the complex transmission and reflection coefficients satisfying $|\mathcal{R}_{\ell m}|^2+|\mathcal{T}_{\ell m}|^2=1$.  This matrix is immediately inverted:
\begin{align}\label{eq:S_matrix_inverse}
    \begin{pmatrix} \mathcal{A}^\text{out}_{\ell m}\\[4pt] \mathcal{A}^\text{in}_{\ell m} \end{pmatrix} = \begin{pmatrix} -\frac{\mathcal{R}_{\ell m}^*}{\mathcal{T}_{\ell m}^*}& \frac{1}{\mathcal{T}_{\ell m}^*} \\[7pt] \frac{1}{\mathcal{T}_{\ell m}} & -\frac{\mathcal{R}_{\ell m}}{\mathcal{T}_{\ell m}}\end{pmatrix}  \begin{pmatrix} \mathcal{B}^\text{out}_{\ell m} \\[4pt] \mathcal{B}^\text{in}_{\ell m} \end{pmatrix} .
\end{align}
From this general form, it is easy to define the standard basis of ``in" and ``up" functions:
The ``in" solution is simply characterized by $\mathcal{B}_{\ell m}^\text{in} = 0$ and $\mathcal{B}_{\ell m}^\text{out} = 1$, in which case $\mathcal{A}_{\ell m}^\text{in} = 1/\mathcal{T}_{\ell m}$ and $\mathcal{A}_{\ell m}^\text{out} = -\mathcal{R}_{\ell m}^*/\mathcal{T}_{\ell m}^*$.
Conversely, the ``up" solution is defined by $\mathcal{A}_{\ell m}^\text{in} = 0$ and $\mathcal{A}_{\ell m}^\text{out} = 1$, in which case $\mathcal{B}_{\ell m}^\text{in} = 1/\mathcal{T}_{\ell m}$ and $\mathcal{B}_{\ell m}^\text{out} =\mathcal{R}_{\ell m}/\mathcal{T}_{\ell m}$.

To relate the solutions to the radial Teukolsky and Regge-Wheeler equations, we now introduce the differential  `Chandrasekhar' operator
\begin{align}\label{eq:Chandrasekhar_operator}
    \mathcal{C}[X](r) &= \,r^2f(r) \left(\frac{\dd }{\dd r_*} + \di \omega\right)\nonumber\\
    &\quad\times\Bigg\{f^{-1}(r) \left(\frac{\dd }{\dd r_*} + \di \omega\right)\Big[r\, X(r)\Big]\Bigg\} 
\end{align}
and find that the Teukolsky homogeneous radial functions are given in terms of the Regge-Wheeler ones by~\cite{Poisson:1993vp,Poisson:1994yf,Poisson:1996ya}
\begin{align}\label{eq:Chandrasekhar_transformation}
R^\text{in}_{\ell m} = \frac{\chi_{\ell m}}{(2 \di m \omega_0)^2}  \mathcal{C}\Big[ X^\text{in}_{\ell m}\Big],% \\
    &&
    R^\text{up}_{\ell m} = \frac{1}{(2 \di m \omega_0)^2}  \mathcal{C}\Big[ X^\text{up}_{\ell m}\Big] \,.
\end{align}
Adapting the normalization of Ref.~\cite{Poisson:1996ya}, we find that  
\begin{align}\label{eq:Q_in_of_A_in}
    Q_{\ell m}^\text{in} = \frac{16(1-2\di m \omega_0)(1-4\di m \omega_0)}{(2\di m\omega_0)^2 \mathcal{T}_{\ell m}}\,   
    \,.
\end{align}
Injecting \eqref{eq:Chandrasekhar_operator} into the formula \eqref{eq:Zlm_of_R}, and performing all the radial derivatives, we find that the $Z_{\ell m}^{\text{in/up}}$ amplitudes can be expressed in terms of the homogeneous solutions to the Regge-Wheeler equation and its successive derivatives. Any radial solution bearing more than two derivatives can be order-reduced using the Regge-Wheeler equation \eqref{eq:RW_equation} itself.  The amplitudes thus only involve the zeroth and first derivatives of the Regge-Wheeler homogeneous radial functions---they read 
\begin{subequations}\label{eq:Z_in_up_of_X}
    \begin{align}
    Z_{\ell m}^{\text{in}} &= 
    \chi_{\ell m} \, \mathcal{T}_{\ell m}
    \Bigg\{\kappa_{\ell m}  X_{\ell m}^\text{in}  + \lambda_{\ell m}  \frac{\dd X_{\ell m}^\text{in}}{\dd r_*} \Bigg\} \,,\\
    Z_{\ell m}^{\text{up}} &= 
    \mathcal{T}_{\ell m}\Bigg\{\kappa_{\ell m}  X_{\ell m}^\text{up}  + \lambda_{\ell m}  \frac{\dd X_{\ell m}^\text{up}}{\dd r_*} \Bigg\}\,,
\end{align}
\end{subequations}
where the homogeneous Regge-Wheeler radial solutions are evaluated for $\omega= m \omega_0$ and $r=r_0$. The coefficients $\kappa_{\ell m}$ and $\lambda_{\ell m} $ depend on $r_0$ and are given explicitly in App.~\ref{app:coeffs_flux}. Note also that these coefficients diverge like \mbox{$(r_0-3)^{-1/2}$} as \mbox{$r_0\rightarrow 3$.}

Plugging the expressions of the amplitudes \eqref{eq:Z_in_up_of_X} back into Eq.~\eqref{eq:Flm_of_Zlm}, we see that the infinity and horizon fluxes in fact have the same functional expression in terms of, respectively, the ``in'' and ``up'' solutions ($X_{\ell m}^{\text{in/up}}$). To present this compactly, we introduce the quantities $(\mathcal{X}_{\ell m}^{\text{in/up}},\mathcal{Y}_{\ell m}^{\text{in/up}},\mathcal{Z}_{\ell m}^{\text{in/up}})$, which are constructed from the homogeneous Regge-Wheeler radial functions as follows:\footnote{The minus sign in the definition of $\mathcal{Z}^{\text{up}}_{\ell m}$ is to ensure that it has the same asymptotics as $\mathcal{Z}^{\text{in}}_{\ell m}$ in the WKB limit---see Eq.~\eqref{eq:cal_XYZ_WKB}.}
\begin{subequations}\label{eq:def_cal_XYZ}
    \begin{align}
    &\mathcal{X}_{\ell m}^{\text{in/up}}(r_0) =  |X_{\ell m}^\text{in/up}(r_0;m\omega_0)|^2 \,,\\
    &\mathcal{Y}_{\ell m}^{\text{in/up}}(r_0) = \left|\frac{\dd X_{\ell m}^\text{in/up}}{\dd r_*}(r_0;m\omega_0)\right|^2 \,,\\
    &\mathcal{Z}_{\ell m}^{\text{in}}(r_0) = X_{\ell m}^{\text{in}*}(r_0;m\omega_0)\,  \frac{\dd X_{\ell m}^\text{in}}{\dd r_*}(r_0;m\omega_0) \,, \\
    &\mathcal{Z}_{\ell m}^{\text{up}}(r_0) = -X_{\ell m}^{\text{up}*}(r_0;m\omega_0)\,  \frac{\dd X_{\ell m}^\text{up}}{\dd r_*}(r_0;m\omega_0) \,.
\end{align}\end{subequations}
We then find that the fluxes through infinity and the horizon explicitly read
\begin{equation}\label{eq:flux_of_X_final}
\boxed{
\begin{aligned}
    &\mathcal{F}^{\infty/\mathcal{H}}_{\ell m} =  \frac{1}{2\pi (m\omega_0)^2} 
    |\chi_{\ell m}|^2 |\mathcal{T}_{\ell m}|^2 \Big\{|\kappa_{\ell m}|^2 \mathcal{X}_{\ell m}^{\text{in/up}} 
    \\
    &\qquad \quad + |\lambda_{\ell m}|^2 \mathcal{Y}_{\ell m}^{\text{in/up}} \pm 2 \operatorname{Re} (\kappa_{\ell m}^* \lambda_{\ell m} \mathcal{Z}_{\ell m}^{\text{in/up}}) \Big\},
\end{aligned}}
\end{equation}
where the upper (lower) sign in $\pm$ corresponds to the flux through infinity (the horizon). In the WKB limit, the moduli of the ``in'' and ``up'' homogeneous radial Regge-Wheeler functions coincide when approaching the light ring,
so the horizon and infinity flux modes similarly coincide in this limit\footnote{The interference terms carrying opposite signs contribute at relative order $\mathcal{O}(m^{-1/2})$ and do not affect this leading equality.
%This statement is true despite the $\pm$ sign because these contributions only enter at subleading order in the large-$m$ expansion.
}, as is well known~\cite{Breuer:1975, Barausse:2021xes,Barack:2026izc}.

With this explicit formula, we can immediately compute the fluxes as soon as we have a solution for the radial Regge-Wheeler equation. This solution can be numerical, or as we shall see, an analytical approximation.

\section{WKB solution to the Regge-Wheeler equations}
\label{sec:WKB}
 
In order to solve the radial Regge-Wheeler equation, we will write it in a form adapted to solving it using the WKB approximation. For this, we introduce the `off-diagonal' integer, $k=\ell-m \ge 0$, and the `distance from the light ring', $\delta = r_0-3$. We also introduce the special tortoise coordinate $x = r -3+ 2 \ln (r-2)$, such that $f =\dd r/\dd x = 1-2/r$. As we will be studying the large-$m$ asymptotics at fixed $k$, we will swap the $(\ell,m)$ indexing for a $(m,k)$ indexing---although this can be formally ambiguous, it should always be clear in context.
Armed with these definitions, the radial Regge-Wheeler equation \eqref{eq:RW_equation} can be rewritten for \mbox{$\omega = m \omega_0$} in a way analogous to the stationary Schrödinger equation:
\begin{align}\label{eq:Schrodinger_equation}
    \frac{\dd^2  X_{m k}}{\dd x^2}+ m^2 Q(x)  X_{m k} =0 \,,
\end{align}
where $Q(x)=\tilde{Q}(r(x))$ and $\tilde{Q}(r)= E-\tilde{V}(r)$ is related to the `energy level' $E=(3+\delta)^{-3}$ and the `potential' $\tilde{V}(r) = V(r)/m^2$ written in Eq.~\eqref{eq:V_def}. It follows that 
\begin{align}  
   \tilde{Q}(r) &= \frac{1}{(3+\delta)^3} -f(r)\left(\frac{(1+\frac{k}{m})(1+\frac{k+1}{m})}{r^2} - \frac{6}{m^2 r^3}\right)\,.
\end{align}
In quantum mechanics, the WKB approximation can be seen as an expansion in $\hbar\ll1$; in our case, this corresponds to $m\gg 1$.  It is thus ideally suited to study infinite sums in which the large-$m$ modes have non-negligible contributions. More precisely, the WKB approximation is valid when the local wavelength varies slowly, namely 
\begin{align}\label{eq:WKB_criterion}
    \left|\frac{Q'(x)}{m\,Q(x)^{3/2}}\right| \ll 1 \,.
\end{align}
Moreover, as the radius of the orbit approaches the light ring, we will have $\delta \ll 1$: this is interpreted as the case where the energy is near the peak of the potential~\cite{Ford:1959, Will:1988zz}. The double limit ($m\gg1$ and $\delta \ll 1$) can be subtle and we will indeed encounter objects involving $m\delta$.

The WKB solution is fixed uniquely using standard boundary conditions at $r_* \rightarrow \pm \infty$ and is valid far from the `classical turning points', which are defined as zeros of $Q(x)$. Near these points, Eq.~\eqref{eq:WKB_criterion} is not satisfied anymore and  the WKB solution blows up. If there is an isolated turning point $x_0$, the standard approach~\cite{BenderOrszag1999} is to linearize $Q(x)$ around that point: the Regge-Wheeler equation can then be solved exactly using Airy functions up to two undetermined constants. These constants are fixed by matching the $x\rightarrow\pm \infty$ behavior of the Airy solutions with the behavior of the WKB solution when approaching the turning point ($x\rightarrow x_0^{\pm}$). The presence of several \textit{well-separated} turning points can be dealt with similarly by successively matching each of the many zones. If the isolated turning point is of second order, namely $Q'(x_0) =0$ and $Q''(x_0) \neq 0$, one needs to take a different approach. In quantum mechanics, this case corresponds to the scattering of a wave on the peak of the potential~\cite{BenderOrszag1999}. The solutions near the turning point are then parabolic cylinder functions, whose $x \rightarrow \pm \infty$ behavior is then matched against the $x\rightarrow x_0^{\pm}$ behavior of the  WKB solution. In the latter case, a classical result is that the square of the moduli of the transmission and reflection coefficients are both exactly equal to $1/2$.

\begin{figure*}
    \centering
    \includegraphics[width=0.9\textwidth]{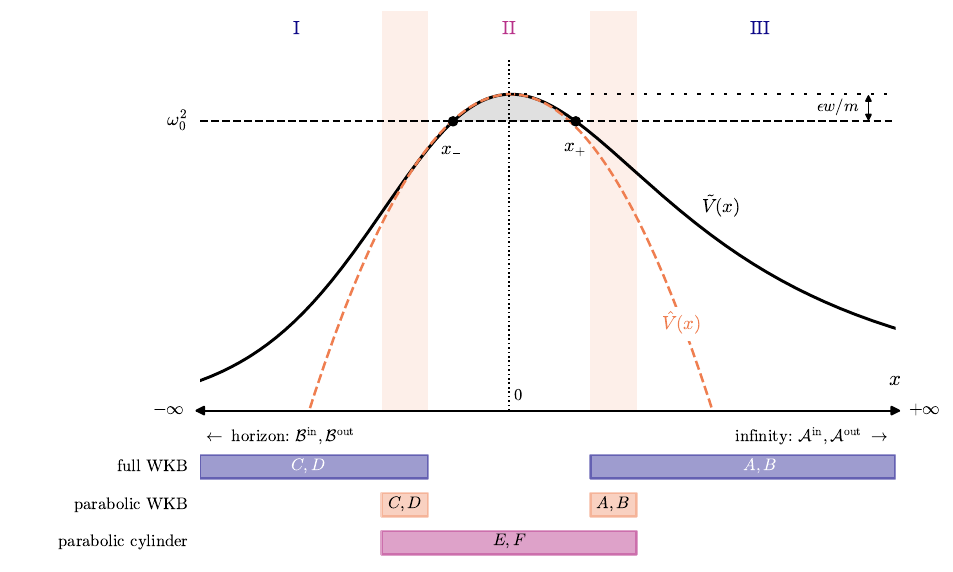}
    \caption{\justifying Illustration of the asymptotic matching between the parabolic cylinder solution and the WKB solutions of the Regge-Wheeler equation~\eqref{eq:Schrodinger_equation}, shown in terms of the tortoise coordinate $x=r-3+2\ln(r-2)$.
   The black curve corresponds to the potential $\tilde V(x)=\omega_0^2-Q(x)$ (solid black line); the orange line is its parabolic approximation $\hat V(x)=\omega_0^2-\hat Q(x)$ (dashed orange line), with $\hat Q(x)=w^2x^2-\epsilon w/m$; see Eq.~\eqref{eq:app:Qhat_def}. Both are drawn exactly for $m=8$, $k=0$ and $\delta=0$. The horizontal dashed line is the energy level $\omega_0^2$, which lies a distance $\epsilon w/m$ below the peak of the potential (sparse dotted line). Its intersections with the potential are the classical turning points $x_\pm$, between which $Q<0$ (gray). Region~I (resp., III) extends towards the horizon $x\to-\infty$ (resp., infinity $x\to+\infty$): its boundary conditions are given by the coefficients $\mathcal{B}^\text{in}_{\ell m}$ and $\mathcal{B}^\text{out}_{\ell m}$ (resp., $\mathcal{A}^\text{in}_{\ell m}$ and $\mathcal{A}^\text{out}_{\ell m}$). This is where the full WKB solutions (blue rectangles) are valid: in region I (resp., III), these solutions are characterized by the coefficients $C$ and $D$ (resp., $A$ and $B$). 
   Region~II surrounds the peak and contains both turning points. This is where the representation in terms of  parabolic cylinder function (characterized by the coefficients $E$ and $F$) is valid, because in this region the parabolic potential $\hat{V}(x)$  approximates well the full potential $\tilde{V}(x)$. 
   %d
   The shaded bands are the overlap zones, where the parabolic WKB solutions are used, obtained by replacing $Q$ with $\hat Q$. They are also characterized by the same $(A,B,C,D)$ as the full WKB solution.  The $x\to\pm\infty$ asymptotics of the parabolic WKB solutions are matched to those of the exact parabolic cylinder solution $E\,D_\nu(t)+F\,D_\nu(-t)$ of region~II, which relates $(E,F)$ to $(A,B,C,D)$.}
    \label{fig:matching_sketch}
\end{figure*}

In our case, we have two classical turning points, 
\begin{align}\label{eq:x_plus_minus}
    x_\pm \approx \pm \sqrt{\frac{\epsilon}{m w}}\,,
\end{align}
where we have introduced the ubiquitous quantities
\begin{subequations}\label{eq:def_w_epsilon}
    \begin{align}
    \label{seq:def_w}
    w &= \sqrt{\frac{Q''(\xmin)}{2}} \approx \frac{1}{27} \,,\\
    \label{seq:def_epsilon}
    \epsilon &= - \frac{m Q(\xmin)}{w} \approx 1+2k+m\delta \,,
\end{align}
\end{subequations}
and where  $x_\text{min}\approx 3/m^2$ is the extremum of the potential---see Eq.~\eqref{seq:app:xmin_precise} for more precise statements. We thus have $\xm <\xmin<\xp$: in the large $m$ limit, the two turning points approach each other and sandwich $\xmin$ around zero. Clearly, the approach adapted to a single turning point does not apply because of the proximity of the two turning points.
One could hope that, since $x_\pm\to0$ as $m\to\infty$, the system reduces to a second order turning point at the top of the potential, i.e.\ that one can neglect the small contribution of $Q(\xmin)$ and simply use $Q\simeq w^2x^2$. But for such a potential, the WKB criterion~\eqref{eq:WKB_criterion} fails for $|x|\lesssim (mw)^{-1/2}$, which is a characteristic length comparable to the distance between the turning points, $|x_+-x_-| \sim 2\sqrt{\epsilon/(mw)}$. The second order turning point method is therefore not relevant here.
%therefore does not apply here.

We therefore follow the approach of Refs.~\cite{Ford:1959,Misner:1972jf, Iyer:1986np}. Here, we sketch the different steps and main results; see Fig.~\ref{fig:matching_sketch} for an illustration of the matching procedure and App.~\ref{app:WKB_details} for the detailed derivation of the matching conditions. 
Near the peak, we approximate the potential by a parabola and solve it using a basis of parabolic cylinder functions. Since we have modified the potential, we also need to introduce a `parabolic WKB solution' which is obtained by applying the WKB approximation to the simplified parabolic potential. We then match the $x \rightarrow \pm \infty$ behavior of the parabolic cylinder solution to the $x \rightarrow \pm \infty$
behavior of the parabolic WKB solution. This fixes the two unknown coefficients of the parabolic cylinder solution to the constants parametrizing the parabolic WKB solution. The latter constants are necessarily equal to the ones parametrizing the full WKB solution and these are fixed by the boundary conditions.  

From the matching, we first obtain two relations between the normalization coefficients $(\mathcal{A}^\text{in}_{m k},\mathcal{A}^\text{out}_{m k},\mathcal{B}^\text{in}_{m k},\mathcal{B}^\text{out}_{m k})$ which can  be framed in the form of an $S$-matrix, as in Eq.~\eqref{eq:S_matrix}.
The complex transmission and reflection coefficients, $\mathcal{T}_{m k}$  and $\mathcal{R}_{m k}$, are given in Eq.~\eqref{eq:app:T_R_expression_WKB}. They have a complicated argument, but the square of their norm simply reads
% %
\begin{equation}
\boxed{
\begin{aligned}
\label{eq:normam_squared_T_R_WKB}
    |\mathcal{T}_{m k}|^2 = \frac{1}{1+\de^{\epsilon\pi}} &&\text{and}&& |\mathcal{R}_{m k}|^2= \frac{1}{1+\de^{-\epsilon\pi}} \,,
\end{aligned}}
\end{equation}
which satisfies the condition $|\mathcal{T}_{m k }|^2+|\mathcal{R}_{m k}|^2=1$. This is a well-known result~\cite{Iyer:1986np, Kemble:1935, Guinn:1989bn}: the quantity $|\mathcal{T}_{m k}|^2$ is referred to in the literature as the `grey-body factor'~\cite{Konoplya:2026qql} and is relevant here because it appears in the flux formula~\eqref{eq:flux_of_X_final}.
\begin{figure*}
    \centering
    \includegraphics[width=\textwidth]{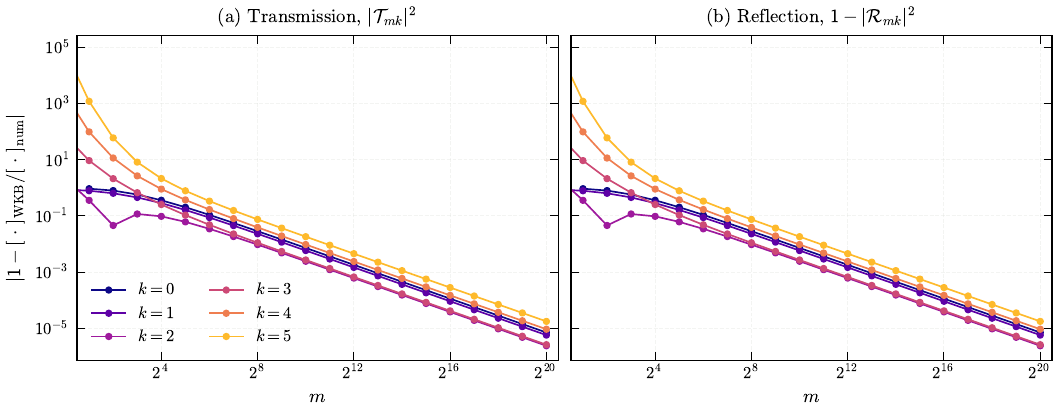}
    \caption{\justifying Relative error of the WKB expressions~\eqref{eq:normam_squared_T_R_WKB} for (a) the transmission probability $|{\cal T}_{mk}|^{2}$ and (b) the complement of the reflection probability, $1 - |{\cal R}_{mk}|^{2}$, with respect to their numerical values at $\delta = 0$, for $k \in [\![0, 5]\!]$ and $m = 2^{n}$ with $1 \leq n \leq 20$.
    We show $1 - |{\cal R}_{mk}|^{2}$ rather than $|{\cal R}_{mk}|^{2}$ because the latter is exponentially close to unity; the two panels coincide since $|{\cal R}_{mk}|^{2} + |{\cal T}_{mk}|^{2} = 1$ holds for the numerical data to all the digits we have computed.}
    \label{fig:error-T2-R2}
\end{figure*}

The matching conditions also entirely specify the parabolic cylinder solution in terms of two normalization coefficients. This allows us to evaluate the basis ``in" and ``up" radial RW functions and their derivatives (with respect to $x$) at $x=0$. The explicit expressions, which we provide in Eq.~\eqref{eq:app:X_Xp_in_up_x_0_WKB}, also exhibit a complicated complex argument.
In view of computing the fluxes, however, we will not require all the information contained in the basis functions, but only those quantities entering Eq.~\eqref{eq:flux_of_X_final}. We find that these quantities  explicitly read  
\begin{equation}\label{eq:cal_XYZ_WKB}
\boxed{
\begin{aligned}
    \mathcal{X}_{mk}^{\text{in/up}} &\!\underset{m\rightarrow \infty}{\sim}\! \frac{\sqrt{m}\,\de^{\frac{\pi \epsilon}{4}}}{2\pi \sqrt{w}(3+\delta)^{3/2}}\, \cosh\left(\frac{\pi \epsilon}{2}\right) \left|\Gamma\left(\frac{1+\di \epsilon}{4}\right)\right|^2,  \\[1em]
    %%%%%%%
     \mathcal{Y}_{mk}^{\text{in/up}} & \!\underset{m\rightarrow \infty}{\sim} \!\frac{2 m^{\frac{3}{2}} \sqrt{w}}{\pi(3+\delta)^{3/2} }\,\de^{\frac{\pi \epsilon}{4}} \cosh\left(\frac{\pi \epsilon}{2}\right) \left|\Gamma\left(\frac{3+\di \epsilon}{4}\right)\right|^2, \\[1em]
     %%%%%%%
     \mathcal{Z}_{mk}^{\text{in/up}}  &\!\underset{m\rightarrow \infty}{\sim}\!    \frac{m(-\di+\de^{\frac{\pi \epsilon}{2}})}{(3+\delta)^{3/2}}.  
\end{aligned}}
\end{equation}

\begin{figure*} 
    \centering
    \vspace{1cm}
    \includegraphics[width=\textwidth]{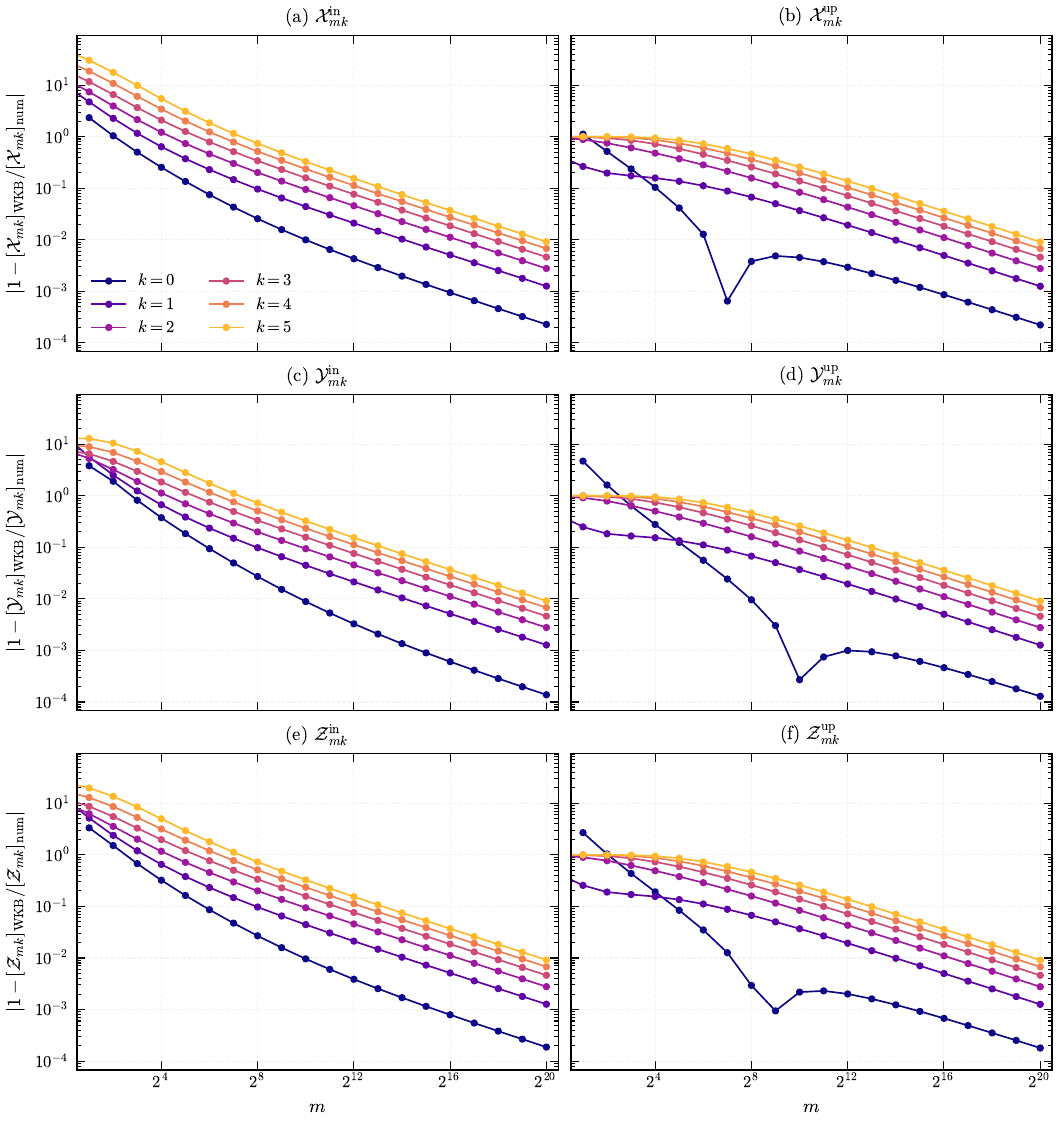}
    \caption{\justifying Relative error of the WKB expressions~\eqref{eq:cal_XYZ_WKB} for the quantities ${\cal X}^{\rm in/up}_{mk}$, ${\cal Y}^{\rm in/up}_{mk}$, and ${\cal Z}^{\rm in/up}_{mk}$ entering the flux expression \eqref{eq:flux_of_X_final}, with respect to numerical values at $\delta = 0$.
    The left panels correspond to the ``in" solution and the right panels to the ``up" solution. 
    The colored circles correspond to $k \in [\![0, 5]\!]$ with $m = 2^{n}$ with $1 \leq n \leq 20$.
    The errors vanish in the large-$m$ limit for all tested values of $k$.}
    \label{fig:error-XYZ}
    \vspace{1cm}
\end{figure*}

\begin{figure*}
\centering
    \includegraphics[width=\textwidth]{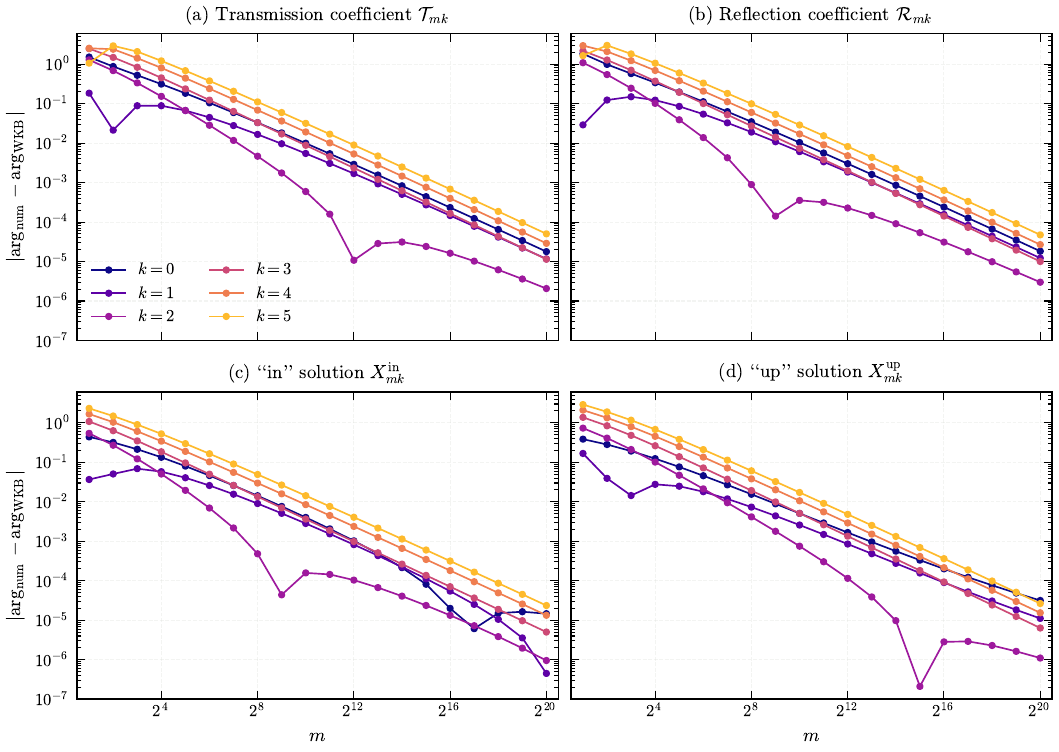}
    \caption{\justifying Absolute error of the WKB phases at $\delta=0$ as a function of $m$. Each panel shows the difference $|\arg_{\rm num}-\arg_{\rm WKB}|$ (modulo $2\pi$) between the numerical value of the complex argument and the analytical prediction of Eq.~\eqref{eq:complex_T_R_X_Xp}, for (a) the transmission coefficient $\mathcal{T}_{mk}$, (b) the reflection coefficient $\mathcal{R}_{mk}$, and the (c) ``in" and (d) ``up" Regge-Wheeler functions evaluated at $x=0$. The colored circles correspond to $k\in[\![0,5]\!]$ and $m=2^n$ with $1\le n\le20$. The errors vanish in the large-$m$ limit for all $k$, which validates both the leading phases $m\widetilde{\mathcal{I}}$ and $m\widetilde{\mathcal{J}}$ and the remaining $m$-independent and logarithmic terms of Eq.~\eqref{eq:complex_T_R_X_Xp}.}
    \label{fig:error-phase}
\end{figure*}

\begin{figure*}
    \centering
    \includegraphics[width=\textwidth]{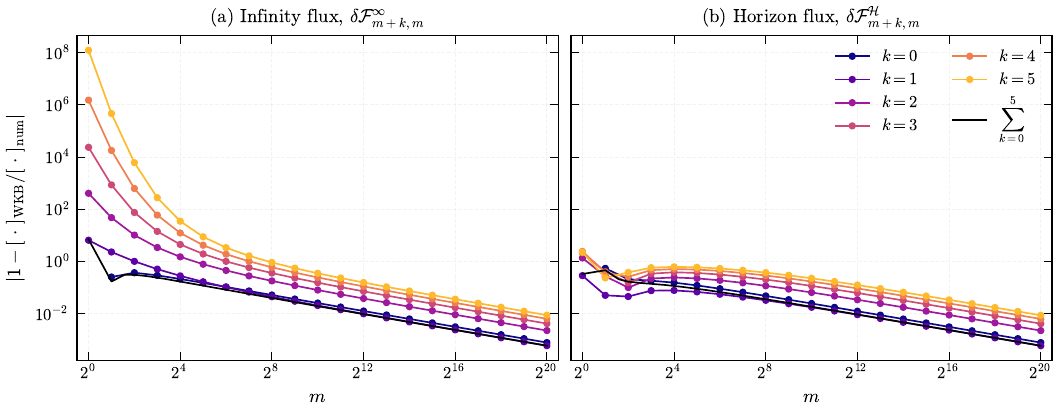}
    \caption{\justifying Relative error of the WKB expression~\eqref{eq:Flm_WKB} for the regularized mode fluxes, taken at $\delta=0$, with respect to their numerical values, for (a) the flux through infinity $\delta\mathcal{F}^{\infty}_{m+k,m}$ and (b) the flux through the horizon $\delta\mathcal{F}^{\mathcal{H}}_{m+k,m}$. 
    The colored circles correspond to $k\in[\![0,5]\!]$ and $m=2^n$ with $n\le20$. 
    The black line shows the error on the sum over $k\in[\![0,5]\!]$ at fixed $m$, for all $m\le2000$ and for $m=2^n$ beyond. 
    The errors decrease in the large-$m$ limit, as expected, and their behavior is consistent with a scaling $\sim m^{-1/2}$.}
    \label{fig:error-flux-m-wkb}
\end{figure*}

Finally, in App.~\ref{subsec:rho_m_delta_0}, we have worked out  simple analytical expressions for the phase dependence of the radial RW basis functions, as well as the transmission and reflection coefficients, in the particular case where $\delta =0$. We thus find that their full (complex) expressions read
\begin{subequations}\label{eq:complex_T_R_X_Xp}
    \begin{align}
    \mathcal{T}_{m k}\Big|_{\delta=0} =& \,\frac{\Gamma\left(\frac{1+\di \epsilon}{2}\right)\de^{- \frac{\epsilon \pi}{4}}}{\sqrt{2\pi}} \nonumber\\*
    &\times\,\de^{-\di\left\{m\left(\widetilde{\mathcal{I}} +\widetilde{\mathcal{J}} \right) + \frac{\epsilon}{2}\ln\left[216\,(2-\sqrt{3})^3\, m\right]\right\}} \,,\\
    %%%%%%%%%%%%%%%%%%%%
    \mathcal{R}_{m k}\Big|_{\delta=0} =& \,\frac{\Gamma\left(\frac{1+\di \epsilon}{2}\right)\de^{ \frac{\epsilon \pi}{4}}}{\sqrt{2\pi}} \nonumber\\*
    &\times
    \,\de^{-\di\left\{2 m \widetilde{\mathcal{J}}+ \frac{\epsilon}{2}\ln\left[216\,(2-\sqrt{3})^4\, m\right] + \frac{\pi}{2}\right\}} \,, \\
    %%%%%%%%%%%%%%%%%%%%
    X_{mk}^\text{in}\Big|_{\substack{x=0\\\delta=0}}&= \frac{\sqrt{\pi}\,m^{\frac{1}{4}}\,\de^{\frac{\epsilon\pi}{8}}}{ \Gamma\left(\frac{3+\di \epsilon}{4}\right)} \nonumber\\*
    &\times
    \,\de^{\di\left\{m\widetilde{\mathcal{J}} +\frac{\epsilon}{4}\ln\left[108\,(2-\sqrt{3})^4\, m\right] - \frac{\pi}{8}\right\}}\,,\\
    %%%%%%%%%%%%%%%%%%%%
    X_{mk}^\text{up}\Big|_{\substack{x=0\\\delta=0}}&= \frac{\sqrt{\pi}\,m^{\frac{1}{4}}\,\de^{\frac{\epsilon\pi}{8}}}{ \Gamma\left(\frac{3+\di \epsilon}{4}\right)} \nonumber\\*
    &\times
    \,\de^{\di\left\{m\widetilde{\mathcal{I}} +\frac{\epsilon}{4}\ln\left[108\,(2-\sqrt{3})^2\, m\right] - \frac{\pi}{8}\right\}}\,,\\
    %%%%%%%%%%%%%%%%%%%%
    {X_{mk}^\text{in}}{\!}'\Big|_{\substack{x=0\\\delta=0}} &= \frac{2\sqrt{\pi}\,m^{\frac{3}{4}}\,\de^{\frac{\epsilon\pi}{8}}}{3\sqrt{3}\,\Gamma\left(\frac{1+\di \epsilon}{4}\right)} \nonumber\\*
    &\times
    \,\de^{\di\left\{m\widetilde{\mathcal{J}} +\frac{\epsilon}{4}\ln\left[108\,(2-\sqrt{3})^4\, m\right] - \frac{3\pi}{8}\right\}}\,,\\
    %%%%%%%%%%%%%%%%%%%%
    {X_{mk}^\text{up}}'\Big|_{\substack{x=0\\\delta=0}} &= -\frac{2\sqrt{\pi}\,m^{\frac{3}{4}}\,\de^{\frac{\epsilon\pi}{8}}}{3\sqrt{3}\,\Gamma\left(\frac{1+\di \epsilon}{4}\right)} \nonumber\\*
    &\times
    \,\de^{\di\left\{m\widetilde{\mathcal{I}} +\frac{\epsilon}{4}\ln\left[108\,(2-\sqrt{3})^2\, m\right] - \frac{3\pi}{8}\right\}}\,,
\end{align}
\end{subequations}
where in this case $\epsilon=1+2k$ and
\begin{subequations}\label{eq:IJ_tilde_expressions}
\begin{align}
    \widetilde{\mathcal{I}}
    &=1+\frac{6\ln(2+\sqrt3)-4\ln2-3}{3\sqrt3},\\
    \widetilde{\mathcal{J}}
    &=1+\frac{6\ln(2+\sqrt3)-8\ln2-6}{3\sqrt3}.
\end{align}
\end{subequations}
\iffalse
\begin{subequations}\label{eq:IJ_tilde_expressions}
\begin{align}
    &\widetilde{\mathcal{I}} = 1  - \frac{1}{3\sqrt{3}}\Bigg\{8 \,\text{arctanh}\!\left(\frac{1-2\sqrt{3}}{3}\right)+6 \ln 2 +3\Bigg\} \,,\\
    %%%%%
   &\widetilde{\mathcal{J}}  = 1  - \frac{1}{3\sqrt{3}}\Bigg\{8 \,\text{arctanh}\!\left(\frac{1-2\sqrt{3}}{3}\right)+10 \ln 2 +6\Bigg\} \,.
\end{align}\end{subequations}
\fi

\section{WKB expressions for the flux modes}
\label{sec:WKB_flux_modes}

We are now ready to inject our WKB solution \eqref{eq:cal_XYZ_WKB} into the flux formula \eqref{eq:flux_of_X_final}. This yields a complicated expression: we are instead interested in the asymptotic behavior of this flux in the double limit $m\rightarrow +\infty$ and $\delta \rightarrow 0^+$. This limit is in fact subtle due to the presence of $m \delta$ contributions, in which these two limits compete. 

We first notice that $\kappa_{\ell m}(\delta) \sim \lambda_{\ell m}(\delta) \sim  1/\sqrt{\delta}$
%and $\lambda_{\ell m}(\delta)  \sim  1/\sqrt{\delta}$
as $\delta \rightarrow0^+$, whereas the Regge-Wheeler functions and their derivatives are finite in that limit. Thus, each $(\ell,m)$ contribution to the flux individually diverges as $1/\delta$. Factoring that divergence away, we find that the combination $\delta \mathcal{F}_{\ell m}$ is finite in the $\delta \rightarrow 0$ limit. Now, we wish to study how the sum over $(\ell,m)$ modes of the $\delta \mathcal{F}_{\ell m}$ modes diverges (logarithmically) in the $\delta \rightarrow 0$ limit. 
We must be careful to not set $\delta=0$ directly, because $\delta$ controls the exponential suppression which ensures finiteness of the sum for $\delta>0$. Instead, we perform the $m \rightarrow + \infty$ limit \textit{at fixed}  $m\delta$. This requires the large-$m$ asymptotic behavior of the spin-weighted spherical harmonics \textit{to subleading order}, which we derived in App.~\ref{app:spherical_harmonic_asymptotics}. It reads
\begin{widetext}
\begin{equation}\label{eq:sYlm_asymptotics}
\boxed{
\begin{aligned}
     {}_sY_{m+k,m}(\pi/2,0) \underset{m\rightarrow+\infty}{\sim}\ & \begin{cases}
         \frac{(-1)^m\sqrt{k!}m^{\frac{1}{4}}}{2^{\frac{k+1}{2}} \pi^{\frac{3}{4}}} \frac{(-1)^{\frac{k}{2}}}{\left(\frac{k}{2}\right)!}\bigg[1- \frac{(2k+1)(8s^2-3)}{16m}+\mathcal{O}\left(\frac{1}{m^2}\right)\bigg] & \qquad\text{if $k$ even}\\
    %%%%%%%
    \frac{(-1)^m\sqrt{k!}}{2^{\frac{k+1}{2}} \pi^{\frac{3}{4}} m^{\frac{1}{4}}} \frac{(-1)^{\frac{k-1}{2}}}{\left(\frac{k-1}{2}\right)!}  (2s) \bigg[1- \frac{(2k+1)(8s^2+1)}{48m}+\mathcal{O}\left(\frac{1}{m^2}\right)\bigg] &\qquad\text{if $k$ odd}
     \end{cases}
\end{aligned}}\,.
\end{equation}
\end{widetext}
The reason we need to go to subleading order is that the contributions from the leading order actually cancel out in $\kappa_{\ell m}$. Defining the `parity'  \mbox{$\frak{p}(k)$} to be equal to $0$ if $k$ is even and $1$ if $k$ is odd, we find that the relevant asymptotic formulas for the fluxes \eqref{eq:flux_of_X_final} read 
\begin{subequations}\label{eq:kappa_lambda_asymptotitcs}\begin{align}
    \hspace{-13pt}\chi_{m+k,m}\,\kappa_{m+k,m} &\sim \frac{\pi^{1/4} \sqrt{k!} (-1)^m \,\di^{k+1} m^{\frac{1}{4}}}{27 \cdot 2^{\frac{k-3}{2}}\lfloor k/2 \rfloor! \sqrt{\delta}} \left(\frac{2}{\sqrt{3m}}\right)^{\frak{p}(k)}\nonumber\\*
    \hspace{-13pt}&\qquad\times\!\left(1+2k+\frac{m\delta}{2}\right)^{1-\frak{p}(k)}, \\
    %%%%%%%%%%%%%
    \hspace{-13pt}\chi_{m+k,m}\,\lambda_{m+k,m} &\sim \frac{\pi^{1/4} \sqrt{k!} (-1)^m \,\di^{k+1} }{9 \cdot 2^{\frac{k-3}{2}}\lfloor k/2 \rfloor! m^{\frac{3}{4}}\sqrt{\delta}}(2\sqrt{3m})^{\frak{p}(k)}.
\end{align}\end{subequations}
Note the residual $m \delta$ dependence: this is because we do \textit{not} further expand in small $\delta$ at fixed $m$. Using this expansion as well as the quantities of Eqs.~\eqref{eq:normam_squared_T_R_WKB}~and~\eqref{eq:cal_XYZ_WKB}, we finally find that the fluxes behave as
\begin{equation}\label{eq:Flm_WKB}
\boxed{
\begin{aligned}    
&\mathcal{F}_{m+k,m}^{\infty/\mathcal{H}}\sim\frac{1}{m \delta}\frac{2^{4 \frak{p}(k) - k}k!}{27 \pi^{3/2} (\lfloor k/2 \rfloor !)^2}  \\
    &\qquad\quad\times  \left(1+2k + \frac{m \delta}{2} \right)^{2[1-\frak{p}(k)]}   \de^{- \frac{\pi}{4} (1+2k+m \delta)} \\
    &\qquad\quad\times \left|\Gamma\left(\frac{1+2\frak{p}(k)+\di (1+2k+m \delta)}{4} \right)\right|^2 \!.
\end{aligned}}
\end{equation}

In the $k=0$ and $k=1$ cases, we compared our expression to Eqs.~(14) and (15) of Ref.~\cite{Breuer:1973kt}; for this comparison, we discard the subleading $\mathcal{O}(1/\sqrt{m})$ contribution in their expression. Up to conventions, their expression agrees with ours in the $k=0$ case, whereas their expression  should be multiplied by a factor $4$ in the $k=1$ case. 

For any $k$, we have also compared our result with (4.28) of Ref.~\cite{Chrzanowski:1974nr}. To perform this comparison, we set $a=0$ in their work and identified $m_\text{crit}=4/(\pi\delta)$. In our expression, we also took the $m\delta \gg 1$ limit to simplify the gamma function, using $ \left|\Gamma(x+\di y)\right|^2 \sim  2 \pi y^{2x-1} \de^{-\pi y}$ for $y \rightarrow +\infty$.
After performing these operations, we found perfect agreement in the case where $k$ is odd. In the case where $k$ is even, we find agreement after correcting an error in their expression: they had mistakenly replaced the quantity $1+2k+m\delta/2$ with $\epsilon=1+2k+m \delta$; note that the difference between these two quantities was correctly accounted for in an earlier paper \cite{Breuer:1973kt}.

\section{Comparison with numerical results on the light ring}
\label{sec:comparison_with_numerical_results}

In order to validate our analytical results, we compare them to new numerical results which we have generated for very large values of $m$.

\subsection{Numerical method}
\label{subsec:numerical_method}
We have solved the Regge-Wheeler equation \eqref{eq:RW_equation} on hyperboloidal slices~\cite{Zenginoglu:2011jz, PanossoMacedo:2022fdi, Leather:2024mls, Leather:2026zhl}, which are horizon penetrating and reach future null infinity.
We introduce the compactified coordinate $\sigma = 2/r$, such that future null infinity is at $\sigma = 0$ and the horizon is at $\sigma = 1$, and rescale the radial function as
\begin{subequations}
    \begin{align}
    X_{\ell m}(r)&=\de^{\di\omega h(\sigma)}\,\chi_{\ell m}(\sigma),
    \label{eq:hyperboloidal_rescaling_RW}\\
    h(\sigma)&=\frac{2}{\sigma}-2\ln\sigma-2\ln(1-\sigma),
    \label{eq:height_function}
\end{align}
\end{subequations}
where $h$ is the height function of the slicing.
The rescaling removes the oscillatory phase of the ``up" solution at infinity and of the ``in" solution at the horizon, so that $\chi$ is smooth and finite at both boundaries.
The Regge-Wheeler equation then takes the form
\begin{multline}
    \sigma^2(1-\sigma)\,\chi''+\left[-4\di\omega+2\sigma+(8\di\omega-3)\sigma^2\right]\chi' \\
    +\left[16\omega^2-\ell(\ell+1)+\left(4-(1-4\di\omega)^2\right)\sigma\right]\chi=0,
    \label{eq:hyperboloidal_RW}
\end{multline}
where a ${}^\prime$ denotes a derivative with respect to $\sigma$ and we have dropped the mode labels.
All coefficients are polynomials in $\sigma$.
The coefficient of $\chi''$ vanishes at $\sigma = 0$ and $\sigma = 1$, and the equation there reduces to a regularity condition relating $\chi'$ and $\chi$.
The ``up" solution is the regular solution at $\sigma = 0$ and the ``in" solution is the regular solution at $\sigma = 1$; we normalize them such that $X^{\rm up}_{\ell m} \rightarrow \de^{\ii \omega r_{*}}$ at infinity and $X^{\rm in}_{\ell m} \rightarrow \de^{-\ii \omega r_{*}}$ at the horizon, as in Sec.~\ref{sec:flux_formalism}.
This is a significant advantage at the large frequencies we require, which reach $\omega \simeq 2 \times 10^{5}$.

We solve Eq.~\eqref{eq:hyperboloidal_RW} with a multidomain Chebyshev collocation method.
Starting from the regularity condition at the boundary $\sigma = 0$ (for ``up") and $\sigma = 1$ (for ``in"), we march towards the particle at $\sigma_{0} = 2 / r_{0}$ through a sequence of subdomains.
On each subdomain the solution is expanded on $N$ Chebyshev-Gauss nodes and solved as an initial value problem, with the value and the first derivative of $\chi$ taken from the previous subdomain.
The width of each subdomain is chosen adaptively from the decay of the Chebyshev coefficients, so that the number of subdomains adjusts to the frequency of the mode.
All calculations are carried out in arbitrary-precision ball arithmetic using the $\texttt{Arb}$ library \cite{2017ITCmp..66.1281J}, with a working precision of 128 bits.
This keeps round-off error well below the truncation error of the collocation scheme for all the modes we compute.

The radial functions $X^{\rm in / up}_{\ell m}$ and their first derivatives at $r_{0}$ follow directly from $\chi^{\rm in / up}$ and their first derivatives at $\sigma_{0}$.
The coefficients ${\cal A}^{\rm in}_{\ell m}$ and ${\cal A}^{\rm out}_{\ell m}$ of the ``in" solution are obtained from the Wronskians $W[X^{\rm in}_{\ell m}, X^{\rm up}_{\ell m}] = 2\ii\omega{\cal A}^{\rm in}_{\ell m}$ and $W[X^{\rm in}_{\ell m}, X^{\rm up *}_{\ell m}] = -2\ii\omega{\cal A}^{\rm out}_{\ell m}$, where $W[f, g] \equiv f \dd g / \dd r_{*} - g \dd f / \dd r_{*}$ is evaluated at $r_{0}$, and give the transmission and reflection coefficients ${\cal T}_{\ell m}$ and ${\cal R}_{\ell m}$ through Eq.~\eqref{eq:S_matrix_inverse}.
We estimate the numerical error of each mode by solving it twice, with $N = 64$ and $N = 48$ nodes per subdomain, and taking the largest relative difference among these quantities.
This difference is below $\sim 4 \times 10^{-23}$ for all the modes that we report.

\subsection{Data on the light ring}
\label{subsec:data_lightring}
We have applied this method to a particle exactly on the light ring, $r_{0} = 3$, for which $\omega = m / (3\sqrt{3})$.
For each mode, we have computed the transmission and reflection coefficients, ${\cal T}_{\ell m}$ and ${\cal R}_{\ell m}$, the complex-valued Regge-Wheeler functions $X^{\rm in / up}_{\ell m}$, as well as their derivatives with respect to $r_{*}$ at $r = r_{0}$.
Injecting these results into Eq.~\eqref{eq:flux_of_X_final} for the fluxes through infinity and the horizon, with the factor $\delta^{-1/2}$ of the coefficients $\kappa_{\ell m}$ and $\lambda_{\ell m}$ removed analytically, we obtain the fluxes multiplied by $\delta$, which are themselves finite at $\delta = 0$.

All of this data is reported in the ancillary file \texttt{data\_light\_ring.csv}, which lists, for each mode, the values of $\ell$, $m$, and $k$, the real and imaginary parts of $X^{\rm in}_{\ell m}$, $\dd X^{\rm in}_{\ell m} / \dd r_{*}$, $X^{\rm up}_{\ell m}$, $\dd X^{\rm up}_{\ell m} / \dd r_{*}$, ${\cal T}_{\ell m}$, ${\cal R}_{\ell m}$, the regularized fluxes $\delta F^{\infty}_{\ell m}$, $\delta F^{\cal H}_{\ell m}$, and the estimate of the numerical error as described above.
The file contains $12\,119$ modes with $k\le5$ and $\ell\ge2$. It is dense for $m\le2000$, where every value of $m$ is included, and sparse beyond, where we include $m=2^n$ and $m=2^n+1$ for $11\le n\le20$.
The dense data are used in Sec.~\ref{sec:flux_lightring} to perform numerical sums over $m$, whereas the sparse data give access to very high values of $m$, up to $m\approx 10^{6}$, in order to demonstrate that the numerical values converge to our analytical predictions.

This convergence is illustrated in Figs.~\ref{fig:error-T2-R2}-\ref{fig:error-phase}.
We first consider the moduli. Figure~\ref{fig:error-XYZ} shows the relative error of the WKB expressions~\eqref{eq:cal_XYZ_WKB} for the quantities $\mathcal{X}^\text{in/up}_{mk}$, $\mathcal{Y}^\text{in/up}_{mk}$ and $\mathcal{Z}^\text{in/up}_{mk}$, which are the combinations of the radial functions that enter the fluxes, and Fig.~\ref{fig:error-T2-R2} shows the relative error of the expressions~\eqref{eq:normam_squared_T_R_WKB} for $|\mathcal{T}_{mk}|^2$ and $|\mathcal{R}_{mk}|^2$. In all cases the error decreases monotonically at large $m$. At $m=2^{20}$ it ranges from $2\times10^{-4}$ for $k=0$ to $9\times10^{-3}$ for $k=5$ for the radial functions, and is below $2\times10^{-5}$ for the transmission and reflection probabilities, which converge faster.
We then turn to the phases.
Figure~\ref{fig:error-phase} shows the difference between the numerical values of the complex arguments of $\mathcal{T}_{mk}$, $\mathcal{R}_{mk}$ and $X^\text{in/up}_{mk}$ and the analytical expressions~\eqref{eq:complex_T_R_X_Xp}. For all values of $k$ this difference decreases as $m$ increases, falling below $10^{-4}$ at $m=2^{20}$, which confirms the phases $\widetilde{\mathcal{I}}$ and $\widetilde{\mathcal{J}}$ of Eq.~\eqref{eq:IJ_tilde_expressions} as well as the subleading logarithmic and constant terms.
Figure~\ref{fig:error-flux-m-wkb} shows the corresponding comparison for the regularized fluxes, whose relative error with respect to the WKB prediction also vanishes in the large-$m$ limit.

\section{Light ring divergence}
\label{sec:flux_lightring}

Now that we have full analytical control over each individual $(\ell,m)$ mode, we wish to perform the sum and analyze its behavior for orbits near the light ring. We will first look at how to study the sum over $m$ at any given value of $k$. We will then perform the sum over $k$, which is exponentially convergent---see Eq.~\eqref{eq:Flm_WKB}. Finally, we will compare our results with the numerical results we generate for this work. 

\subsection{Summing over $m$ for any given $k$}
\label{subsec:summing_m}
First we define
\begin{align}\label{eq:Fk_def}
    \mathcal{F}_k^{\infty/\mathcal{H}}=\sum_{m=1+[k=0]}^\infty \mathcal{F}_{m+k,m}^{\infty/\mathcal{H}}\,,
\end{align}
such that the total fluxes are given by 
\begin{align}\label{eq:F_as_sum_over_k}
    \mathcal{F}^{\infty/\mathcal{H}} = \sum_{k=0}^\infty \mathcal{F}_k^{\infty/\mathcal{H}}  \,.
\end{align}
In the lower bound of Eq.~\eqref{eq:Fk_def}, note that we have used the Iverson bracket: $[k=0]$ is equal to $1$ if $k=0$ and $0$ otherwise.

We now want to study the behavior of the total flux at the light ring, i.e., as $\delta \rightarrow 0^+$. Each individual $\mathcal{F}_{m+k,m}$ mode diverges like $1/\delta$ as $\delta\rightarrow 0$, so we will focus on the product $\delta \mathcal{F}_{m+k,m}$, which is finite at the light ring. 
First, we acknowledge that the WKB results are only valid for large enough values of $m$. Thus, we divide the flux sum into: (i) ``low $m$'' modes, which are computed by solving the Regge-Wheeler equation numerically; and (ii) ``high $m$'' modes, which are computed analytically using WKB. 
For each value of $k$, the flux sum is split at some switching point $m_*$ which is yet to be determined:\footnote{In contrast to Ref.~\cite{Barack:2026izc}, this approach allows us to numerically evaluate $\delta\mathcal{F}_{m+k,m}^{\infty/\mathcal{H}}$ at $\delta = 0$ for each $m$ without requiring an extrapolation based on values for the flux computed at $\delta >0$.}
\begin{align}
    \delta\mathcal{F}_k^{\infty/\mathcal{H}} &\simeq %=
    \sum_{m=1+[k=0]}^{m_*}\!\left[\delta\mathcal{F}_{m+k,m}^{\infty/\mathcal{H}}\right]^\text{num} \nonumber\\*
&\quad+\sum_{m=m_*+1}^{\infty}\!\!\!\!\left[\delta\mathcal{F}_{m+k,m}\right]^\text{WKB}. 
\label{eq:deltaFk_switch}
\end{align}

In order to compute the analytical WKB sum, we recognize that it has the structure of a \textit{harmonic sum}. Following Ref.~\cite{Flajolet:1995}, and adding a hat to any notation introduced in that reference, we notice that our sum can indeed be written in the form
\begin{align}
    \hat{G}_k(\hat{x}) &\equiv\sum_{m=m_*+1}^{\infty}\left[\delta {\mathcal{F}}_{m+k,m} \right]^\text{WKB} \nonumber \\*
    &=   \sum_{m= m_*+1}^{\infty} \hat{\lambda}_m \hat{g}_k(\hat{\mu}_m \hat{x}),
\end{align}
where in our case $\hat\lambda_m = m^{-1}$, $\hat\mu_m=m$, $\hat{x}  = \delta$ and  
\begin{align}
    \hat{g}_k(\hat{u})&=\frac{2^{4 \frak{p}(k) - k}  k!}{27 \pi^{3/2} (\lfloor k/2 \rfloor !)^2} \left(1+2k + \frac{\hat{u}}{2} \right)^{2[1-\frak{p}(k)]} \nonumber\\*
    &\times\de^{-\frac{\pi (1+2k+\hat{u})}{4}} \nonumber\\*
    &\times \left|\Gamma\left(\frac{1+2\frak{p}(k)+\di (1+2k+\hat{u})}{4} \right)\right|^2  \!.
    \label{eq:ghat_def}
\end{align}
The asymptotic behavior of harmonic sums is ideally studied using Mellin transforms. For any function $\hat{f}(\hat{x})$, we denote its Mellin transform with a star ($\star$): 
\begin{align}
    \hat{f}^\star(\hat{s}) &\equiv \int_0^\infty\dd \hat{u} \,\hat{u}^{\hat{s}-1}\,\hat{f}(\hat{u})\,.
\end{align}
The Mellin transform of a harmonic sum has a nice separability property:
\begin{align}
    \hat{G}_k^\star(\hat{s})= \hat{\Lambda}_k(\hat{s}) \hat{g}_k^\star(\hat{s}) \,,
\end{align}
where 
\begin{align}
    \hat{\Lambda}_k(\hat{s}) \equiv \sum_{m=m_*+1}^\infty \hat{\lambda}_m (\hat{\mu}_m)^{-\hat{s}}\,.
\end{align}
In our case, we immediately recognize the Riemann zeta function:
\begin{align}
    \hat{\Lambda}_k(\hat{s})  = \sum_{m=m_*+1}^\infty m^{-\hat{s}-1} =  \zeta(\hat{s}+1) - \sum_{m=1}^{m_*}{m^{-\hat{s}-1}} \,.
\end{align}

The motivation for introducing the Mellin transform is that the poles of $\hat{G}_k^\star(\hat{s})$ in the complex plane determine the asymptotic $\hat{x} \rightarrow 0$ behavior of the original harmonic sum~\cite{Flajolet:1995}. First, we recall that 
\begin{align}
    \hat{\Lambda}_k(\hat{s}) \underset{\hat{s}\rightarrow0^+}{\sim} \frac{1}{\hat{s}}-\psi(m_*+1) + o(1) \,,
\end{align}
where $\psi$ is the digamma function. Regarding the $\hat{s}\rightarrow 0$ behavior of  $\hat{g}_k^\star(\hat{s})$, we first notice that the exponential damping means that no singularity can arise from the large-$\hat{u}$  tail of the integral. The singularity only comes from the $\hat{u}\rightarrow 0$ limit, so we split the integral as follows:
\begin{align}
    \hat{g}_k^\star(\hat{s}) &= \hat{g}_k(0) \int_0^1\dd \hat{u} \,\hat{u}^{\hat{s}-1}\, +  \int_0^1\dd \hat{u} \,\hat{u}^{\hat{s}-1}\,\Big[\hat{g}_k(\hat{u})-\hat{g}_k(0)\Big] \nonumber\\*
    &\qquad+ \int_1^\infty\dd \hat{u} \,\hat{u}^{\hat{s}-1}\,\hat{g}_k(\hat{u}) \,.
\end{align}
The first integral is computed exactly and leads to a $1/\hat{s}$ divergence, whereas the second and third integrals are finite as $\hat{s}\rightarrow 0$. Thus, the singular structure reads
\begin{align}
    \hat{g}_k^\star(\hat{s}) &\underset{\hat{s}\rightarrow 0}{\sim}  \frac{A_k}{\hat{s}}+B_k + o(1),
\end{align}
where
\begin{subequations}
    \begin{align}
    A_k &= \hat{g}_k(0),\\
    B_k &= \int_0^1 \frac{\dd \hat{u}}{\hat{u}} \Big[\hat{g}_k(\hat{u})-\hat{g}_k(0)\Big] + \int_1^\infty\frac{\dd \hat{u}}{\hat{u}} \,\hat{g}_k(\hat{u}) \,.
    \label{seq:Bk_def}
\end{align}
\end{subequations}
The integrals in $B_k$ are evaluated numerically, as to our knowledge they are not known in closed form. Their numerical values for $k \in [\![0,5 ]\!]$ are reported in Table~\ref{tab:coeffs_k}.

The singular structure of $\hat{G}_k^\star(\hat{s})$ finally reads
\begin{align}
    \hat{G}^\star_k(\hat{s}) \sim \frac{A_k}{\hat{s}^2} + \frac{B_k-A_k \psi(m_*+1)}{\hat{s}}+ o(\hat{s}^{-1})
\end{align}
Using Theorem 5 of \cite{Flajolet:1995}, we then find that our initial harmonic sum has the singular structure
\begin{align}
    \hat{G}_k(\hat{x}) = -A_k \ln \hat{x} + B_k-A_k \psi(m_*+1) + o(1).
\end{align}
Translating back to our notation and adding up the divergent behavior of the ``low $m$'' and ``high $m$'' pieces of the sum, we finally find that
\begin{align}
    \mathcal{F}_k^{\infty/\mathcal{H}} \underset{\delta\rightarrow 0^+}{\sim} \frac{\alpha_k \ln (1/\delta)+\beta_k^{\infty/\mathcal{H}}(m_*)}{\delta}+o(\delta^{-1}) \,.
\end{align}
The coefficient of the logarithm solely arises from the large-$m$ asymptotics, which is controlled using WKB. It is universal, i.e., it is the same for the flux at infinity and down the horizon. It reads  
\begin{equation}\label{eq:alpha_k_formula}
\boxed{
\begin{aligned}
    \alpha_k &= A_k =   \frac{2^{4 \frak{p}(k) - k}  \left(1+2k \right)^{2[1-\frak{p}(k)]} k!\,\de^{-\frac{\pi (1+2k)}{4}} }{27 \pi^{3/2} (\lfloor k/2 \rfloor !)^2}   \\
    &\qquad\qquad\qquad\times\left|\Gamma\left(\frac{1+2\frak{p}(k)+\di (1+2k)}{4} \right)\right|^2.
\end{aligned}}
\end{equation}

The coefficient of the pole is different for the infinity and horizon fluxes and depends on the switch value $m_*$:
\begin{align}
    \beta_k^{\infty/\mathcal{H}}(m_*)  &= \sum_{m=1+[k=0]}^{m_*}\left[\delta\mathcal{F}_{m+k,m}^{\infty/\mathcal{H}}\right]^\text{num} \nonumber\\
    &\qquad+   B_k -  A_k\psi(m_*+1) \,.
\end{align}
We expect that $\beta_k^{\infty/\mathcal{H}}(m_*)$ becomes increasingly accurate as $m_*$ gets larger, since the WKB approximation is only valid for large values of $m$. Thus, we expect $\beta_k^{\infty/\mathcal{H}}(m_*)$ to converge to the true value as \mbox{$m_* \rightarrow \infty$.} Note that the WKB contribution  \mbox{$B_k-A_k\psi(m_*+1)$} diverges like \mbox{$- \alpha_k \ln m_*$} as \mbox{$m_* \rightarrow \infty$}. Thus, in order to ensure a finite limit for $\beta_k^{\infty/\mathcal{H}}(m_*)$, the sum of numerical fluxes must necessarily diverge like $+\alpha_k \ln m_*$. In order to sidestep this subtle cancellation, we use \mbox{$\psi(m_*+1) = \sum_{m=1}^{m_*}m^{-1}- \gamma_\text{E}$.} Rearranging, we find
\begin{align}
    \beta_k^{\infty/\mathcal{H}}(m_*) &=  \sum_{m=1+[k=0]}^{m_*}\Bigg\{\left[\delta\mathcal{F}_{m+k,m}^{\infty/\mathcal{H}}\right]^\text{num}-\frac{\alpha_k}{m} \Bigg\} \nonumber \\
    &\qquad +  B_k+ A_k \big(\gamma_\mathrm{E} - [k=0]\big),
\end{align}
where now the numerical sum is convergent. 

Numerically, we only have access to fluxes for \mbox{$m_* \lesssim 2000$}, and the convergence is slow, which makes it difficult to read off a limit. However, we can approximate $\beta_k^{\infty/\mathcal{H}}(m_*)$ very well by a power series in $(m_*)^{-1/2}$. More precisely, we fit the data corresponding to $m\in [2,m_*^\text{fit}]$ with the function
\begin{align}
    \Big[\beta_k^{\infty/\mathcal{H}}\Big]_\text{fit}(m_*) \equiv \sum_{N=0}^{N_\text{max}} \frac{K^{\infty/\mathcal{H}}_N(N_\text{max}, m_*^\text{fit})}{(m_*)^{N/2}}
\end{align}
using the method of least squares. 
We find that the value of the limit $K^{\infty/\mathcal{H}}_0(N_\text{max}, m_*^\text{fit})$ is stable with respect to the order of the fit, the upper end $m_*^{\textrm{fit}}$ of the fitted range and its lower end.
We take as our estimate for $\beta^{\infty/{\cal H}}_{k}$ the value obtained with $N_{\rm max} = 10$ and $m_*^{\textrm{fit}} = 2000$: these are presented in Table \ref{tab:coeffs_k} for $k \in [\![0,5]\!]$, together with the constants $B_{k}$ of Eq.~\eqref{seq:Bk_def}.
We estimate the uncertainty as half of the spread of $K^{\infty/{\cal H}}_{0}$ when varying $6 \leq N_{\rm max} \leq 14$, $500 \leq m_*^{\textrm{fit}} \leq 2000$ and the lower end of the fitted range between $m = 2$ and $m = 20$.
The main source of error comes from $k = 0$, for which we find $\Delta \beta^{\infty}_{0} \simeq \Delta \beta^{\cal H}_{0} \simeq 4 \times 10^{-7}$; the numerical error of the individual modes is negligible in comparison.
Since the uncertainties for different values of $k$ arise from the same fitting procedure, we add them linearly when summing over $k$.

\begin{table*}[t!]
    \centering
    \renewcommand{\arraystretch}{1.4}
    \setlength{\tabcolsep}{4pt}          % tighter columns (default 6pt)
    \resizebox{\textwidth}{!}{%
    \begin{tabular}{|c|c|c|c|c|c|c|}
    \hline
         $k$ & 0 & 1 & 2 & 3 & 4 & 5 \\
         \hline\hline
         $\alpha_k$ 
         & $1.851117197 \times 10^{-2}$ 
         & $2.511311843 \times 10^{-3}$ 
         & $1.834736415 \times 10^{-4}$ 
         & $1.106919522 \times 10^{-5}$ 
         & $6.154745799 \times 10^{-7}$ 
         & $3.249837297 \times 10^{-8}$ \\ \hline
         $\beta_k^\infty$
         & $+8.5362(4) \times 10^{-3}$
         & $-6.7129(1) \times 10^{-3}$
         & $-8.4528(5) \times 10^{-4}$
         & $-6.218(1) \times 10^{-5}$
         & $-3.9012(7) \times 10^{-6}$
         & $-2.245(2) \times 10^{-7}$ \\ \hline
         $\beta_k^\mathcal{H}$
         & $-3.47659(4) \times 10^{-2}$
         & $-5.7548(4) \times 10^{-4}$
         & $+4.254(3) \times 10^{-5}$
         & $+1.390(1) \times 10^{-5}$
         & $+1.780(3) \times 10^{-6}$
         & $+1.686(4) \times 10^{-7}$ \\ \hline
         $B_k$
         & $-1.181812977 \times 10^{-2}$
         & $-2.296346001 \times 10^{-3}$
         & $-1.772790944 \times 10^{-4}$
         & $-1.088442541 \times 10^{-5}$
         & $-6.113399693 \times 10^{-7}$
         & $-3.249905394 \times 10^{-8}$ \\ \hline
    \end{tabular}}
    \caption{\justifying Coefficients entering the divergent part of the fluxes for $k\in[\![0,5]\!]$. The coefficients $\alpha_k$ and $B_k$ follow from the WKB analysis, see Eqs.~\eqref{eq:alpha_k_formula} and~\eqref{seq:Bk_def}. The coefficients $\beta_k^{\infty/\mathcal{H}}$ combine the numerical fluxes at $\delta=0$ for $m\le2000$ with the WKB tail; the number in parentheses is the uncertainty on the last digit, estimated from the stability of the fit.}
    \label{tab:coeffs_k}
\end{table*}

\subsection{Summing over $k$}

We first focus on the logarithmic coefficient, and more specifically on its behavior as $k \rightarrow \infty$. We recall that  
\begin{align}
    \left|\Gamma(x+\di y)\right|^2 \underset{y\rightarrow + \infty}\sim  2 \pi y^{2x-1} \de^{-\pi y} \,.
\end{align}
Using this identity and the Stirling formula for the factorials, we find that 
\begin{align}
    \alpha_k \underset{k\rightarrow \infty}{\sim} \frac{8}{27 \pi}(2k+1)\de^{-\frac{(2k+1)\pi}{2}} \,,
\end{align}
regardless of the parity of $k$. Due to this exponential decay, the sum 
\begin{align}
    \alpha = \sum_{k=0}^\infty\alpha_k
\end{align}
is convergent, as expected. If the sum is truncated, keeping only $k \le k_\text{max}$, the absolute error \mbox{$\Delta \alpha \equiv \sum_{k=k_\text{max}+1}^\infty \alpha_k$} can be approximated by
\begin{align}
    \Delta \alpha  &\approx \frac{8}{27 \pi}\sum_{k=k_\text{max}+1}^\infty (2k+1)\de^{-\frac{(2k+1)\pi}{2}} \\
    &\approx  \frac{8\,\de^{-\pi(2k_{\max}+1)/2}}{27\pi(\de^\pi-1)} \left(2k_{\max}+3+\frac{2}{\de^\pi-1}\right)\,.\nonumber
\end{align}
We will choose $k_\text{max} = 5$, such that $ \Delta \alpha \approx 2 \cdot  10^{-9}$; we find numerically that this truncation is also suitable for the $\beta^{\infty/\mathcal{H}}$. We obtain the final values:
\begin{equation}\label{eq:alpha_beta_values}
\boxed{
\begin{aligned}
    \alpha = \sum_{k=0}^\infty\alpha_k &\approx +0.021217676364\ldots \\
    \beta^\infty = \sum_{k=0}^\infty\beta^\infty_k &\approx +0.0009117(6) \\
    \beta^\mathcal{H} = \sum_{k=0}^\infty  \beta^\mathcal{H}_k &\approx -0.0352830(5)
\end{aligned}}
\end{equation}
The $\alpha$ coefficient agrees with the numerical estimate of Eq.~(3) of Ref.~\cite{Barausse:2021xes}. Indeed, that work predicts \mbox{$k_1  \approx 0.12 \pm 0.01$.} We identify $\gamma = 1/\sqrt{3 \delta}$ (where $\gamma$ is defined in their paper and $\delta$ follows our conventions) and reexpand in small $\delta$, keeping only the logarithmic term. Then their constraint translates to $0.0183 \lesssim  \alpha \lesssim  0.0217$, which is consistent with our result.

Finally, note that there exists an integral representation of $\alpha$, without any sum over $k$; we derive this in App.~\ref{app:integral_alpha}. Note however that the sum representation converges extremely rapidly and remains the best choice for numerical estimates.

\begin{figure*}
    \includegraphics[width=\textwidth]{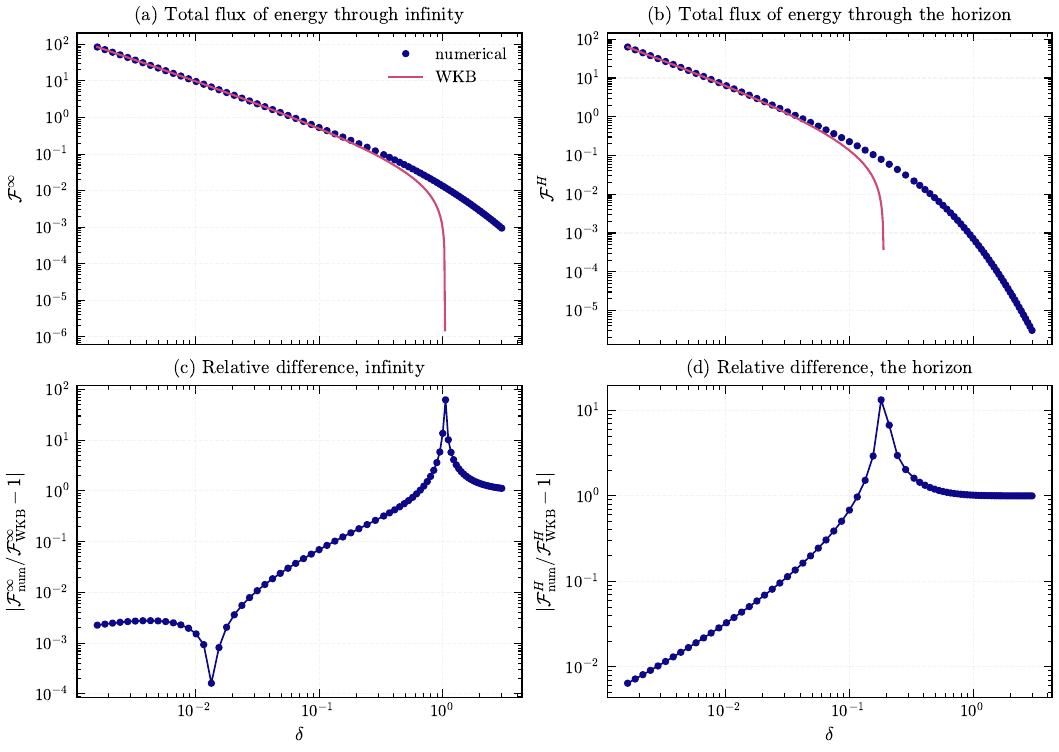}
    \caption{\justifying Comparison of the total fluxes of energy (through infinity and the horizon) obtained numerically and from the analytical prediction of Eq.~\eqref{eq:F_infty_H_total_alpha_beta}. The residual of the horizon flux has a clear falloff. The residual of the infinity flux goes through zero, and appears to then converge with a slower falloff. This could potentially be explained by missing subleading terms which we do not control.}
    \label{fig:comparison-error-total-flux} 
\end{figure*}

\subsection{Comparison with numerics}
\label{subsec:comparison_numerics}

We now compare the asymptotic expansion~\eqref{eq:F_infty_H_total_alpha_beta}, with the coefficients of Eq.~\eqref{eq:alpha_beta_values}, against numerical values of the total fluxes for orbits at $\delta > 0$.
This comparison is demanding on the numerical side.
The mode fluxes are only damped as $\de^{-\pi m \delta / 2}$, so the number of modes that must be summed grows like $1/\delta$ as the light ring is approached.
One of the authors previously produced a dataset of fluxes in Ref.~\cite{Barack:2026izc} for $\delta \in (0,3]$ by summing over $2 \le \ell \le 500$ and $-\ell \le m \le \ell$.
To probe the expansion closer to the light ring, we have generated a new dataset\footnote{The new dataset supersedes the previous one. In addition to its larger mode content, it is more accurate: the modes with \mbox{$\ell-m\geq2$} of the earlier dataset were affected by a minor normalization error, which we have corrected here. This changes the total fluxes at the ISCO, $r_0=6$, by a relative $4\times10^{-5}$ for the flux through infinity and $6\times10^{-4}$ for the flux through the horizon.} with the code described in Sec.~\ref{subsec:numerical_method}, now applied to the Teukolsky equation.

We compute the fluxes mode by mode from Eq.~\eqref{eq:Flm_of_Zlm}, with the amplitudes $Z^{\rm in/up}_{\ell m}$ obtained from the homogeneous solutions of the radial Teukolsky equation through Eq.~\eqref{eq:Zlm_of_R}.
We use the same hyperboloidal slicing and compactification as for the Regge-Wheeler equation, and rescale the Teukolsky radial function as
\begin{align}
    R_{\ell m}(r)&=\frac{\sigma}{2}\left[\frac{4(1-\sigma)}{\sigma^2}\right]^{2}\de^{\di\omega h(\sigma)}\,\psi_{\ell m}(\sigma),
    \label{eq:hyperboloidal_rescaling}
\end{align}
with $h$ given in Eq.~\eqref{eq:height_function}.
In addition to the oscillatory phase, this rescaling removes the power-law behavior of the ``up" solution at infinity and of the ``in" solution at the horizon.
The homogeneous Teukolsky equation then takes the form
\begin{multline}
    \sigma^2(1-\sigma)\,\psi''+\left[-4\di\omega-2\sigma+(8\di\omega-1)\sigma^2\right]\psi'\\*
    +\bigl[16\omega^2+8\di\omega-(\ell-1)(\ell+2)\\*
    +(1+16\omega^2)\sigma\bigr]\psi=0,
    \label{eq:hyperboloidal_teukolsky}
\end{multline}
which has the same structure as Eq.~\eqref{eq:hyperboloidal_RW} and is solved in exactly the same way.
The amplitudes then follow from $\psi^{\rm in/up}_{\ell m}$ and their first derivatives at $\sigma_0$.
We again estimate the numerical error of each mode from the relative difference between the fluxes obtained with $N=64$ and $N=48$.
Any mode for which this difference exceeds $10^{-20}$ is recomputed at 192 bits with a tighter tolerance on the subdomain widths; no mode required this.

We only compute modes with $m\geq1$ and use the $m\to-m$ symmetry, as in Eq.~\eqref{eq:Flm_of_Zlm}.
The mode sum is truncated adaptively.
For each $\ell$, we start from $m=\ell$ and decrease $m$ until four consecutive modes each contribute less than $10^{-20}$ of the sum at that $\ell$, for both the infinity and the horizon flux.
The sum over $\ell$ is terminated once the remainder, estimated from a geometric fit to the last ten values of $\ell$, falls below $10^{-18}$ of the total.

The dataset contains 78 radii in the range $3.001613\leq r_0\leq6$, i.e., $1.6\times10^{-3}\lesssim\delta\leq3$, and is provided in the ancillary file \texttt{flux\_UCO.csv}.
At the radius closest to the light ring the sum extends to $\ell_{\rm max}=15911$ and contains $3.2\times10^{5}$ modes with $m\geq1$; in total we have computed $2.5\times10^{6}$ modes.
The value of $\ell_{\rm max}$ grows like $1/\delta$, as expected from the exponential damping of the modes.
The estimated relative error on the total fluxes, obtained by summing the errors of the individual modes, is below $10^{-16}$ at all radii.
The numerical error is therefore negligible in the comparison below.

The comparison is presented in Fig.~\ref{fig:comparison-error-total-flux}.
The upper panels show the numerical fluxes through infinity and the horizon together with Eq.~\eqref{eq:F_infty_H_total_alpha_beta}, and the lower panels show their relative difference.
The asymptotic expansion captures the divergence of both fluxes as $\delta\to0^+$.
At the radius closest to the light ring, $\delta=1.6\times10^{-3}$, the relative difference is $2.3\times10^{-3}$ for the flux through infinity and $6.5\times10^{-3}$ for the flux through the horizon.
We stress that no information from the data at $\delta>0$ enters Eq.~\eqref{eq:F_infty_H_total_alpha_beta}: the coefficient $\alpha$ is known in closed form, and $\beta^{\infty/\mathcal{H}}$ are determined from the fluxes at $\delta=0$.
The agreement is therefore a genuine test of the expansion.

\section{Conjecture for Kerr fluxes}
\label{sec:Kerr}

We now consider the case of a particle on an unstable circular orbit near the light ring of a Kerr black hole.  We keep $M=1$ and choose a positive orbital frequency, so that the spin $a$ is positive for prograde orbits and negative for retrograde orbits. For $|a|<1$, we denote $r_0$ as the Boyer-Lindquist radius of the particle's orbit.

Chrzanowski and Misner \cite{Chrzanowski:1974nr} had obtained an expression for the flux modes in the case of a Kerr black hole---see their Eq.~(4.28). As pointed out in Sec.~\ref{sec:WKB_flux_modes}, our Schwarzschild result recovers their spectrum in the large-$m \delta$ expansion after correcting the even $k$ source factor. Here, rather than redoing a full derivation in Kerr, we use their spin-dependent prefactor and cutoff to make an educated guess.

We propose that the Kerr mode fluxes admit the following WKB expansion
\begin{align}    
&\mathcal{F}_{m+k,m}^{\infty/\mathcal{H}}(a) \sim\Phi(a)\frac{\hat g_k(m\delta(a))}{m\delta(a)},  
\label{eq:Flm_WKB_Kerr}
\end{align}
where $\hat g_k$ was defined in Eq.~\eqref{eq:ghat_def} and the global prefactor Chrzanowski and Misner derived reads
\begin{align}
    \Phi(a)&=\frac{54\sqrt3(r_\odot-1)}{r_\odot^{3/2}(r_\odot+3)^2},
 \label{eq:Kerr_Phi}
\end{align}
with the corresponding light ring radius~\cite{Bardeen:1972fi}
\begin{equation}
    r_{\odot}(a)
    =2\left[1+\cos\left(\frac23\arccos(-a)\right)\right].
    \label{eq:Kerr_light_ring}
\end{equation}
The scale $\delta(a)$ entering the WKB formula follows from the cutoff $m_{\rm crit}$ derived in Eq.~(2.35b) of \cite{Chrzanowski:1974nr}: expressed in terms of the orbital radius, their result gives \mbox{$4/(\pi m_{\rm crit})\sim\delta(a)$} near the light ring, where 
\begin{subequations}
    \begin{align}
\delta(a)&=\Psi(a)\bigl[r_0-r_\odot(a)\bigr],
 \\\Psi(a)&=\frac{4\sqrt{3r_\odot}}{(r_\odot-1)(r_\odot+3)}.
 \label{eq:Kerr_Psi}
\end{align}
\end{subequations}
Their barrier parameter $1+2k+4m/(\pi m_{\rm crit})$ therefore becomes 
\begin{equation}
    \epsilon(a) = 1 +2 k + m \delta(a).
    \label{eq:kerr_epsilon}
\end{equation}

In the Schwarzschild case ($a=0$), we recover $r_\odot=3$ and $\Phi(0)=\Psi(0)=1$, such that $\delta(0)=r_0-3$. We thus recover  our original definition for $\epsilon$ [Eq.~\eqref{seq:def_epsilon}] as well as the Schwarzschild expression for the flux [Eq.~\eqref{eq:Flm_WKB}].
Additionally, we note the special values of the global prefactor in the extremal Kerr cases: $\Phi(-1)=(81\sqrt{3})/196\approx0.715797$ and $\Phi(1)=0$. We also find that $\Phi(a)$ reaches a maximum for $a_\star=9/(5\sqrt{5})\approx0.804984$, with \mbox{$\Phi(a_\star)=(25/24)\sqrt{5/3}\approx1.34479$.} The factor $\Psi(a)$ instead increases monotonically with $a$: \mbox{$\Psi(-1)=8\sqrt{3}/21\approx0.659829$}, $\Psi(0)=1$ and $\Psi(a) \rightarrow +\infty$ as $a\to1^-$.

Assuming the correctness of Eq.~\eqref{eq:Flm_WKB_Kerr}, summing over modes follows identically to Sec.~\ref{sec:flux_lightring}. We then expect the flux to behave like
\begin{align}
    \mathcal{F}^{\infty/\mathcal{H}}(a) \sim \frac{\alpha(a)\ln\big[1/\delta(a)\big]+ \beta^{\infty/\mathcal{H}}(a)}{\delta(a)} + o\left(\frac{1}{\delta(a)}\right),
    \label{eq:Kerr_fluxWKB}
\end{align}
where $\alpha(a) =\Phi(a) \sum_k \hat{g}_k(0) = \Phi(a) \alpha(0)$. The $\beta^{\infty/\mathcal{H}}(a)$ coefficients require dense  numerical data on the Kerr light ring, which is beyond the scope of this work.

\begin{figure*}
    \centering
    \includegraphics[width=\linewidth]{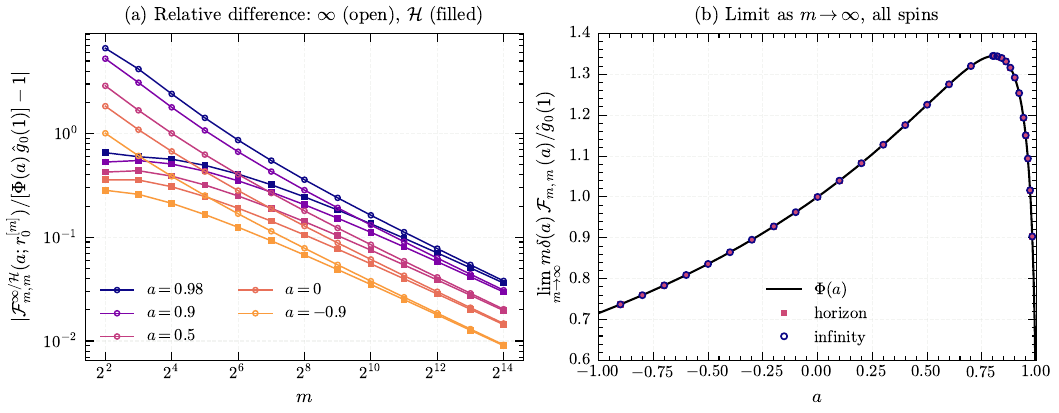}
    \caption{\justifying Test of the Kerr mode formula, Eq.~\eqref{eq:kerr_mode_limit}, for the modes $\ell = m$ ($k = 0$).
    For each spin and each $m = 2^{n}$ with \mbox{$2 \le n \le 14$}, the orbit is placed at $r^{[m]}_{0} = r_\odot(a)+1/[m\Psi(a)]$, such that $m\delta(a) = 1$, and we compute $\mathcal{F}_{m+k,m}^{\infty/\mathcal{H}}(a;r_0^{[m]})/\hat{g}_k(1)$, which Eq.~\eqref{eq:kerr_mode_limit} predicts tends to $\Phi(a)$.
    \textit{Panel (a).} The relative difference $|\mathcal{F}_{m,m}^{\infty/\mathcal{H}}/[\Phi(a)\hat g_0(1)]-1|$ as a function of $m$ for five values of the spin, for the fluxes through infinity (open circles) and through the horizon (filled squares).
     \textit{Panel (b).}  The limit $m\to\infty$, obtained by polynomial extrapolation in $m^{-1/2}$ from the five largest values of $m$, for 30 values of the spin in the range $-0.9\le a\le0.98$.
    The solid line is $\Phi(a)$; the extrapolated values agree with it to $10^{-4}$.}
    \label{fig:Kerr_Phi_modes}
\end{figure*}

Our goal is now to validate these analytical expressions against numerical results. The code described in Sec.~\ref{sec:comparison_with_numerical_results} applies equally to the Kerr geometry: the radial Teukolsky equation is solved on hyperboloidal slices with the same multidomain collocation method, while the spin-weighted spheroidal harmonics and their eigenvalues are obtained with a spectral decomposition onto spin-weighted spherical harmonics.

We first test these predictions numerically on a mode-by-mode basis. To this end, we introduce a sequence of radii \mbox{$r_0^{[m]} = r_\odot(a)+1/(m \Psi(a))$} and define \mbox{$\delta^{[m]}(a) = \Psi(a)\big[r_0^{[m]}-r_\odot(a)\big]$} such that for all $m$, \mbox{$m\delta^{[m]}(a) =1$} holds exactly. Using this last identity, it follows from Eq.~\eqref{eq:Flm_WKB_Kerr} that
\begin{align}
    \frac{\mathcal{F}_{m+k,m}^{\infty/\mathcal{H}}(a;r_0^{[m]})}{\hat{g}_k(1)} \underset{m\rightarrow \infty}{\longrightarrow} \Phi(a) \,.
    \label{eq:kerr_mode_limit}
\end{align}
We find that the numerical fluxes indeed converge as expected in this limit---this is illustrated in Fig.~\ref{fig:Kerr_Phi_modes}  for the $k=0$ case.

\begin{figure*}
    \centering
    \includegraphics[width=\linewidth]{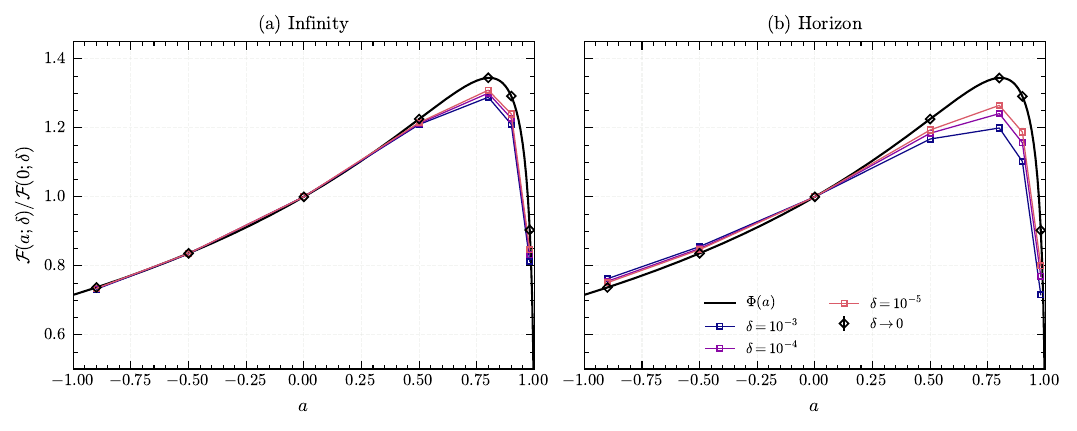}
    \caption{\justifying Ratio of the Kerr to the Schwarzschild total flux, ${\cal F}(a;\delta)/{\cal F}(0;\delta)$, at equal $\delta$, through infinity (a) and through the horizon (b), compared with $\Phi(a)$ (solid black). 
    Open squares are the totals at $\delta\in\{10^{-3}, 10^{-4}, 10^{-5}\}$, each summed over modes computed at that orbit. 
    Diamonds are the limit $\delta\to0$, obtained by fitting for the $(c_1,c_2,c_3)$ in the ansatz \eqref{eq:ansatz_Kerr_fit} %$P+c\,x+e\,\delta$, with $x=1/[\ln(1/\delta)+\beta(0)/\alpha]$, 
    %to the ratio
    based on numerical data for  $\delta\in\{10^{-3}, 3\times10^{-4},10^{-4}, 10^{-5}\}$.     
    The fitted $c_1$ agrees with $\Phi(a)$ to better than $5.4\times10^{-4}$ relative error at every spin.}
    \label{fig:Kerr_Phi_total}
\end{figure*}

We then want to probe the total flux as predicted in Eq.~\eqref{eq:Kerr_fluxWKB}. Since it is divergent at the light ring, the comparison is sensitive to the manner in which we approach it. For each value of $a$, we compare the flux for a radius $r_0 = r_\odot(a) + \delta(a)/\Psi(a)$ to the Schwarzschild flux at a \textit{corresponding} radius $r_0^\text{Schw}(a)= 3+\delta(a)$. This ensures that the quantity $\delta$ in the Schwarzschild case is equal to $\delta(a)$ in the Kerr case. From Eq.~\eqref{eq:Kerr_fluxWKB} we learn that
\begin{align}\label{eq:flux_ratio_Kerr_Schw}
    \frac{\mathcal{F}^{\infty/\mathcal{H}}_\text{Kerr}\big(a;r_0(a)\big)}{\mathcal{F}^{\infty/\mathcal{H}}_\text{Schw}\big(r_0^\text{Schw}(a)\big)} &\sim  \Phi(a)+\frac{\beta^{\infty/\mathcal{H}}(a)-\beta^{\infty/\mathcal{H}}(0) \Phi(a)}{\rho},  
\end{align}
where we have introduced
\begin{align}\label{eq:rho_def}
    \rho(a)\equiv\alpha(0) \ln(1/\delta(a))+\beta(0) \,,
\end{align}
where we have suppressed the $\infty$ or $\mathcal{H}$ labels for brevity.
We numerically compute the left-hand side of Eq.~\eqref{eq:flux_ratio_Kerr_Schw} for seven spins, $a\in\{-0.9,-0.5,0,0.5,0.8,0.9,0.98\}$, at $\delta(a)\in\{10^{-3}, 3\times10^{-4}, 10^{-4}, 10^{-5}\}$. 
At each of these orbits, all modes with $m\le200$ and $\ell-m\le9$ are computed, while at larger $m$ the modes are sampled logarithmically up to $m\delta(a)\approx10$ and the sum over $m$ is obtained by interpolation; on orbits where every mode was computed, this procedure reproduces the total flux to $10^{-7}$.
It is numerically challenging to get any closer to the light ring, so we rely on extrapolation, which requires a model for the subleading terms ignored in Eq.~\eqref{eq:flux_ratio_Kerr_Schw}. We have tested a few schemes: empirically, we found that the one that gave the best agreement consists in modeling the left-hand side of Eq.~\eqref{eq:flux_ratio_Kerr_Schw} with the ansatz 
\begin{align}\label{eq:ansatz_Kerr_fit}
    c_1 + \frac{c_2}{\rho(a)}+c_3\,\delta(a)
\end{align}
and fitting for $(c_1,c_2,c_3)$. We then compare the fitted value of $c_1$ to the theoretical prediction $\Phi(a)$.  In principle one would expect a $\sqrt{\delta(a)}$ dependence as well, but when adding such a term, we found that the fit led to an associated coefficient that was very small, so we have omitted it here. We also expect higher order terms in the expansion in $1/\rho(a)$, which all formally  fall off slower than $\delta(a)$. However, these led to worse agreement so we have discarded them. 
%Adding higher powers of $1/\rho(a)$, or other powers of $\delta(a)$, to this ansatz did not improve the agreement.
The results are reported in Fig.~\ref{fig:Kerr_Phi_total}.

A complete derivation of Eq.~\eqref{eq:Flm_WKB_Kerr} requires expanding the circular source and matching the local radial solutions to the boundary conditions at infinity and the horizon, which is left for future work. In particular, establishing rigorously the value of $\alpha$ in Eq.~\eqref{eq:Kerr_fluxWKB} also requires control of the full harmonic sum, particularly its large $k$ tail and the region $m\delta(a)\to0$ responsible for the logarithm. Moreover, the coefficients $\beta^{\infty/\mathcal H}(a)$ depend on the finite multipoles as well as the nonlogarithmic contribution of the large-$m$ sum. Finally, our conjecture concerns fixed subextremal spin $|a|<1$; the extremal limit requires a separate analysis~\cite{Wilkins:1972rs}. We therefore reserve the complete treatment of the Kerr case for future work.  
\section{Discussion}

In this work, we have revisited how to analytically solve the homogeneous radial Regge-Wheeler equation for a massive particle near the light ring of a Schwarzschild black hole. We obtain explicitly, for the first time, the complex argument of these solutions. Using a numerical Teukolsky solver, we have verified that the numerical solutions clearly asymptote to our analytical predictions. Then, using the Teukolsky formalism and a Chandrasekhar transformation, we obtain predictions for the mode decomposition of the energy fluxes through infinity and the horizon. We have also validated these results numerically and corrected several small mistakes in the literature of the 1970s~\cite{Breuer:1973kt,Chrzanowski:1974nr}. Based on these works, we made some predictions in the case of a Kerr black hole, which we validated numerically. Finally, we studied the divergent behavior of the total sum as the particle approaches the light ring by recognizing a harmonic sum and using a Mellin transform. We obtain an analytical expression for the logarithmic coefficient $\alpha$, which is common to the infinity and horizon fluxes. The distinct subleading pole coefficients $\beta^\infty$ and $\beta^{\mathcal H}$ are estimated by combining numerical modes on the light ring with the analytical large-$m$ tail.
%We thus obtain analytical predictions not only for the logarithmic divergence, but also numerical predictions for the subleading power-law divergence. 
In parallel, we computed highly accurate numerical results for the total flux on unstable circular orbits which improve the data provided with Ref.~\cite{Barack:2026izc}. This data provides an independent test of the predicted divergent behavior near the light ring, which fully confirms our predictions.
%We have checked that our data agrees well with our predictions for the divergent behavior near the light ring.

This work has highlighted the efficiency of analytical methods in regions of parameter space where numerical methods become very expensive. This was well known in the weak field~\cite{Warburton:2024xnr}, but here it is illustrated even in the strong field. An obvious extension of this work is that of a particle around a Kerr black hole. The setup can be straightforwardly extended by replacing the Regge-Wheeler equation by a Sasaki-Nakamura equation, and the Chandrasekhar transformation thus becomes a Sasaki-Nakamura transformation. As our preliminary results indicate, we believe that the final form of the flux should only undergo simple modification for nonzero spin $(a\neq0)$. This could be particularly useful for studying near-extremal Kerr black holes, which remain a challenge for numerical methods. A study of Reissner-Nordström black holes could also be of (academic) interest~\cite{Kokkotas:1988fm}.
Another direction is to improve the convergence rates of these results by pushing the expansion to higher order, e.g. by controlling  the $1/\sqrt{\delta}$ and the non-divergent contributions. This would require, among other things, a higher-order WKB treatment along the lines of \cite{Iyer:1986np}. Moreover, numerical studies have identified quasinormal oscillations, often called ``wiggles'', in waveforms following periapsis passage in eccentric bound orbits~\cite{Thornburg:2019ukt,Rifat:2019fkt,Nasipak:2019hxh}. We believe that the analytic WKB solutions derived here could help to understand analytically how these oscillations are excited, and investigate similar features in waveforms from ultrarelativistic scattering orbits near the light ring~\cite{Hopper:2017qus}.

Away from the light ring, WKB methods remain applicable, but the matching scheme must be adapted. For $\delta=\mathcal{O}(1)$ the two classical turning points are well separated and the potential between them is no longer well approximated by a parabola. The matching to parabolic cylinder functions used in this work must then be replaced by the standard treatment of two isolated turning points, in which Airy functions are matched to the eikonal solutions on either side of the potential barrier.  Of course, such an expansion cannot access the low-$m$ modes, which then dominate the flux. It does, however, provide analytical control over the large-$m$ tail of the mode sum.
We expect this control to be useful in several contexts.
In any frequency-domain calculation of the fluxes the mode sum must be truncated, and the neglected remainder is usually estimated by extrapolating the last computed modes, as we did in Sec.~\ref{subsec:comparison_numerics}.
An analytical expression for the tail would replace this extrapolation by a controlled remainder.
This is most valuable where the sum converges slowly, namely for orbits close to the separatrix, for highly eccentric orbits, and during the whirl phase of zoom-whirl and scattering orbits, whose fluxes are governed by the unstable circular orbits studied here.
For unbound orbits the spectrum is continuous, and the same barrier-penetration problem controls the high-frequency part of the radiated spectrum.

The large-$m$ tail also enters second order self-force calculations.
There, the source of the field equations is quadratic in the first order perturbation, such that each of its modes receives contributions from first order modes of arbitrarily high $\ell$~\cite{Miller:2016hjv}.
The slow convergence of this coupling is a known bottleneck, and closed-form expressions for the first order radial functions at large $\ell$, including their phases, could be used to sum its tail analytically.
%Finally, the large-$m$ behavior obtained here can be combined with post-Newtonian or numerical information on the low-$m$ modes to construct resummed models of the fluxes that remain accurate up to the light ring, in the spirit of Ref.~\cite{Barack:2026izc}.

Finally, although unstable circular orbits are not a configuration we expect to naturally encounter in nature, understanding their behavior can help give us a better analytical control of the fluxes as functions of the frequency. High-order post-Newtonian results for the fluxes on circular orbits in the extreme mass-ratio limit~\cite{Fujita:2012cm,Fujita:2014eta} are quite accurate in the weak field, but exhibit a slow and oscillatory behavior near the innermost stable circular orbit (ISCO) at $r_0=6$, which worsens when entering the unstable circular orbit regime. Understanding the asymptotics both in the weak and strong field limits could lead to resummations that improve the post-Newtonian expansion, making it highly accurate for all radii. Such ideas were already demonstrated in the resummation of enhancement functions of the eccentricity, where controlling both the $e \rightarrow 0$ and $e \rightarrow 1$ expansion can lead to extremely accurate analytical representation for all values of eccentricity~\cite{Trestini:2025yyc,Loutrel:2016cdw,Munna:2019fjz,Forseth:2015oua,Munna:2020juq}. Although such resummations should initially be studied in the extreme-mass ratio limit, they would be most useful if extended to the comparable mass regime where fewer coefficients are available~\cite{Blanchet:2023bwj} and where numerical methods (which are restricted to numerical relativity) are much more computationally expensive. Furthermore, for scattering observables, the analytic coefficient $\alpha$ and the estimated coefficients $\beta^{\infty/\mathcal H}$ characterize the ultrarelativistic behavior of the coefficient of the logarithmic enhancement of scattering losses near the separatrix. They therefore supply strong field input for resummations of post-Minkowskian radiative observables, including horizon absorption~\cite{Goldberger:2020wbx,Jones:2023ugm,Cipriani:2026myb,Bautista:2024emt}, and provide a benchmark for extending weak field scattering predictions towards capture~\cite{Barack:2026izc}.

\label{sec:discussion}

\vspace{-10pt}
\section*{Note}

While this work was near completion~\cite{Trestini:2026GdRtalk}, Barausse independently obtained many identical or very similar results~\cite{Barausse:2026kvp}. The two works were
carried out without knowledge of each other. One important difference in the approach is that Barausse studied the case of a \textit{massless} particle \textit{on} the light ring then argued how to extend the result to the case of a massive particle on     a timelike geodesic. In this work, we have assumed a massive particle \textit{near} the light ring all along, and looked at its behavior as it approaches the light ring. The expressions for $|X_{\ell m}(0)|$ and $|\partial_{r_*}X_{\ell m}(0)|$ are identical, but we 
also report the complex phase of these functions as well as their expression for $\delta>0$. In particular, $\epsilon= 2\eta_j^\text{Barausse} + m\delta$, such that the two coincide only when $\delta=0$.

We thus control not only the logarithmic divergence (which agrees with Barausse's result through \mbox{$\alpha_k= 2 \kappa_k^\text{Barausse}/3$} and \mbox{$\alpha= \kappa^\text{Barausse}/3$}), but also the subleading $\beta^\mathcal{H}$ and $\beta^\infty$, which have not been computed in that reference. These subleading coefficients were essential in fully reproducing the divergent behavior of the fluxes observed numerically; see Fig.~\ref{fig:comparison-error-total-flux}.

Finally, the two works differ in their numerical validation.
Barausse has developed some Runge-Kutta integrators for the Zerilli/Regge-Wheeler and
Teukolsky equations, which he validated against the \texttt{Black Hole Perturbation Toolkit}  
(reporting checks up to \mbox{$\ell = 12800$} in Schwarzschild and $\ell = 800$ in Kerr).
Here we have used a different, dedicated code, described in Sec.~\ref{sec:comparison_with_numerical_results}, which solves the radial Regge-Wheeler and Teukolsky equations on hyperboloidal slices with a multidomain spectral method in arbitrary-precision arithmetic.
This allowed us to obtain computationally demanding data at high precision in two complementary regimes.
Exactly on the light ring, $\delta=0$, we have computed the Regge-Wheeler functions, their derivatives and the transmission and reflection coefficients for $k\leq5$ up to $m=2^{20}\approx10^{6}$, with an estimated relative error below $10^{-22}$; this data determines $\beta^{\infty/\mathcal{H}}$ and confirms not only the moduli but also the complex phases of Eq.~\eqref{eq:complex_T_R_X_Xp}.
Away from the light ring, we have summed the fluxes down to $\delta\approx1.6\times10^{-3}$, where the mode sum extends to $\ell\approx1.6\times10^{4}$ and contains more than $3\times10^{5}$ modes with $m\geq1$, with an estimated relative error on the total fluxes below $10^{-16}$.

\vspace{-10pt}
\section*{Acknowledgments}
 
DT acknowledges the support of the ERC Consolidator/UKRI Frontier Research Grant GWModels (selected by the ERC and funded by UKRI [grant number EP/Y008251/1]) as well as financial support from the Centre national d’études spatiales (CNES), 
France (ROR: \url{https://ror.org/04h1h0y33}) within the framework of the LISA mission.
RG  acknowledges the support of the Royal Society University Research Fellowship ``High-precision effective field theory for extreme-mass-ratio inspirals'' [grant number URF\textbackslash R1\textbackslash 261173].
BL gratefully acknowledges funding from the European Union’s Horizon Europe research and innovation programme under the Marie Sklodowska-Curie grant agreement No. 101209791.
RG acknowledges OpenAI for providing access to advanced AI models through the ``ChatGPT for Academic Researchers'' programme. DT and BL acknowledge the use of Claude Pro (`Opus' model). The writing and computations were performed by the authors. AI tools were used (i) to assist in some of the mathematical proofs of the appendix; (ii) to help spot and resolve mistakes in the computations; and (iii) during the final stages of paper preparation, to assist with revisions and literature searches. 
This work makes use of the Black Hole Perturbation Toolkit~\cite{BHPToolkit}.

\appendix

\section{Coefficients entering the flux formulas}
\label{app:coeffs_flux}

The coefficients entering Eq.~\eqref{eq:Zlm_of_R} read
\begin{subequations}\begin{align}
  {}_{0}b_{\ell m}  &= \sqrt{\frac{(\ell-1)\ell(\ell+1)(\ell+2)\,r_0}{4\,(r_0-3)}}\;\,
                       {}_{0}Y_{\ell m}\!\left(\tfrac{\pi}{2},0\right) , \\[4pt]
  {}_{-1}b_{\ell m} &= \sqrt{\frac{(\ell-1)(\ell+2)}{r_0-3}}\;\,
                       {}_{-1}Y_{\ell m}\!\left(\tfrac{\pi}{2},0\right) , \\[4pt]
  {}_{-2}b_{\ell m} &= \sqrt{\frac{1}{r_0\,(r_0-3)}}\;\,
                       {}_{-2}Y_{\ell m}\!\left(\tfrac{\pi}{2},0\right) .
\end{align}\end{subequations}
The coefficients appearing in Eqs.~\eqref{eq:Z_in_up_of_X}~and~\eqref{eq:flux_of_X_final} are
\begin{widetext}
\begin{subequations}
\begin{align}
    \kappa_{\ell m}(r_0)&= \frac{\pi \omega_0^2}{32\,m\,\sqrt{r_0-3}(1-2\di m \omega_0)(1-4\di m \omega_0)}\nonumber\\*
   &\quad\times\Bigg\{ r_0^{2}\sqrt{\ell (\ell-1)(\ell+1)(\ell+2)}\,\Bigg[2 m (r_0-3)\sqrt{r_0}+2\,\di\,m^{2} r_0-\di (r_0-2)\left(\ell^2 r_0+\ell r_0-6\right)\Bigg]{\,}_0 Y_{\ell m}\left(\frac{\pi}{2},0\right) \nonumber\\*
   &\quad  -2\,r_0^{5/2}\sqrt{\ell^2+\ell-2}\,\Bigg[\ell^2\left(\di\,m \sqrt{r_0}+r_0-2\right)+\ell\left(\di\,m \sqrt{r_0}+r_0-2\right)-2 m^2\Bigg]{\,}_{-1}Y_{\ell m}\left(\frac{\pi }{2},0\right) \nonumber\\*
   &\quad  +\di\,r_0\Bigg[\ell (\ell-1)(\ell+1)(\ell+2)\, r_0^2+2\,\di\left(\ell^2+\ell-2\right) m\, r_0^{3/2}+12 r_0-24\Bigg]{\,}_{-2}Y_{\ell m}\left(\frac{\pi }{2},0\right)\Bigg\}, \\*
   \lambda_{\ell m}(r_0)&= \frac{\pi  \omega_0^2}{32\,m\,\sqrt{r_0-3}(1-2\di m \omega_0)(1-4\di m \omega_0)}\Bigg\{ 2r_0^{4}\sqrt{\ell (\ell-1)(\ell+1)(\ell+2)}\,\bigg[m\sqrt{r_0}-\di (r_0-3)\bigg]{\,}_0 Y_{\ell m}\left(\frac{\pi }{2},0\right)  \nonumber\\*
   &\quad-2r_0^{4}\sqrt{\ell^2+\ell-2}\,\bigg[\ell (\ell+1) \sqrt{r_0}+2\,\di\,m\bigg]{\,}_{-1}Y_{\ell m}\left(\frac{\pi }{2},0\right)  +2\di\,r_0^{3}\bigg[\left(\ell^2+\ell-2\right) r_0+6\bigg]{\,}_{-2}Y_{\ell m}\left(\frac{\pi }{2},0\right)\Bigg\}.
\end{align}\end{subequations}
\end{widetext}

\section{Derivation of the WKB solution}
\label{app:WKB_details}

In this Appendix, we provide details on the derivation of the WKB solution to the Regge-Wheeler equation and how we perform the asymptotic matching.

\subsection{Study of the potential}

Since $\delta$ only enters $Q$ through the constant \mbox{$\omega_0^2=(3+\delta)^{-3}$}, the position $\xmin$ of the minimum of $Q$ and its curvature are independent of $\delta$. As $m\to\infty$,
\begin{subequations}
\begin{align}\label{seq:app:xmin_precise}
    \xmin &= \frac{3}{m^2}-\frac{3(1+2k)}{m^3}+\mathcal{O}\left(\frac{1}{m^4}\right),\\*
    Q''(\xmin) &= \frac{2}{729}+\mathcal{O}\left(\frac{1}{m}\right),
\end{align}
\end{subequations}
and 
\begin{align}
    Q(\xmin)= \frac{1}{(3+\delta)^{3}}-\frac{1}{27}-\frac{1+2k}{27m}+\mathcal{O}(m^{-2}) \,.
\end{align}
We define
\begin{subequations}
 \begin{align}
    w &\equiv \sqrt{\frac{Q''(\xmin)}{2}} = \frac{1}{27}+\mathcal{O}\left(\frac{1}{m}\right),\\
    \epsilon &\equiv -\frac{m\,Q(\xmin)}{w} \\
    &\,= 1+2k + m\left[1-\left(1+\frac{\delta}{3}\right)^{-3}\right]+\mathcal{O}\left(\frac{1}{m},\delta\right), \nonumber
\end{align}   
\end{subequations}
where the remainder is uniform in $m$. 
Throughout this Appendix we assume $m\gg1$, $\delta\ll1$ and $m\delta^{3/2}\ll1$, which includes the regime \mbox{$m\delta=\mathcal O(1)$.}
The remaining modes, with $m\delta^{3/2 }\gtrsim 1$, do not contribute to
the flux at the order we consider because they are suppressed by
$\de^{-\pi m\delta/2}\lesssim \de^{-\pi/(2\sqrt\delta)}$. 
Since $1-(1+\delta/3)^{-3}=\delta+\mathcal{O}(\delta^2)$, we have
$\epsilon = 1+2k+m\delta+\mathcal{O}(1/m,\delta,m\delta^2)$, where
$m\delta^2\ll\sqrt\delta$ in the regime $m\delta^{3/2}\ll1$ considered here.

The classical turning points $x_\pm$, defined by $Q(x_\pm)=0$ with $x_-\le 0\le x_+$, are then
\begin{align}\label{eq:app:x_pm_expansion}
  x_\pm = \pm\sqrt{\frac{\epsilon}{mw}}+\frac{\epsilon}{m}+\mathcal{O}\left(\Big(\frac{\epsilon}{m}\Big)^{3/2}\right).
\end{align}
We then naturally define $r_\pm = r(x_\pm)$, and find that
\begin{align}\label{eq:app:r_pm_expansion}
    r_\pm = 3 \pm \sqrt{\frac{3\epsilon}{m}}+\frac{4\epsilon}{3m}+\mathcal{O}\left(\Big(\frac{\epsilon}{m}\Big)^{3/2}\right).
\end{align}
Since $\xmin=\mathcal{O}(m^{-2})$, we set $\xmin=0$ from now on. We then solve the RW equation in three regions: region~I towards the horizon ($x\to-\infty$), region~II around the peak ($|x|\ll1$, containing $x_\pm$), and region~III towards infinity ($x\to+\infty$), as detailed in the sections below.

%%%%%%%%%%%%%%%%%%
%%%%%%%%%%%%%%%%%%

\subsection{Exact parabolic solution in region II}

Region II is centered around the peak of the potential at $x\approx0$ and includes the two classical turning points. In this region we can approximate the potential by a parabola:
\begin{align}\label{eq:app:Qhat_def}
       \hat{Q}(x) \equiv Q(\xmin)+\frac{1}{2}Q''(\xmin)\,x^2 = w^2x^2-\frac{\epsilon w}{m}\,.
\end{align}
With this simplified potential, the RW equation can be mapped to Weber's equation~\cite{Weber:1869},
\begin{align}\label{eq:app:Weber_equation}
    \frac{\dd^2 y}{\dd t^2}+\left(\nu+\frac{1}{2}-\frac{1}{4}t^2\right)y(t)=0\,,
\end{align}
by choosing $y= X_{\ell m}$, $t=\sqrt{2mw}\,\de^{-\di \pi/4}x$ and \mbox{$\nu=-(1+\di \epsilon)/2$}. We then find that the most general homogeneous solution is
\begin{align}\label{eq:app:X_II_ansatz}
     X_\text{II}(x) = E \,D_\nu(t)+F \,D_\nu(-t),
\end{align}
where $D_\nu(t)$ are parabolic cylinder functions and $(E,F)$ are arbitrary constants that are determined by matching.

We now want to study the asymptotics of this solution as $x\rightarrow \pm \infty$.
For this, we will first require the asymptotic behavior of the parabolic cylinder functions~\cite{BenderOrszag1999}:
\begin{subequations}
\label{eq:app:parabolic_cylinder_asymptotics}
    \begin{align}
        D_\nu(z)&\sim z^\nu\de^{-z^2/4},
        \qquad \qquad \qquad \! \text{if }\arg z=-\frac{\pi}{4},\\*
            D_\nu(z)
            &\sim z^\nu\de^{-z^2/4} \nonumber\\*
            &\, -\frac{\sqrt{2\pi}\de^{\di\pi\nu}}{\Gamma(-\nu)}
            z^{-\nu-1}\de^{z^2/4}, \quad \text{if }\arg z=\frac{3\pi}{4}.
    \end{align}
\end{subequations}
%
\iffalse
\begin{subequations}\label{eq:app:parabolic_cylinder_asymptotics}
    \begin{align}
    D_\nu(z) &\underset{|z|\rightarrow\infty}{\sim} \exp\left(-\frac{1}{4}z^2 + \nu \ln z\right) \quad\text{if}
    \arg z = - \frac{\pi}{4}\\
    D_\nu(z) &\underset{|z|\rightarrow\infty}{\sim} \exp\left(-\frac{1}{4}z^2 + \nu \ln z\right)\quad  \text{if}  \arg z = + \frac{3\pi}{4}\nonumber\\
    &\quad- \frac{\sqrt{2\pi}}{\Gamma(-\nu)} \de^{\di \pi\nu}\exp\left(\frac{1}{4}z^2 - (\nu+1)\ln z\right)
\end{align}
\end{subequations}
\fi
%
For $x>0$, we have $\arg t=-\pi/4$ and $\arg (-t)=3\pi/4$, whereas for $x<0$, we instead have $\arg t=3\pi/4$ and $\arg (-t)=-\pi/4$. We thus obtain the asymptotic behavior
\begin{widetext}
\begin{subequations}\label{eq:app:X_II_asymptotics}
\begin{align}
     X_{\ell m}^\text{II}(x) \underset{x\rightarrow +\infty}{\sim} & \frac{\exp\left(\frac{\di m w x^2}{2}-\frac{\di \epsilon}{2}\ln x\right)}{\sqrt{x}} \Bigg\{\frac{E}{(2mw)^{1/4}} \de^{-\epsilon \pi/8+\di \pi/8-(\di \epsilon/4) \ln (2mw)}+\frac{F}{(2mw)^{1/4}} \de^{3\epsilon \pi/8-3\di \pi/8-(\di \epsilon/4) \ln (2mw)}\Bigg\}\nonumber\\
    &+ \frac{\exp\left(-\frac{\di m w x^2}{2}+\frac{\di \epsilon}{2}\ln x\right)}{\sqrt{x}}\frac{F\sqrt{2\pi}}{(2mw)^{1/4}\,\Gamma\left(\frac{1+\di\epsilon}{2}\right)} \de^{\epsilon \pi/8+\di \pi/8 + (\di \epsilon/4)\ln (2m w)}, \\
     X_{\ell m}^\text{II}(x) \underset{x\rightarrow -\infty}{\sim} & \frac{\exp\left(\frac{\di m w x^2}{2}-\frac{\di \epsilon}{2}\ln (-x)\right)}{\sqrt{-x}} \Bigg\{\frac{E}{(2mw)^{1/4}} \de^{3\epsilon \pi/8-3\di \pi/8-(\di \epsilon/4) \ln (2mw)}+\frac{F}{(2mw)^{1/4}} \de^{-\epsilon \pi/8+\di \pi/8-(\di \epsilon/4) \ln (2mw)}\Bigg\}\nonumber\\
    &+ \frac{\exp\left(-\frac{\di m w x^2}{2}+\frac{\di \epsilon}{2}\ln (-x)\right)}{\sqrt{-x}}\frac{E\sqrt{2\pi}}{(2mw)^{1/4}\,\Gamma\left(\frac{1+\di\epsilon}{2}\right)} \de^{\epsilon \pi/8+\di \pi/8 + (\di \epsilon/4)\ln (2mw)}.
\end{align}\end{subequations}
\end{widetext}

\subsection{WKB solution in regions I and III}

Regions I and III are located far from the classical turning points $x_{\pm}$, where the WKB solutions are valid. For $m\gg 1$, standard WKB methods lead to the `eikonal' solution to the RW equations:
\begin{subequations}\label{eq:app:X_I_III_ansatz}
\begin{align}
     &\hspace{-6pt}X_{\ell m}^\text{III}(x)= A \left[Q(x)\right]^{-1/4}\exp\left(-\di m \int_{\xp}^x \dd x' \sqrt{Q(x')}\right)\nonumber\\
     &\,\,\, +B \left[Q(x)\right]^{-1/4}\exp\left(\di m \int_{\xp}^x \dd x' \sqrt{Q(x')}\right), \\
     &\hspace{-6pt}X_{\ell m}^\text{I}(x)= C \left[Q(x)\right]^{-1/4}\exp\left(\di m \int^{\xm}_x \dd x' \sqrt{Q(x')}\right)\nonumber\\
     &\,\,\, +D \left[Q(x)\right]^{-1/4}\exp\left(-\di m \int^{\xm}_x \dd x' \sqrt{Q(x')}\right).
\end{align}\end{subequations}
The arbitrary constants $(A,B,C,D)$ are determined by matching to the chosen boundary conditions \eqref{eq:X_asymptotic}. For this, we use that   
$ Q(x) \underset{x\rightarrow \pm\infty}{\sim} \omega_0^2 + \mathcal{O}(x^{-2})$. Following  Bender and Orszag~\cite{BenderOrszag1999}, we  find that
\begin{subequations}\label{eq:app:integrals_sqrt_Q}\begin{align}
    m\int_{\xp}^x \dd x' \sqrt{Q(x')}&\underset{x\rightarrow +\infty}{\sim} m\omega_0 x- \rho_m^{(+)} \,, \\*
    m\int^{\xm}_x \dd x' \sqrt{Q(x')}&\underset{x\rightarrow -\infty}{\sim} -m\omega_0 x  -  \rho_m^{(-)}  \,,
\end{align}\end{subequations}  
where the $x$-independent contributions read
\begin{subequations}\label{eq:app:rho_def}\begin{align}
    &\hspace{-13pt}\rho_m^{(+)}(r_0) \equiv m\omega_0 \xp  + m\int_{\xp}^{+\infty} \dd x' \!\left[\omega_0-\sqrt{Q(x')}\right],\\
     &\hspace{-13pt}\rho_m^{(-)}(r_0) \equiv  -m\omega_0 \xm+m\int^{\xm}_{-\infty} \dd x' \!\left[\omega_0-\sqrt{Q(x')} \right].
\end{align}\end{subequations}
We will not need to control $\rho_m^{(\pm)}$ to compute the fluxes, it suffices to know that it is a constant with respect to $x$. In Sec.~\ref{subsec:rho_m_delta_0}, we will see how to compute $\rho_m^{(\pm)}$ in the limiting case where $r_0=3$, i.e., $\delta=0$.

We then find that asymptotic behaviors of WKB solutions read
\begin{subequations}\label{eq:app:X_I_III_asymptotics}
    \begin{align}
    &X_{\ell m}^\text{III}(x)\underset{x\rightarrow +\infty}{\sim} \frac{A \,\de^{-\di (m\omega_0 x - \rho_m^{(+)} )}  + B \, \de^{\di   (m\omega_0 x  - \rho_m^{(+)} )}}{\sqrt{\omega_0}}       \,, \\
     &X_{\ell m}^\text{I}(x)\underset{x\rightarrow-\infty}{\sim} \frac{C\,\de^{-\di  (m\omega_0 x +  \rho_m^{(-)})} +D\,\de^{\di (m\omega_0 x + \rho_m^{(-)})} }{\sqrt{\omega_0}}  \,.
\end{align}\end{subequations}
Comparing with standard boundary conditions of Eq.~\eqref{eq:X_asymptotic},
and accounting for the relation $x = r_*-3+2\ln2$ between the tortoise coordinates, we find that the arbitrary constants must satisfy 
\begin{align}\label{eq:app:matching_ABCD_calABinout}
    A &= \mathcal{A}^\text{in} \sqrt{\omega_0} \,  \de^{-\di \,\widetilde{\rho}_m^{(+)} } \,, &&&
    B&= \mathcal{A}^\text{out}\sqrt{\omega_0} \,  \de^{\di \,\widetilde{\rho}_m^{(+)} } \,,\nonumber \\
    C &= \mathcal{B}^\text{out} \sqrt{\omega_0}  \,\de^{\di \,\widetilde{\rho}_m^{(-)} } \,,&&&
    D&= \mathcal{B}^\text{in} \sqrt{\omega_0} \,   \de^{ -\di \,\widetilde{\rho}_m^{(-)} }\,,
\end{align}
where we have introduced 
\begin{subequations}\label{eq:app:rho_tilde_def}
    \begin{align}
    \widetilde{\rho}_m^{(+)}&=\mathcal{\rho}_m^{(+)}-m\omega_0(2 \ln2-3) \,,\\
    \widetilde{\rho}_m^{(-)}&=\mathcal{\rho}_m^{(-)}+m\omega_0(2 \ln2-3)\,.
\end{align}
\end{subequations}

\subsection{Parabolic WKB solution}

We define a `parabolic WKB solution' as the WKB solution obtained when using the parabolic potential $\hat{Q}(x)$ rather than the full potential $Q(x)$. The parabolic WKB solution will overlap with the full WKB in the buffer zones (between I and II or between II and III) and will exhibit unphysical divergence as $x\rightarrow \pm \infty$. As we shall see, these divergences have the same form as the asymptotics of the exact parabolic solution, and can hence be matched. The parabolic WKB solution is thus defined as
\begin{subequations}\label{eq:app:Xhat_I_III_ansatz}\begin{align}
    &\hspace{-10pt}\hat{X}_{\ell m}^\text{III}(x)= A \left[\hat{Q}(x)\right]^{-1/4}\exp\left(-\di m \int_{\xp}^x \dd x' \sqrt{\hat{Q}(x')}\right)\nonumber\\
    &\,+B \left[\hat{Q}(x)\right]^{-1/4}\exp\left(\di m \int_{\xp}^x \dd x' \sqrt{\hat{Q}(x')}\right), \\
    &\hspace{-10pt}\hat{X}_{\ell m}^\text{I}(x)= C \left[\hat{Q}(x)\right]^{-1/4}\exp\left(\di m \int^{\xm}_x \dd x' \sqrt{\hat{Q}(x')}\right)\nonumber\\
    &\, +D \left[\hat{Q}(x)\right]^{-1/4}\exp\left(-\di m \int^{\xm}_x \dd x' \sqrt{\hat{Q}(x')}\right).
\end{align}\end{subequations}
Notice that the $(A,B,C,D)$ coefficients are the same as in the full WKB solution \eqref{eq:app:X_I_III_ansatz} to allow for overlap in the buffer region. 

It is immediate to see that $\hat{Q}(x) \sim (w x)^2$ as \mbox{$x\rightarrow \pm \infty$}. We now study the behavior of the integrals, and find after the change of variables $s=x'\sqrt{mw/\epsilon}$ that 
\begin{align}\label{eq:app:rescaling_integral_sqrt_Qhat}
    &\int_{a}^b \dd x'\sqrt{\hat{Q}(x')}= \frac{\epsilon}{m}\int_{a\sqrt{mw/\epsilon}}^{b\sqrt{mw/\epsilon}}\dd s\sqrt{s^2-1} \nonumber \\*  &\qquad= \frac{\epsilon}{m}\left[\Xi\left(b\sqrt{\frac{mw}{\epsilon}}\right)-\Xi\left(a\sqrt{\frac{mw}{\epsilon} }\right)\right],
\end{align}
where we have introduced the antiderivative
\begin{align}\label{eq:app:def_Xi}
    \Xi(s) = \frac{s\sqrt{s^2-1}}{2}+ \frac{1}{4}\ln\left(\frac{s-\sqrt{s^2-1}}{s+\sqrt{s^2-1}}\right),
\end{align}
such that \mbox{$\Xi'(s)=\sqrt{s^2-1}$} for \mbox{$s>1$} and \mbox{$s<-1$.} We have \mbox{$\Xi(1)=0$} and \mbox{$\Xi(s)\sim (\pm1/2)\left(s^2-\ln(2|s|)-1/2\right)$} as \mbox{$s\rightarrow\pm \infty$.} We also notice that 
\begin{align}\label{eq:app:Xi_pm_1}
    \Xi\left(x_\pm\sqrt{mw/\epsilon}\right)\underset{m\rightarrow \infty}{\longrightarrow} \Xi(\pm1)=0 \,.
\end{align}
Thus, we find that
\begin{subequations}\label{eq:app:integral_sqrt_Qhat_asymptotics}\begin{align}
    &\int_{\xp}^x \dd x'\sqrt{\hat{Q}(x')} \nonumber\\*
    &\quad \underset{x\rightarrow\infty}{\sim} \frac{w x^2}{2}-\frac{\epsilon}{2m}\Bigg[\ln\left(2x\sqrt{\frac{mw}{\epsilon}}\right)+\frac{1}{2}%+2 \xi_+
    \Bigg], \\
    &\int^{\xm}_x \dd x'\sqrt{\hat{Q}(x')} \nonumber\\*
    &\quad \underset{x\rightarrow-\infty}{\sim} \frac{w x^2}{2} -\frac{\epsilon}{2m}\Bigg[\ln\left(-2x\sqrt{\frac{mw}{\epsilon}}\right)+\frac{1}{2}%-2 \xi_-
    \Bigg]. 
\end{align}
\end{subequations}
The asymptotic behavior of these functions then finally reads
\begin{subequations}\label{eq:app:Xhat_I_III_asymptotics}\begin{align}
    &\hat X_{\ell m}^\text{III}(x)\!\!\underset{x\rightarrow+\infty}{\sim}  \!\!\frac{\exp\left[-\frac{\di mw}{2}x^2+ \frac{\di \epsilon}{2}\ln x\right]}{\sqrt{wx}}  A \,  \de^{\frac{\di \epsilon}{4}[\ln(4mw/\epsilon)+1%+4\xi_+
    ]} \nonumber\\
    &\qquad\qquad\ \  +\frac{\exp\left[\frac{\di m w}{2}x^2- \frac{\di \epsilon}{2}\ln x\right]}{\sqrt{w x}}  B \, \de^{-\frac{\di \epsilon}{4}[\ln(4 m w/\epsilon)+1
    ]},  \\
    &\hat X_{\ell m}^\text{I}(x) \!\!\underset{x\rightarrow-\infty}{\sim} \!\!   \frac{\exp\left[\frac{\di m w}{2}x^2- \frac{\di \epsilon}{2}\ln(-x)\right]}{\sqrt{- w x}}  C \,\de^{-\frac{\di \epsilon}{4}[\ln(4mw/\epsilon)+1%-4\xi_-
    ]} \nonumber\\
    &\qquad\qquad\  +\frac{\exp\left[-\frac{\di m w}{2}x^2+ \frac{\di \epsilon}{2}\ln(-x)\right]}{\sqrt{-w x}}  D \, \de^{\frac{\di \epsilon}{4}[\ln(4mw/\epsilon)+1%-4\xi_-
    ]}.
\end{align}\end{subequations}

\subsection{Asymptotic matching}

We now wish to relate $(E,F)$ to $(A,B,C,D)$, or equivalently, to $(\mathcal{A}_{\ell m}^\text{in},\mathcal{A}_{\ell m}^\text{out},\mathcal{B}_{\ell m}^\text{in},\mathcal{B}_{\ell m}^\text{out})$, via asymptotic matching. For this, we will ask that the $x \rightarrow \pm\infty$ expansions of the parabolic solution in zone II coincide with the  $x \rightarrow \pm\infty$ expansions in zones I and III of the WKB solution \textit{for which we replaced $Q(x)$ by $\hat{Q}(x)$}~\cite{Misner:1972jf,Iyer:1986np}. We see that the parabolic solution at the peak and the parabolic WKB solution have the same asymptotics, which allows us to match them. We thus obtain four links between the coefficients $(E,F)$ and the coefficients $(A,B,C,D)$. 
Since the latter can be expressed in terms of $(\mathcal{A}_{\ell m}^\text{in}, \mathcal{A}_{\ell m}^\text{out},\mathcal{B}_{\ell m}^\text{in},\mathcal{B}_{\ell m}^\text{out})$, the intermediate $(A,B,C,D)$ coefficients can be eliminated. Since there are six parameters and four constraints, we recover that there are two degrees of freedom to specify boundary conditions. For example, one can specify $(\mathcal{A}_{\ell m}^\text{in},\mathcal{A}_{\ell m}^\text{out})$ and deduce the other coefficients, in particular $(\mathcal{B}_{\ell m}^\text{in},\mathcal{B}_{\ell m}^\text{out})$. We have reduced the problem to determining a unitary $S$-matrix associated to the black hole, i.e. determining the  complex-valued transmission and reflection coefficients $\mathcal{T}_{mk}$ and $\mathcal{R}_{mk}$. We recover (3.33) of~\cite{Iyer:1986np} and find that the relation can be written as a unitary S-matrix
\begin{align}\label{eq:app:S_matrix}
    \begin{pmatrix} \mathcal{B}^\text{out}_{mk} \\[4pt] \mathcal{B}^\text{in}_{mk} \end{pmatrix} = \begin{pmatrix} \frac{\mathcal{R}_{m k}}{\mathcal{T}_{m k}}& \frac{1}{\mathcal{T}_{m k}^*} \\[7pt]
    \frac{1}{\mathcal{T}_{m k}} & \frac{\mathcal{R}_{m k}^*}{\mathcal{T}_{m k}^*}\end{pmatrix}  \begin{pmatrix} \mathcal{A}^\text{out}_{mk} \\[4pt] \mathcal{A}^\text{in}_{mk} \end{pmatrix} \ ,
\end{align}
where we have identified the complex transmission and reflection coefficients  
\begin{subequations}\label{eq:app:T_R_expression_WKB}
    \begin{align}\label{seq:app:T_expression_WKB}
    \mathcal{T}_{m k} &= \frac{\Gamma\left(\frac{1+\di \epsilon}{2}\right)\de^{- \frac{\epsilon \pi}{4}}}{\sqrt{2\pi}}\de^{-\di \left(\widetilde{\rho}_m^{(+)} +\widetilde{\rho}_m^{(-)}\right)
    +\frac{\di\epsilon}{2} -\frac{\di\epsilon}{2}\ln\left(\!\frac{\epsilon}{2}\!\right)} \,,\\
    \label{seq:app:R_expression_WKB}
    \mathcal{R}_{m k} &= \frac{\Gamma\left(\frac{1+\di \epsilon}{2}\right)\de^{ \frac{\epsilon \pi}{4}}}{\sqrt{2\pi}}\de^{-2\di \widetilde{\rho}_m^{(-)}
    +\frac{\di\epsilon}{2} -\frac{\di\epsilon}{2}\ln\left(\!\frac{\epsilon}{2}\!\right) - \frac{\di \pi}{2}}  \,.
\end{align}
\end{subequations}
Taking their norm squared, we find
\begin{align}\label{eq:app:absT2_absR2_expression_WKB}
    | \mathcal{T}_{m k}|^2 = \frac{1}{1+\de^{\epsilon\pi}} &&\text{and}&& | \mathcal{R}_{m k}|^2= \frac{1}{1+\de^{-\epsilon\pi}} \,,
\end{align}
which agree with (3.37) of \cite{Iyer:1986np}. Note that we also recover exactly $|\mathcal{R}_{m k}|^2+|\mathcal{T}_{m k}|^2=1$, and the latter relation helps us invert the relation \eqref{eq:app:S_matrix}:
\begin{align}\label{eq:app:S_matrix_inverse}
    \begin{pmatrix} \mathcal{A}^\text{out}_{mk} \\[4pt] \mathcal{A}^\text{in}_{mk}\end{pmatrix} = \begin{pmatrix} -\frac{\mathcal{R}_{m k}^*}{\mathcal{T}_{m k}^*}& \frac{1}{\mathcal{T}_{m k}^*} \\[7pt] \frac{1}{\mathcal{T}_{m k}} & -\frac{\mathcal{R}_{m k}}{\mathcal{T}_{m k}}\end{pmatrix}  \begin{pmatrix} \mathcal{B}^\text{out}_{mk} \\[4pt] \mathcal{B}^\text{in}_{mk} \end{pmatrix}.
\end{align}

The matching conditions also yield the expression of the coefficients $\{E,F\}$ in terms of any two of the  coefficients $\{\mathcal{A}^\text{in}_{mk}, \mathcal{A}_{mk}^\text{out},\mathcal{B}_{mk}^\text{in},\mathcal{B}_{mk}^\text{out}\}$, which entirely specifies the Regge-Wheeler function near the peak. The  simplest form reads  
\begin{subequations}\label{eq:app:matching_E_F_of_calABin}\begin{align}
    E &= \frac{m^{\frac{1}{4}} \,\Gamma\left(\frac{1+\di \epsilon}{2}\right) \de^{-\frac{\epsilon\pi}{8}}}{(2w)^{\frac{1}{4}}\sqrt{\pi}\,r_0^{\frac{3}{4}} }\nonumber\\
    &\qquad\times\de^{\di\{- \widetilde{\rho}_m^{(-)}
    + \frac{\epsilon}{4} -\frac{\epsilon}{4}\ln\left(\!\frac{\epsilon}{2}\!\right) - \frac{\pi}{8}\}} \,\mathcal{B}_{mk}^\text{in},  \\
    F &= \frac{ m^{\frac{1}{4}} \,\Gamma\left(\frac{1+\di \epsilon}{2}\right)  \de^{-\frac{\epsilon\pi}{8}}}{(2w)^{\frac{1}{4}}\sqrt{\pi} \,r_0^{\frac{3}{4}}}\nonumber\\
    &\qquad\times\de^{\di\{-\widetilde{\rho}_m^{(+)}
    + \frac{\epsilon}{4} -\frac{\epsilon}{4}\ln\left(\!\frac{\epsilon}{2}\!\right) - \frac{\pi}{8} \}} \,\mathcal{A}_{mk}^\text{in}.
\end{align}\end{subequations}

\subsection{Standard ``in" and ``up" solutions}

A standard basis of solutions to the homogeneous radial RW equation is the pair of so-called ``in" and ``up" solutions, which are uniquely specified by their asymptotics.

The ``in" solution is defined by the conditions $\mathcal{B}^\text{in} = 0$ and $\mathcal{B}^\text{out}=1$, namely
\begin{subequations}\label{eq:app:Xin_asymptotic_condition}
\begin{align}
     X_{\ell m}^\text{in}(x) &\underset{x\rightarrow +\infty}{\sim} \mathcal{A}_{\ell m}^\text{in} \,\de^{-\di m \omega_0 r_*}+\mathcal{A}_{\ell m}^\text{out}\, \de^{\di m \omega_0 r_*},  \\*
     X_{\ell m}^\text{in}(x) &\underset{x\rightarrow -\infty}{\sim} \de^{-\di m \omega_0 r_*} \,, 
\end{align}\end{subequations}
where we recall that $r_* = x +3-2 \ln 2$.
From Eqs.~\eqref{eq:app:S_matrix_inverse} and \eqref{eq:app:matching_E_F_of_calABin}, we find that this implies
\begin{subequations}\label{eq:app:matching_E_F_for_in_solution}
\begin{align}
    E_\text{in}&=0,  \\
    F_\text{in} &=  \frac{2^{1/4} \de^{\frac{\pi\epsilon}{8}}{m}^{\frac{1}{4}}}{w^{\frac{1}{4}}\,r_0^{\frac{3}{4}}} \de^{\di\left[ \widetilde{\rho}_m^{(-)} 
    - \frac{\epsilon}{4} + \frac{\epsilon}{4}\ln\left(\!\frac{\epsilon}{2}\!\right) - \frac{\pi}{8}\right]}.
\end{align}\end{subequations}
Note that for the ``in" solution, \mbox{$\mathcal{A}_{\ell m}^\text{in}=1/\mathcal{T}_{mk}$} and \mbox{$\mathcal{A}_{\ell m}^\text{out}=-\mathcal{R}_{mk}^*/\mathcal{T}_{mk}^*$.} 

The ``up" solution is defined by the conditions \mbox{$\mathcal{A}_{\ell m}^\text{in} = 0$} and \mbox{$\mathcal{A}_{\ell m}^\text{out}=1$.} 
\begin{subequations}\label{eq:app:Xup_asymptotic_condition}
\begin{align}
     X_{\ell m}^\text{up}(x) &\underset{x\rightarrow +\infty}{\sim}  \de^{\di m \omega_0 r_*},  \\
     X_{\ell m}^\text{up}(x) &\underset{x\rightarrow -\infty}{\sim}\mathcal{B}_{\ell m}^\text{in} \de^{\di m \omega_0 r_*}+\mathcal{B}_{\ell m}^\text{out}\de^{-\di m \omega_0 r_*} \,. 
\end{align}\end{subequations}
From Eqs.~\eqref{eq:app:S_matrix} and \eqref{eq:app:matching_E_F_of_calABin}, we find that this implies 
\begin{subequations}\label{eq:app:matching_E_F_for_up_solution}
\begin{align}
    E_\text{up}&=  \frac{2^{1/4} \de^{\frac{\pi\epsilon}{8}}{m}^{\frac{1}{4}}}{w^{\frac{1}{4}}\,r_0^{\frac{3}{4}}}  \de^{\di\left[ \widetilde{\rho}_m^{(+)}  
    - \frac{\epsilon}{4} + \frac{\epsilon}{4}\ln\left(\! \frac{\epsilon}{2}\!\right) - \frac{\pi}{8}\right]}, \\
    F_\text{up}&= 0.
\end{align}
\end{subequations}
Also note that, for the up solution, \mbox{$\mathcal{B}_{\ell m}^\text{in}=1/\mathcal{T}_{mk}$} and \mbox{$\mathcal{B}_{\ell m}^\text{out}=\mathcal{R}_{mk}/\mathcal{T}_{mk}$.}

In both cases, it is the transmitted wave  that is normalized to $1$. We now want to evaluate our ``in" and ``up" solutions at the peak of the potential. For this, we first recall that the parabolic cylinder functions and their derivatives simplify at $x=0$: 
\begin{align}\label{eq:app:parabolic_cylinder_of_zero}
    D_{\nu}(0) = \frac{2^{\frac{\nu}{2}}\sqrt{\pi}}{\Gamma\left(\frac{1-\nu}{2}\right)} &&\text{and}&& D_{\nu}'(0)= -\frac{2^{\frac{1+\nu}{2}}\sqrt{\pi}}{\Gamma\left(-\frac{\nu}{2}\right)},
\end{align}
where in our case we recall that $\nu=-(1+\di \epsilon)/2$.
We find that evaluating the basis functions and their derivatives (with respect to $x$) at $x=0$ yields 
\begin{subequations}\label{eq:app:X_Xp_in_up_x_0_WKB}
    \begin{align}
     X_{mk}^\text{in}(x=0)&=  \frac{m^{\frac{1}{4}}}{w^{\frac{1}{4}}r_0^{\frac{3}{4}}}\frac{\sqrt{\pi}\de^{\frac{\epsilon\pi}{8}}}{ \Gamma\left(\frac{3+\di \epsilon}{4}\right)}\nonumber\\
    &\times
    \de^{\di\{\widetilde{\rho}_m^{(-)}  
     - \frac{\epsilon}{4} +\frac{\epsilon}{4}\ln\left(\!\frac{\epsilon}{4}\!\right) -  \frac{\pi}{8}\}},\\
     X_{mk}^\text{up}(x=0)&= \frac{m^{\frac{1}{4}}}{w^{\frac{1}{4}}r_0^{\frac{3}{4}}}\frac{\sqrt{\pi}\de^{\frac{\epsilon\pi}{8}}}{\Gamma\left(\frac{3+\di \epsilon}{4}\right)}\nonumber\\
    &\times
     \de^{\di\{\widetilde{\rho}_m^{(+)}  
     - \frac{\epsilon}{4} +\frac{\epsilon}{4}\ln\left(\!\frac{\epsilon}{4}\!\right) - \frac{\pi}{8} \}},\\
    {X_{mk}^\text{in}}{\!}'(x=0) &= \frac{m^{\frac{3}{4}} w^{\frac{1}{4}}}{r_0^{\frac{3}{4}}}\frac{2\sqrt{\pi}\de^{\frac{\epsilon\pi}{8}}}{\Gamma\left(\frac{1+\di \epsilon}{4}\right)}
     \nonumber\\
    &\times\de^{\di\{\widetilde{\rho}_m^{(-)}  
     - \frac{\epsilon}{4} +\frac{\epsilon}{4}\ln\left(\!\frac{\epsilon}{4}\!\right) - \frac{3\pi}{8} \}}, \\
    {X_{mk}^\text{up}}'(x=0)&= -\frac{m^{\frac{3}{4}} w^{\frac{1}{4}}}{r_0^{\frac{3}{4}}}\frac{2\sqrt{\pi}\de^{\frac{\epsilon\pi}{8}}}{\Gamma\left(\frac{1+\di \epsilon}{4}\right)}
    \nonumber\\
    &\times\de^{\di\{\widetilde{\rho}_m^{(+)}  
     - \frac{\epsilon}{4} +\frac{\epsilon}{4}\ln\left(\!\frac{\epsilon}{4}\!\right) -\frac{3\pi}{8}\}}.
\end{align}\end{subequations}
Within this WKB approximation, the values and derivatives of the ``in'' and ``up'' solutions at $x=0$ satisfy
\begin{equation}
\label{eq:app:Xin_over_Xup_x_0_WKB}
    \begin{aligned}
        \frac{X_{mk}^{\mathrm{in}}(0)}
             {X_{mk}^{\mathrm{up}}(0)}
        &=
        -\frac{(X_{mk}^{\mathrm{in}})'(0)}
              {(X_{mk}^{\mathrm{up}})'(0)}
        \\
        &=
        \frac{F_{\mathrm{in}}}{E_{\mathrm{up}}}
        =
        \de^{\di\left(
            \widetilde{\rho}_m^{(-)}
            -\widetilde{\rho}_m^{(+)}
        \right)}.
    \end{aligned}
\end{equation}
The constant Wronskian then reads
\begin{align}\label{eq:app:Wronskian_WKB}
   W_{mk}(x=0) &\equiv \Big[ X_{mk}^\text{in}\, X_{mk}^{\text{up}\,\prime} - X_{mk}^{\text{in}\,\prime}\, X_{mk}^\text{up}\Big](x=0) \nonumber\\*
   &= \frac{4\di\pi\, m\, \de^{\epsilon \pi/4}}{\Gamma\left(\frac{1+\di \epsilon}{4}\right)\Gamma\left(\frac{3+\di \epsilon}{4}\right)r_0^{3/2}}\nonumber \\*
   \,&\qquad\times\de^{\di \left[ \widetilde{\rho}_m^{(+)} +\widetilde{\rho}_m^{(-)} +\frac{\epsilon}{2}\ln\left(\!\frac{\epsilon}{4}\!\right)-\frac{\epsilon}{2}\right]} \,.
\end{align}

\subsection{Determination of $\rho_m^{(\pm)}$ when $\delta =0$}

\label{subsec:rho_m_delta_0}
The explicit study of $\rho_m$ is quite complicated and is not needed for the flux computation. Here, we will determine its large-$m$ behavior analytically in the limiting case where $\delta=0$. %In this case, the
Its expansion as $m\rightarrow \infty$ reads
\begin{align}
    \rho_m^{(\pm)} &= \sqrt{m\epsilon}\pm\frac{\epsilon}{3\sqrt{3}} \nonumber \\
    &\quad\pm m\int_{x_\pm}^{\pm\infty} \dd x'\left[\frac{1}{3\sqrt{3}}-\sqrt{Q(x')}\right] + o(1), 
    \label{eq:rho_m_pm_expression_delta_0}
\end{align}
where we have replaced $x_\pm$ by its expression in  Eq.~\eqref{eq:app:x_pm_expansion}.

Expanding the integrand in Eq.~\eqref{eq:rho_m_pm_expression_delta_0} for large $m$, we find the $\mathcal{O}(m)$ piece is integrable as $x\rightarrow0$, but not the $\mathcal{O}(m^0)$ piece. The integral is well defined because $x_-<0<x_+$, but the large-$m$ limit is not, because $x_\pm\rightarrow 0$ as $m\rightarrow \infty$. We thus introduce a buffer scale $x_b^{(\pm)}$ which is chosen such that $m^{-1/2} \ll |x_b^{(\pm)}|  \ll m^{-1/3}$. It is related to the scale $r_b^{(\pm)}$ through $x_b^{(\pm)} = r_b^{(\pm)} -3+2\ln(r_b^{(\pm)}-2)$.

In the inner regions $x_+<x<x_b^{(+)}$ and $x_b^{(-)}<x<x_-$, we replace $Q(x)$ by its parabolic approximation $\hat{Q}(x)$, which is accurate there. The integrals can then be performed in closed form thanks to Eq.~\eqref{eq:app:rescaling_integral_sqrt_Qhat}, with $\Xi(\pm1)=0$ at the turning points.  Since $|x_b^{(\pm)}|\gg\sqrt{\epsilon/m}$, we expand the result using $\Xi(S)\simeq\pm\big[S^2/2-1/4-\frac12\ln(2|S|)\big]$ for $S=x_b^{(\pm)}\sqrt{mw/\epsilon}\rightarrow\pm\infty$, replace $x_\pm$ by its expansion~\eqref{eq:app:x_pm_expansion} (including both terms), and neglect terms that vanish in the overlap region $\sqrt{\epsilon/m}\ll|x_b^{(\pm)}|\ll m^{-1/3}$. We find that
\begin{align}
    &\pm m\int_{x_\pm}^{x_b^{(\pm)}} \dd x' \left[\frac{1}{3\sqrt{3}}-\sqrt{\hat{Q}(x')}\right]\nonumber\\
    &= - \sqrt{m \epsilon} \mp \frac{\epsilon}{3\sqrt3} \pm \frac{m\,x_b^{(\pm)}}{3\sqrt3} - \frac{m\,\big(x_b^{(\pm)}\big)^2}{54} \nonumber\\
    &\qquad + \frac{\epsilon}{4} + \frac{\epsilon}{4} \ln\left(\frac{4m}{27\epsilon}\right) + \frac{\epsilon}{2} \ln \big|x_b^{(\pm)}\big| \,,
\end{align}
where one should pay attention to the global sign and ordering of the bounds. We can already see that the $\sqrt{m \epsilon}$ cancels against the one in Eq.~\eqref{eq:rho_m_pm_expression_delta_0}.

We then turn to the outer region. Changing variables from $x$ to $r$, we perform the large-$m$ expansion of the integrand at fixed $r\neq3$, which reads
\begin{align}
    \frac{1}{1-2/r}\left[\frac{1}{3\sqrt3}-\sqrt{Q(r)}\right]
    &= \sigma(r)
    +\frac{\epsilon}{2m}\lambda(r) + \mathcal{O}\left(\frac{1}{m^2}\right)
\end{align}
where
\begin{subequations}\begin{align} 
    \sigma(r) &= \frac{1}{1-2/r} \left[\frac{1}{3\sqrt{3}}-\sqrt{\frac{1}{27}-\frac{1-2/r}{r^2}} \right] \,,\\
    \lambda(r)&= \frac{1}{r^2}  \left(\frac{1}{27}- \frac{1-2/r}{r^2}\right)^{-1/2} \,.
\end{align}   
\end{subequations}
Notice, as mentioned at the beginning of the section, that $\lambda(r)$ is not integrable at $r=3$.
To perform the integral explicitly, we introduce the antiderivatives
\begin{subequations}\begin{align}\label{eq:app:def_Sigma}
    \Sigma(r) &= \frac{1}{3\sqrt{3}}\bigg\{ r-3+2\ln(r-2)  \nonumber\\
    &\qquad\qquad\qquad-\text{sg}(r-3) \Big[\Theta(r)- \Theta(3) \Big] \bigg\} \,,\\
    \Lambda(r) &= - \text{sg}(r-3) \nonumber\\
    &\qquad \times\ln\left(\frac{6\sqrt{3} \sqrt{r(r+6)} +12r+18}{|r-3|}\right) \,,
\end{align}\end{subequations}
where
\begin{align}
    \Theta(r) &= 8 \,\text{arctanh}\left(\frac{2}{3}\sqrt{1+\frac{6}{r}} - \frac{1}{3}\!\right) \nonumber\\
    &\qquad + 2 \ln(r-2) + \sqrt{r(r+6)} \,. 
    \label{eq:app:def_Theta}
\end{align}
These are defined such that $\Sigma'=\sigma$ and $\Lambda'=\lambda$ on $]2,3[$ and $]3,+\infty[$. Additionally, $\Sigma$ is $C^1$ at $r=3$, but $\Lambda$ is not even continuous as it diverges logarithmically there. 
We then find that the outer integrals are given by
\begin{subequations}\begin{align}
    &m\int_{r_b^{(+)}}^{+\infty} \frac{\dd r'}{1-2/r'}\left[\frac{1}{3\sqrt{3}}-\sqrt{Q(r')}\right]\nonumber \\*
    &=m\Big[  \Sigma(+\infty) - \Sigma(r_b^{(+)}) \Big]+ \frac{\epsilon}{2} \big[\Lambda(+\infty) - \Lambda(r_b^{(+)})\big]\nonumber\\*
    &= m\Sigma(+\infty) +  \frac{\epsilon}{2} \Lambda(+\infty) - \frac{m|x_b^{(+)}|}{3\sqrt3} + \frac{m\big(x_b^{(+)}\big)^2}{54} \nonumber \\*
    &\quad+ \frac{\epsilon}{2}\big[2 \ln 2 + 4 \ln 3 -\ln |x_b^{(+)}|\big]  ,\\*
    &m\int^{r_b^{(-)}}_{2} \frac{\dd r'}{1-2/r'}\left[\frac{1}{3\sqrt{3}}-\sqrt{Q(r')}\right]  \nonumber \\*
    &=  m\Big[\Sigma(r_b^{(-)})-\Sigma(2) \Big] + \frac{\epsilon}{2} \big[\Lambda(r_b^{(-)})-\Lambda(2) \big] \nonumber \\*
    &= - m \Sigma(2) - \frac{\epsilon}{2} \Lambda(2) - \frac{m|x_b^{(-)}|}{3\sqrt3} + \frac{m\big(x_b^{(-)}\big)^2}{54} \nonumber\\*
    &\quad + \frac{\epsilon}{2}\big[2 \ln 2 + 4 \ln 3 -\ln |x_b^{(-)}|\big]  ,
\end{align}\end{subequations}
where terms that vanish in $m$  have been neglected.

Adding together the inner and outer pieces, we gladly find that all contributions that depend on $x_b^{(\pm)}$ cancel out. We define $\mathcal{I}=\Sigma(+\infty)$ and $\mathcal{J}=-\Sigma(2)$ and we use $\Lambda(+\infty)=-\ln[6(2+\sqrt3)]$ and $\Lambda(2)=\ln[6(2+\sqrt3)^2]$. We then finally find that the expression for $\rho_m^{(\pm)}$ is  given, for $\delta =0$, by
\begin{subequations}\label{eq:app:rho_expansion}\begin{align}
    \rho_m^{(+)} &= m \mathcal{I}+\frac{\epsilon}{4}\left\{\ln\left[\frac{432\,(2-\sqrt3)^2\,m}{\epsilon}\right]+1\right\}+ o(1),\\
    \rho_m^{(-)} &= m\mathcal{J}+\frac{\epsilon}{4}\left\{\ln\left[\frac{432\,(2-\sqrt3)^4\,m}{\epsilon}\right]+1\right\} + o(1).
\end{align}\end{subequations}
where $\epsilon=1+2k$ and  
\begin{subequations}\label{eq:app:I_J_expressions}
\begin{align}
    \mathcal{I}
    &=1+\frac{6\ln(2+\sqrt3)-2\ln2-6}{3\sqrt3},\\
    \mathcal{J}
    &=1+\frac{6\ln(2+\sqrt3)-10\ln2-3}{3\sqrt3}.
\end{align}
\end{subequations}
Numerically, we find the values \mbox{$ \mathcal{I}\approx 1.099199$} and \mbox{$\mathcal{J} \approx 0.609379$}.

Finally, we can add the correction \eqref{eq:app:rho_tilde_def} linked to the change in tortoise coordinate. We find
\begin{subequations}\label{eq:app:rho_tilde_expansion}\begin{align}
    \widetilde{\rho}_m^{(+)} &= m \widetilde{\mathcal{I}}+\frac{\epsilon}{4}\left\{\ln\left[\frac{432\,(2-\sqrt3)^2\,m}{\epsilon}\right]+1\right\}+ o(1),\\
    \widetilde{\rho}_m^{(-)} &= m\widetilde{\mathcal{J}}+\frac{\epsilon}{4}\left\{\ln\left[\frac{432\,(2-\sqrt3)^4\,m}{\epsilon}\right]+1\right\} + o(1),
\end{align}\end{subequations}
where, in terms of Eqs.~\eqref{eq:app:I_J_expressions}, we have
\begin{subequations}\label{eq:app:IJ_tilde_3_expressions}
\begin{align}
    \widetilde{\mathcal{I}}
    &=\mathcal{I}-\frac{2\ln2-3}{3\sqrt3},\\
    \widetilde{\mathcal{J}}
    &=\mathcal{J}+\frac{2\ln2-3}{3\sqrt3}.
\end{align}
\end{subequations}

\section{Asymptotics of spin-weighted spherical harmonics}
\label{app:spherical_harmonic_asymptotics}

An explicit expression for ${}_sY_{\ell m}(\pi/2,0)$ in modern conventions is given in Eq.~(2.15) of \cite{Poisson:1993vp}. In that expression, we first replace $\ell=m+k$ in view of  studying the large $m$ asymptotics at fixed $k$ and $s$. The bounds of the sum involve `minimum' and `maximum' functions, but for large enough $m$, they simply reduce to $p_0 = m-s$ and $p_1= m-s+k$. We can then perform the following change of variables in the sum: $r=p-(m-s)$. We then obtain the following expression for ${}_sY_{\ell m}(\pi/2,0)$, making it explicit that the number of terms in the sum does not depend on $m$:
\begin{widetext}
  \begin{align}
    {}_sY_{\ell m}(\pi/2,0) &= \frac{(-1)^{m+k}\sqrt{(m+k+s)!(m+k-s)!(2m+k)!k!(2m+2k+1)}}{2^{m+k+1}\sqrt{\pi}}\nonumber\\
    &\times \sum_{r=0}^{k}\frac{(-1)^{r}}{r!(k-r)!(r+m-s)!(m+s+k-r)!}\,,
\end{align}
\end{widetext}
First, using the Stirling formula
\begin{align}\label{eq:app:stirling}
    \ln n! &= n \ln \left(\frac{n}{\de}\right) + \frac{1}{2} \ln (2\pi n) \nonumber\\*
    &\quad+\sum_{N=2}^{N_\text{max}} \frac{(-1)^N B_N}{N(N-1) n^{N-1}} + \mathcal{O}\left(\frac{1}{n^{N_\text{max}}}\right)\,,
\end{align}
where $B_N$ is the Bernoulli number (e.g. $B_2 =1/6 $),
we find that the term outside the sum has the following asymptotic behavior as $m \rightarrow \infty$:
\begin{align}
    &\frac{(-1)^{m\!+\!k}\!\sqrt{(m\!+\!k\!+\!s)!(m\!+\!k\!-\!s)!(2m\!+\!k)!k!(2m\!+\!2k\!+\!1)}}{2^{m\!+\!k\!+\!1}\sqrt{\pi}} \nonumber\\*
    &=(-1)^{m+k}2^{(1-k)/2}\,\pi^{1/4}\sqrt{k!}   \,\de^{-2m}m^{2m+3k/2+5/4} \nonumber\\*
    &\qquad\times\Bigg(1+\frac{30 k^2+54 k+24 s^2+17}{48 m} + \mathcal{O}\left(\frac{1}{m^2}\right)\Bigg) 
    \,.
\end{align}

We now wish to estimate the behavior of the sum: for this, we anticipate that it is not enough to study the leading orders of the asymptotic behavior of the summand because these leading terms will vanish upon summation. Instead, we study the summand to all orders in the large $m$ expansion. Using the Stirling formula~\eqref{eq:app:stirling}, we find that it behaves as
\begin{align}\label{eq:app:combination_expanded_as_exp}
    &\frac{1}{(r+m-s)!(m+s+k-r)!}\nonumber \\*
    &\qquad\sim \frac{\de^{2m}}{2\pi \,m^{2m+k+1}} \exp\left(\sum_{a=1}^{\infty} \frac{\varphi_a(r)}{m^a}\right) \,,
\end{align}
where $\varphi_a(r)$ is expressed in terms of the Bernoulli polynomials $B_a(x)$ as
\begin{align}
    \varphi_a(r) &= \frac{(-1)^a}{a(a+1)}\bigg\{B_{a+1}(1-s+r)\nonumber\\*
    &\qquad \qquad \qquad +B_{a+1}(k+s+1-r)\bigg\}.
    \label{eq:app:varphi_as_poly_in_r}
\end{align}
Using the addition theorem for Bernoulli polynomials---see \href{https://dlmf.nist.gov/24.4.E12}{(24.4.12)} of Ref.~\cite{NIST:DLMF}---we can explicitly write $\varphi_a(r)$ as a polynomial in $r$, namely
\begin{align}
    \varphi_a(r) = \sum_{b=0}^{a+1} \psi_{a,b} r^b \,,
\end{align}
with
\begin{align}
    &\hspace{-10pt}\psi_{a,b} = \frac{(-1)^a}{a(a+1)} \frac{(a+1)!}{b!(a+1-b)!} \nonumber\\*
    &\times \bigg\{B_{a+1-b}(1-s)+(-1)^b B_{a+1-b}(k+s+1)\bigg\}.
\end{align}
Notice that $\psi_{a,a+1} = -\tfrac{2}{a(a+1)}\neq 0$ if $a$ is odd and $\psi_{a,a+1} = 0$ if $a$ is even. Similarly, $\psi_{a,a}= \tfrac{k+1}{a}\neq0$ if $a$ even and $\psi_{a,a}= \frac{k+2s}{a}\neq0$ if $a$ odd. Therefore, $\varphi_a(r)$ is a polynomial in $r$ of degree $\text{deg }  \varphi_a(r)=a+\frak{p}(a)$.
Upon reexpansion, and exchanging sums and products, we find that we can express the sum in terms of partitions:
\begin{align}
\hspace{-10pt}\exp\left(\sum_{a=1}^{\infty} \frac{\varphi_a(r)}{m^a}\right) &= \sum_{i=0}^\infty\frac{1}{m^i} \sum_{\substack{\kappa_1,\dots,\kappa_i \ge 0 \\ 
\sum_{b=1}^{i} b\,\kappa_b = i}} 
\prod_{a=1}^i \frac{\varphi_a(r)^{\kappa_a}}{\kappa_a ! } \,.
\end{align}
We will come back to this complicated expression later. For now, we only use the fact that $\varphi_a(r)$ is polynomial in $r$ of degree $a+\frak{p}(a)$, therefore the whole coefficient of $m^{-i}$ in this expression is a polynomial in $r$. For each partition $(\kappa_1, \cdots, \kappa_i)$, the corresponding term is  of degree
\begin{align}\label{eq:app:degree}
    \sum_{a=1}^i \Big(a+\frak{p}(a)\Big)\kappa_a = i + \sum_{b= 0}^{\lfloor \frac{i-1}{2} \rfloor} \kappa_{2b+1} \,.
\end{align}
The maximum degree of this polynomial is thus $2 i$, which is reached for the partition $(i,0,\cdots,0)$. Thus, we learn that our summand takes the form
\begin{align}\label{eq:app:combination_expanded_as_sum}
    &\frac{1}{(r+m-s)!(m+s+k-r)!}\nonumber \\*
    &\qquad= \frac{\de^{2m}}{2\pi \,m^{2m+k+1}} \sum_{i=0}^\infty\frac{1}{m^i}\sum_{j=0}^{2i}r^{j}\alpha_{ij}(k,s) \,,
\end{align}
where $\alpha_{00}(k,s)=1$. The $\alpha_{ij}(k,s)$ coefficients have very complicated expressions in general, but we will only require a subset of them: we keep their calculation for the end. Applying the sum on $r$ and inverting the order of the sums, we find
\begin{align}
   &\sum_{r=0}^{k}\frac{(-1)^{r}}{r!(k-r)!(r+m-s)!(m+s+k-r)!} \nonumber\\*
    &\qquad=  \frac{(-1)^k\de^{2m}}{2\pi \,m^{2m+k+1}} \sum_{i=0}^\infty\frac{1}{m^i}\sum_{j=0}^{2i}\alpha_{ij}(k,s) \begin{Bmatrix}
        j\\k
    \end{Bmatrix},
\end{align}
where we have introduced the Stirling numbers of the second kind, defined in \href{https://dlmf.nist.gov/26.8}{(26.8.6)} of \cite{NIST:DLMF} as 
\begin{align}
    \begin{Bmatrix}
        j\\k
    \end{Bmatrix}
    \equiv (-1)^k \sum_{r=0}^k \frac{(-1)^r r^j}{r! (k-r)!} \,.
\end{align}
The Stirling numbers vanish as soon as $k>j$, so the terms in the $j$-sum are nonvanishing only if $j\in [\![k,2i]\!]$. This means that the whole $j$-sum is nonvanishing only when $k\le 2i$, which in turn restricts the $i$-sum to the cases $i \ge \lceil k/2 \rceil$. 

Keeping only nonvanishing terms and reindexing the $j$ sum, we find
\begin{align}
    &\sum_{r=0}^{k}\frac{(-1)^{r}}{r!(k-r)!(r+m-s)!(m+s+k-r)!} \nonumber\\*
    &\qquad
    \sim  \frac{(-1)^k\de^{2m}}{2\pi \,m^{2m+k+1+\lceil k/2 \rceil}}  \nonumber\\*
    &\qquad \times\sum_{i=0}^\infty\frac{1}{m^i} \!\!\sum_{j=0}^{\frak{p}(k)+2i}\alpha_{\lceil k/2 \rceil+i,\, j+k}(k,s) \begin{Bmatrix}
        k+j\\k
    \end{Bmatrix},
\end{align}
where we have introduced 
\begin{align}
    \frak{p}(k)=2\lceil k/2 \rceil-k = \begin{cases}
        0 & \text{if $k$ is even}\\
        1 & \text{if $k$ is odd}
    \end{cases} \,.
\end{align}
We can then multiply with the prefactor, and we find that at  the required subleading order, the spin-weighted harmonics read
\begin{widetext}
\begin{align}\label{eq:app:sYlm_expansion_beta_gamma}
     {}_sY_{m+k,m}(\pi/2,0) 
     &\sim \frac{(-1)^m\sqrt{k!}m^{1/4-\frak{p}(k)/2}}{2^{\frac{k+1}{2}} \pi^{\frac{3}{4}}}  \Bigg(\beta_k(s)
    +\frac{\gamma_k(s)}{m}\Bigg)\Bigg[1+ \frac{1}{m}\Bigg(\frac{17}{48}+\frac{9 k}{8}+\frac{5 k^2}{8}+\frac{s^2}{2}\Bigg)+\mathcal{O}\left(\frac{1}{m^2}\right)\Bigg],
\end{align}
\end{widetext}
where we have defined
\begin{subequations}\label{eq:app:beta_gamma_inTermsOf_alpha}
    \begin{align}
    \beta_k(s) &\equiv \sum_{j=0}^{\frak{p}(k)}\alpha_{\lceil k/2 \rceil,\, j+k}(k,s) \begin{Bmatrix}
        k+j\\k
    \end{Bmatrix}, \\
    \gamma_k(s) &\equiv \sum_{j=0}^{\frak{p}(k)+2}\alpha_{\lceil k/2 \rceil +1,\, j+k}(k,s) \begin{Bmatrix}
        k+j\\k
    \end{Bmatrix} .
\end{align}
\end{subequations}
We thus learn that at subleading order in $m$, we only require $\alpha_{i,\,2i-n}(k,s)$ for $n\in [\![ 0,3]\!]$. Inspecting Eq.~\eqref{eq:app:degree}, we see that there are only three partitions that allow the degree to be greater than or equal to $2i-3$: $(i,0, \cdots,0)$, $(i-2,1, 0, \cdots,0)$, and $(i-3,0,1, 0, \cdots,0)$. Thus, we have
\begin{align}
&\exp\left(\sum_{a=1}^{\infty} \frac{\varphi_a(r)}{m^a}\right) = \sum_{i=0}^\infty\frac{1}{m^i}  \Bigg\{\frac{\varphi_1(r)^i}{i!} + \frac{\varphi_1(r)^{i-2}\varphi_2(r)}{(i-2)!}\nonumber\\*
&\qquad\qquad\qquad\qquad\qquad+ \frac{\varphi_1(r)^{i-3}\varphi_3(r)}{(i-3)!} +\mathcal{O}(r^{2i-4})\Bigg\}\,.
\end{align}
All we have left to do is reexpand $\{\varphi_1(r),\varphi_2(r),\varphi_3(r)\}$ as polynomials in $r$ using Eq.~\eqref{eq:app:varphi_as_poly_in_r}, 
compare this expression to Eqs.~\eqref{eq:app:combination_expanded_as_exp} and \eqref{eq:app:combination_expanded_as_sum}
and discard all but the first few dominant orders in $r$. We then have all the elements to read off the required coefficients:
\begin{subequations}\begin{align}
  \alpha_{i,\,2i}(k,s)   &= \frac{(-1)^{i}}{i!}, \\
  \alpha_{i,\,2i-1}(k,s) &= \frac{(-1)^{i-1}(k+2s)}{(i-1)!},\\
  \alpha_{i,\,2i-2}(k,s) &= \frac{(-1)^{i}}{(i-1)!}\nonumber\\
      &\hspace{-34pt}\times
   \Bigg[(2i-1)\left(s+\frac{k}{2}\right)^2 +\frac{3k^2+6ik+2i^2}{12}\Bigg],\\
  \alpha_{i,\,2i-3}(k,s) &= \frac{2(-1)^{i-1}}{(i-2)!}\left(s+\frac{k}{2}\right)\nonumber\\
      &\hspace{-34pt}\times\Bigg[\frac{2i-1}{3}\left(s+\frac{k}{2}\right)^2+\frac{3k^2+6ik+2i^2}{12}\Bigg].
\end{align}\end{subequations}
Using these expressions, as well as \href{https://dlmf.nist.gov/26.8}{(26.8.4)} and \href{https://dlmf.nist.gov/26.8}{(26.8.16)} of Ref.~\cite{NIST:DLMF} to simplify the Stirling numbers of the second kind, we find that Eq.~\eqref{eq:app:beta_gamma_inTermsOf_alpha} reduces to
\begin{subequations}
\begin{align}
    \beta_k(s)
    &=
    \frac{(-1)^{\lfloor k/2\rfloor}}{\lfloor k/2\rfloor!}  (2s)^{\frak{p}(k)},
    \\
    \gamma_k(s)
    &=
    -\beta_k(s)
    \begin{cases}
        \begin{aligned}[t]
            &\frac58 k^2+\frac34 k+\frac16 +(k+1)s^2
        \end{aligned}
        & \text{if $k$ even},
        \\[6pt]
        \begin{aligned}[t]
            &\frac58 k^2+\frac76 k+\frac38  +\frac{k+2}{3}s^2 
        \end{aligned}
        & \text{if $k$ odd}.
    \end{cases}
\end{align}
\end{subequations}
Injecting these expressions into \eqref{eq:app:sYlm_expansion_beta_gamma}, we finally recover the subleading asymptotics expansion of the spin-weighted spherical harmonics; see Eq.~\eqref{eq:sYlm_asymptotics} in the main text. In the case $s=0$, and at leading order, we recover the general expression given in (B15) of \cite{Chrzanowski:1974nr}. 

\section{Integral representation of $\alpha$}
\label{app:integral_alpha}

In this appendix, we derive an integral representation of $\alpha$. Using the integral representation of the gamma function, for real-valued $x$ and $y$, we have
\begin{align}
    |\Gamma(x+\di y)|^2 = \int_0^{\infty} \dd t \int_0^{\infty} \dd t' \,t^{x + \di y - 1} t'^{x - \di y - 1} \de^{-(t+t')} \,.
\end{align}
Performing the change of variables \mbox{$\tau = \tfrac{1}{2} \ln(t/t')$} 
and $\tau' = t+t'$, we find
\begin{align}
    |\Gamma(x+\di y)|^2 = 2\,\Gamma(2x) \int_{-\infty}^{\infty} \dd \tau \,\frac{\de^{2\di y \tau}}{(2 \cosh \tau)^{2x}} \,. 
\end{align}
We use this formulation to re-express $\alpha_k$:
\begin{align}
    \alpha_k &=  \frac{2^{4 \frak{p}(k) - k+1}  \left(1+2k \right)^{2[1-\frak{p}(k)]} k!\,\de^{-\frac{\pi (1+2k)}{4}} }{27 \pi^{3/2} (\lfloor k/2 \rfloor !)^2}  \\
    &\qquad\times\,\Gamma\left(\frac{1+2\frak{p}(k)}{2}\right) \int_{-\infty}^{\infty} \dd \tau \,\frac{\de^{\frac{\di (1+2k)\tau}{2}}}{(2 \cosh \tau)^{ \frac{1+2\frak{p}(k) }{2}}}\,. \nonumber
\end{align}
It is now convenient to sum separately over even and odd values of $k$. Permuting the sum and the integral, we find
\begin{subequations}\begin{align}
    \sum_{k \text{ even}} \alpha_k &= \frac{2 \de^{-\frac{\pi}{4}}}{27 \pi } \int_{-\infty}^{\infty} \dd \tau \,\frac{\de^{\frac{\di\tau}{2}}}{\sqrt{2 \cosh \tau}} \nonumber \\*
    &\qquad \times\sum_{n = 0}^{\infty}\frac{ \left(1+4n \right)^{2} (2n)!}{4^{n} (n!)^2}  \de^{(2 \di \tau- \pi)n} \,,\\
    \sum_{k \text{ odd}} \alpha_k &=  \frac{8 \de^{-\frac{3\pi}{4}}   }{27 \pi } \int_{-\infty}^{\infty} \dd \tau  \,\frac{\de^{\frac{3\di\tau}{2}}}{(2 \cosh \tau)^{ \frac{3}{2}}} \nonumber \\*
    &\qquad\times \sum_{n=0}^\infty   \frac{ (2n+1)!}{4^{n}(n!)^2}   \de^{(2\di \tau -\pi) n}\,.   
\end{align}\end{subequations}
These sums can be performed exactly:
\begin{subequations}\begin{align}
    \sum_{n = 0}^{\infty}\frac{ \left(1+4n \right)^{2} (2n)!}{4^{n} (n!)^2}  \xi^n &= \frac{1+10 \xi + \xi^2}{(1-\xi)^{5/2}}\,, \\
    \sum_{n=0}^\infty   \frac{ (2n+1)!}{4^{n}(n!)^2}    \xi^n &= \frac{1}{(1-\xi)^{3/2}}\,,
\end{align}\end{subequations}
where we have in our case $\xi =  \de^{2\di \tau -\pi} $. Summing the odd and even contributions together, we find
\begin{align}
    &\hspace{-10pt}\alpha =  \frac{2 \de^{-\frac{\pi}{4}}}{27 \pi } \int_{-\infty}^{\infty} \dd \tau \Bigg\{\frac{\de^{\frac{\di\tau}{2}}}{\sqrt{2 \cosh \tau}} \frac{1+10 \de^{2\di \tau -\pi} + \de^{2(2\di \tau -\pi)}}{(1-\de^{2\di \tau -\pi})^{5/2}} \nonumber \\
    &\qquad\qquad\qquad\,\,+ \frac{\de^{\frac{3\di \tau}{2}}}{(2 \cosh \tau)^{3/2}} \frac{4 \de^{-\frac{\pi}{2}} }{(1-\de^{2\di \tau -\pi})^{3/2}} \Bigg\}\,,
\end{align}
Despite its appearance, this expression is real valued as required because the imaginary part of the integrand is odd in $\tau$ and integrates to zero. We can thus only keep the real part of the integrand and reduce the integration domain to positive $\tau$. After some lengthy algebra, we find
\begin{equation}
\boxed{
\begin{aligned}
    \alpha &= \frac{2 \sqrt{2} \,   }{27 \pi} \int_{0}^{+\infty} \!\!\!\dd \tau \Bigg\{ \!
   \frac{2^{1/4} \cos \big(3 \mathcal{A}\big)}{\mathcal{B}^{3}
  \mathcal{C}^{3}} +\frac{\cos
   (\mathcal{A})}{2^{1/4}\,\mathcal{B}} \left(\frac{3}{\mathcal{C}^{5}}-\frac{1}{\mathcal{C}}\right)   \!\Bigg\} \!\!\!\!\!\!\!
\end{aligned}}\,,
\end{equation}
where 
\begin{subequations}\begin{align}
    \mathcal{A}(\tau) &= \frac{\tau}{2}+\frac{1}{2} \arctan
 \left(\frac{ \sin (2
   \tau)}{\de^{\pi }- \cos (2 \tau)}\right)\,,\\
   \mathcal{B}(\tau) &=  \Big(\cosh \pi-\cos (2 \tau)\Big)^{1/4}\,, \\
   \mathcal{C}(\tau) &= \sqrt{\cosh \tau}\,.
\end{align}\end{subequations}

\bibliography{references} 

\end{document}